\documentstyle[11pt,aaspp4]{article}

\catcode`\@=11
\long\def\@makefntext#1{
\protect\noindent \hbox to 3.2pt {\hskip-.9pt
$^{{\ninerm\@thefnmark}}$\hfil}#1\hfill}

\def\@makefnmark{\hbox to 0pt{$^{\@thefnmark}$\hss}}

\def\ps@myheadings{\let\@mkboth\@gobbletwo
\def\@oddhead{\hbox{}
\rightmark\hfil\ninerm\thepage}
\def\@oddfoot{}\def\@evenhead{\ninerm\thepage\hfil
\leftmark\hbox{}}\def\@evenfoot{}
\def\sectionmark##1{}\def\subsectionmark##1{}}

\@ifundefined{epsfbox}{\@input{epsf.sty}}{\relax}

\def\eps@scaling{.95}
\def\epsscale#1{\gdef\eps@scaling{#1}}

\def\plotone#1{\centering \leavevmode
    \epsfxsize=\eps@scaling\columnwidth \epsfbox{#1}}

\def\plotfiddle#1#2#3#4#5#6#7{\centering \leavevmode
    \vbox to#2{\rule{0pt}{#2}}
    \includegraphics{#1}}

\def\spose#1{\hbox to 0pt{#1\hss}}
\def\simlt{\mathrel{\spose{\lower 3pt\hbox{$\mathchar"218$}}
     \raise 2.0pt\hbox{$\mathchar"13C$}}}
\def\simgt{\mathrel{\spose{\lower 3pt\hbox{$\mathchar"218$}}
     \raise 2.0pt\hbox{$\mathchar"13E$}}}
\def\lsim{\rlap{$<$}{\lower 1.0ex\hbox{$\sim$}}}
\def\gsim{\rlap{$>$}{\lower 1.0ex\hbox{$\sim$}}}
\def\etal{et al.}
\def\kms{km~s$^{-1}$}
\def\refs{\leftskip=.3truein\parindent=-.3truein}
\def\unrefs{\leftskip=0.0truein\parindent=20pt}

\newcommand{\Lya}{\mbox{Ly$\alpha$}}

\newcommand{\mgii}{\mbox{Mg {\sc ii}}}
\newcommand{\nv}{\mbox{N {\sc v}}}
\newcommand{\civ}{\mbox{C {\sc iv}}}
\newcommand{\ciii}{\mbox{C {\sc iii}}}

\newcommand{\aliii}{\mbox{Al {\sc iii}}}
\newcommand{\siii}{\mbox{Si {\sc ii}}}

\newcommand{\siiv}{\mbox{Si {\sc iv}}}

\def \m.	{\rlap{$.$}^{\rm m}}
\def \s.	{\rlap{$.$}^{\rm s}}
\def \am.	{\rlap{$.$}'}
\def \as.	{\rlap{$.$}''}

\newcounter{sectionc}\newcounter{subsectionc}\newcounter{subsubsectionc}
\renewcommand{\section}[1] {\vspace{0.6cm}\addtocounter{sectionc}{1}
\setcounter{subsectionc}{0}\setcounter{subsubsectionc}{0}\noindent
	{\bf\thesectionc. #1}\par\vspace{0.4cm}}
\renewcommand{\subsection}[1] {\vspace{0.6cm}\addtocounter{subsectionc}{1}
	\setcounter{subsubsectionc}{0}\noindent
	{\it\thesectionc.\thesubsectionc. #1}\par\vspace{0.4cm}}
\renewcommand{\subsubsection}[1]
{\vspace{0.6cm}\addtocounter{subsubsectionc}{1}
	\noindent {\rm\thesectionc.\thesubsectionc.\thesubsubsectionc.
	#1}\par\vspace{0.4cm}}

\newcounter{appendixc}
\newcounter{subappendixc}[appendixc]
\newcounter{subsubappendixc}[subappendixc]

\renewcommand{\appendix}[1] {\vspace{0.6cm}
        \refstepcounter{appendixc}
        \setcounter{figure}{0}
        \setcounter{table}{0}
        \setcounter{equation}{0}
        \renewcommand{\thefigure}{\Alph{appendixc}.\arabic{figure}}
        \renewcommand{\thetable}{\Alph{appendixc}.\arabic{table}}
        \renewcommand{\theappendixc}{\Alph{appendixc}}
        \renewcommand{\theequation}{\Alph{appendixc}.\arabic{equation}}
        \noindent{\bf Appendix \theappendixc #1}\par\vspace{0.4cm}}

\newcounter{itemlistc}
\newcounter{romanlistc}
\newcounter{alphlistc}
\newcounter{arabiclistc}

\newcommand{\fcaption}[1]{
        \refstepcounter{figure}
        \setbox\@tempboxa = \hbox{\tenrm Fig.~\thefigure. #1}
        \ifdim \wd\@tempboxa > 6in
           {\begin{center}
        \parbox{6in}{\tenrm\baselineskip=12pt Fig.~\thefigure. #1 }
            \end{center}}
        \else
             {\begin{center}
             {\tenrm Fig.~\thefigure. #1}
              \end{center}}
        \fi}

\newcommand{\tcaption}[1]{
        \refstepcounter{table}
        \setbox\@tempboxa = \hbox{\tenrm Table~\thetable. #1}
        \ifdim \wd\@tempboxa > 6in
           {\begin{center}
        \parbox{6in}{\tenrm\baselineskip=12pt Table~\thetable. #1 }
            \end{center}}
        \else
             {\begin{center}
             {\tenrm Table~\thetable. #1}
              \end{center}}
        \fi}

\def\@citex[#1]#2{\if@filesw\immediate\write\@auxout
	{\string\citation{#2}}\fi
\def\@citea{}\@cite{\@for\@citeb:=#2\do
	{\@citea\def\@citea{,}\@ifundefined
	{b@\@citeb}{{\bf ?}\@warning
	{Citation `\@citeb' on page \thepage \space undefined}}
	{\csname b@\@citeb\endcsname}}}{#1}}

\newif\if@cghi
\def\cite{\@cghitrue\@ifnextchar [{\@tempswatrue
	\@citex}{\@tempswafalse\@citex[]}}
\def\citelow{\@cghifalse\@ifnextchar [{\@tempswatrue
	\@citex}{\@tempswafalse\@citex[]}}
\def\@cite#1#2{{$\null^{#1}$\if@tempswa\typeout
	{IJCGA warning: optional citation argument
	ignored: `#2'} \fi}}

\def\fnt#1#2{\footnotetext{\kern-.3em
	{$^{\mbox{\sevenrm #1}}$}{#2}}}

 1
 1
 1

\font\tenrm=cmr10

\font\ninerm=cmr9

\received{ }
\revised{REVISION DATE}
\accepted{ }
\journalid{VOL}{JOURNAL DATE}
\articleid{START PAGE}{END PAGE}
\paperid{ID}
\ccc{CODE}
\slugcomment{accepted by ApJ Supplement}
\lefthead{Williger, Hazard, Baldwin \& McMahon}
\righthead{Large-Scale Structure at $z \sim 2.5$}

\begin{document}
\pagenumbering{arabic}

\title{Large-Scale Structure at {\boldmath $z \sim 2.5$}}
\author{G.M. Williger\altaffilmark{1,2}, C. Hazard\altaffilmark{3,4},
J.A. Baldwin\altaffilmark{1} \& R.G. McMahon\altaffilmark{4}}

\altaffiltext{0}{$^1$Cerro Tololo Inter-American Observatory, Casilla
603, La Serena, Chile.  Operated by the Association of Universities for
Research in Astronomy (AURA), Inc., under cooperative agreement with
the National Science Foundation.}
\altaffiltext{0}{$^2$present address: Max-Planck-Institut f\"ur Astronomie,
K\"onigstuhl 17, D-69117, Heidelberg, Germany}
\altaffiltext{0}{$^3$Dept. of Physics \& Astronomy, Univ. of Pittsburgh,
Pittsburgh, PA 15260}
\altaffiltext{0}{$^4$Institute of Astronomy, Madingley Road, Cambridge CB3
0HA, England}


\abstract{
We have made a statistically complete, unbiased survey of \civ\ systems
toward a region of high QSO density near the South Galactic Pole using
25 lines of sight spanning $1.5<z<2.8$.  Such a survey makes an
excellent probe of large-scale structure at early epochs.  We find
evidence for structure on the $15-35h^{-1}$ proper Mpc scale ($H_0
\equiv 100$ \kms\ Mpc${-1}$) as determined by the two point
\civ-\civ\ absorber correlation function, and reject the null
hypothesis that \civ\ systems are distributed randomly on such scales
at the $\sim 3.5\sigma$ level.  The structure likely reflects the
distance between two groups of absorbers subtending $\sim~ 13 \times 5
\times 21h^{-3}$ and $\sim 7 \times 1 \times 15h^{-3}$ Mpc$^3$ at
$z\sim 2.3$ and $z \sim 2.5$ respectively.  There is also a marginal trend for
the association of high rest equivalent width \civ\ absorbers and QSOs
at similar redshifts but along different lines of sight.  The total
number of \civ\ systems detected is consistent with that which would be
expected based on a survey using many widely separated lines of
sight.  Using the same data, we also find 11 \mgii\ absorbers in a complete
survey toward 24 lines of sight; there is no evidence for
\mgii-\mgii\ or \mgii-QSO clustering, though the sample size is likely
still small to detect such structure if it exists.
}

\vfill\eject
\section{Introduction}

QSO absorption lines provide the means to study large-scale structure
far beyond the $z \simlt 0.5$ limits of current galaxy surveys.   They
are a natural extension of pencil beam galaxy surveys  (Broadhurst
\etal\ 1990) which claim to find structure at $z < 0.5$, and which have
sparked interest in ``foam" models for large scale structure (e.g. Icke
\& van de Weygaert 1987, Coles 1990, van de Weygaert 1991, SubbaRao \&
Szalay 1992).  However, regions of sufficient surface density of
bright, high-$z$ QSOs which can provide enough systems to outline structure in
three dimensions at high redshift and to discriminate between some of
the various models (Kaiser \& Peacock 1991) are rare.  In this paper we
investigate the region with the highest known surface density
of bright, high redshift QSOs, mapping structure at $z>1.5$
using \civ\ absorption systems.

Earlier work of this type explored the common \civ\ absorption in the
vicinity of Tol~1037-2703 and Tol~1038-2712 (Jakobsen \& Perryman
1992).  The 17 CIV systems at $1.8 \simlt z \simlt 2.2$ extend over
nearly a degree, outlining a sheet-like structure $5-10 \times
40-50h^{-2}$Mpc$^2$ seen nearly edge-on ($H_0 = 100h$ \kms\ Mpc$^{-1}$,
$q_0=0.5$ and $\Lambda=0$ are assumed throughout).
A similar study was
made toward PKS~0237-233.  Foltz et al. (1993) selected it {\it a
priori}\, on the evidence of a \civ\ cluster (Heisler, Hogan \& White 1989)
found in a survey spanning
many widely distributed lines of sight
by Sargent, Boksenberg, \& Steidel (1988, hereafter SBS).  There appears
to be a spatial overdensity of CIV systems over $1.58 \simlt z \simlt
1.67$ along the lines of sight to 6 high $z$ QSOs with $12-16$
\civ\ systems spanning up to $65h^{-1}$Mpc, but most of the absorption
is seen only toward two QSOs.

The new QSO field which forms the basis for the present work was found
by Hazard et al. (1995) from an objective prism survey in the South
Galactic Cap.  There are 25 confirmed QSOs within a $\sim 1$ deg$^2$ region
having redshifts $1.5 \simlt z \simlt 3.4$.  These QSOs are more closely
spaced, brighter and at higher redshift than the QSO fields studied by
Jakobsen \& Perryman (1992) and Foltz et al. (1993).  Therefore, it is
possible to search a relatively longer baseline for \civ\ systems at
higher spectral and spatial resolution than in previous studies.  The
results are then compared with general \civ\ properties determined from
SBS and Steidel (1990).  Only two of the QSOs in this SGP region have
previously been surveyed for \civ\ absorption (Steidel 1990).  A study
of this region will indicate whether or not such \civ\ absorber
associations as those in the Tol~$1037-2703$/$1038-2712$ and
PKS~$0237-233$ fields are common.  It is also useful for studying the
distribution of \Lya\ forest lines along closely spaced lines of sight.
We present results for a
spectroscopic survey of these QSOs to search for \civ\ absorbers as
indications of large scale structure.  The observations are described
in \S~2.  The absorption system selection and descriptions of
individual spectra are in \S~3.  The results of various
cross-correlations are described and discussed in \S~4, with
conclusions in \S~5.

\section{Observations and Reductions}
\label{sec-observations}

	The observations were made on the CTIO 4m telescope over two observing
sessions covering 1992 September 25-27 and 1993 October 14-17 respectively,
under conditions of variable transparency with $1.2-3.0$ arcsec
seeing. The data were taken using the Argus multifiber spectrograph in
conjunction with a Reticon CCD. Two 632 line/mm gratings were used to
permit data coverage spanning 1120 \AA\ in a single exposure, using
three different bandpasses
(3835--4954, 4594--5713, 4810--5926 \AA ).
Nine separate Argus pointings were used, all
centered around RA 00$^{\rm h}$ 42$^{\rm m}$,  dec $-26^\circ 40'$.  This
permitted observations of
all previously confirmed QSOs with $z > 1.5$ in the Hazard et al. field to
be observed in the unvignetted 48 arcmin diameter Argus field of view.
The remaining fibers were used to observe additional QSOs and QSO candidates
from
the Hazard et al. list as well as a number of faint QSOs with $z\simgt 3$
from the deep multi-color surveys of Warren, Hewett, \& Osmer (1991).
Coordinates for the fiber positioning were measured with the Automated
Plate Measuring machine at
Cambridge, England (Kibblewhite et al. 1984).  The data were extracted
optimally using a routine adapted from that used in Rauch
\etal\ (1992).   The variance for each pixel was determined based on
photon counting statistics from the object, sky and readout noise.
Final spectra for each object were formed by adding the sky-subtracted,
extinction corrected, flux calibrated spectra from each frame, using
inverse variance weighting.  The resolution is $\sim 2$\AA .
In all, spectra were obtained
for a total of 26 QSOs. Their positions, the best estimate that we have
for the emission line redshifts and the relevant exposure times are
given in Table 1. The surface distribution of the QSOs
is illustrated in Figure 1, while the individual spectra are presented
in Figure 2.

\section{Spectra and Absorption Systems} \label{sec-absorption}

\subsection{Individual Spectra}

Although the primary purpose of this study is to study large-scale
structure using \civ\ absorption systems, we note a few other items of
interest concerning the spectra.  Overall, we do not detect any
candidate damped \Lya\ absorbers.  We consider spectra ranges for the
11 QSOs where the signal-to-noise ratio is sufficient for a 4$\sigma$
detection of a \Lya\ line of rest equivalent width $W_0=5.0$\AA , up to
3000 \kms\ from \Lya\ emission.  Using the statistics of Lanzetta et
al.  (1991), with redshift density $d{\cal N}/dz = N_0 (1+z)^\gamma $,
$N_0 = 0.16\pm 0.03$, $\gamma = 0.3\pm1.4$, we expect 0.2, 1.3, 7.8
damped \Lya\ systems for $N_0=0.16$ and $\gamma = -1.1, 0.3, 1.7$.  We
conclude that the region is not anomalously under-dense in damped \Lya\
systems.  We find one obvious BAL in the sample of 26 QSOs, consistent
with a frequency of occurrence of $\sim 10$\% ;
these objects are worthy of further
study with greater resolution and spectral coverage.  We
call attention to the following spectra.

Q$0041-2658$ ($z_{em}=2.457$): There is a complex absorption
feature to the blue of \civ\ $\lambda\lambda 1548,1550$ which
corresponds in extent and shape to an absorption feature at the
corresponding position of \nv\ $\lambda \lambda 1238,1242$ as well as
\Lya\ absorption.  However, the line ratio at 4255--4275
\AA\ (near where associated \nv\ $\lambda\lambda 1238,1242$ would be
expected) is consistent with that of \mgii\ $\lambda\lambda
2796,2803$.  Also, no \siiv\ $\lambda\lambda 1393,1402$ nor
\siii\ $\lambda 1260$ are found at $z=2.4264,2.4383$.  The automated
search procedure may not give consistently good results over such
regions, and so only the general troughs are noted.

Q$0042-2656$ ($z_{em}=3.33$): There is a high equivalent width
\Lya\ line at $z=3.3303$ (rest equivalent width $W_0=1.65$\AA ) which
may be associated with the QSO.  However, a search for metals at the
same redshift using Morton's (1991) line list only reveals a match for
\siii\ $\lambda 1304$.  This would be coincident with \civ\ $\lambda
1548$ at $z=2.6489$, the presence of which is strongly supported by
\Lya\ absorption with rest equivalent width $W_0=1.39$\AA .

Q$0042-2714$ ($z_{em}=2.36$): There is strong \Lya\ absorption just
blueward of \Lya\ emission, which may indicate strong associated absorption.
Unfortunately, wavelength coverage is insufficient to confirm this with
\civ\ $\lambda 1550$.

Q$0043-2647$ ($z_{em}=2.12$):  This is an extreme example of the BAL
phenomenon.  We have excluded the spectrum blueward of 4800 \AA\ from
our analysis, with the region from $4250-4800$ \AA\ dominated by \civ\
and \siiv\ features, with additional BAL absorption further to the blue.
We have also not attempted to deblend any of the lines in the region
$4580-4800$ \AA .

\centerline{\it Absorption Line Measurements}

The main objective of this work is an investigation of clustering
using intervening \civ\ absorbers. However, because our sample contains
so many high redshift QSOs, it also provides much valuable data of
\Lya\ forest absorption systems. Therefore, we have not confined our
analysis to the red of \Lya\ but when possible have also produced lists
of \Lya\ forest lines. The presence of high equivalent width
\Lya\ forest lines is also useful to check independently the validity
of suspected metal absorption systems.  Absorption lines were searched
for with automated software using the statistically based procedure
described in Young et al. (1979).  All features significant at the
4$\sigma$ level were noted.  For metal line doublets, features with
both components at the $\geq 3\sigma$ significance are used, subject to
their being consistent with laboratory wavelength and equivalent width
ratios.  This is more significant than an isolated $3\sigma$ absorption
line.  Standard IRAF\footnote{IRAF is distributed by National Optical
Astronomy Observatories} routines were used interactively to deblend a
subset of metal line complexes ($24$\% of all metal doublets).  We note
that Q$0041-2638$ ($z_{em}=3.053$) and Q$0042-2627$ ($z_{em}=3.289$)
were observed by Steidel, providing an overlap with our linelists for
$5056-5827$ and $5120-5793$ \AA\ respectively.
Quantitatively, of the 33 lines in Steidel's data in those two
wavelength intervals, we list lines or blends for 29 of
them.\footnote{We detect a line toward Q$0041-2638$ at
$5077.52\pm0.46$\AA\ with observed equivalent width
$0.31\pm0.08$\AA\ at $3.7\sigma$ significance.} Excluding the three
lines from Steidel which we do not detect and 7 which are noticeably
blended in our lower resolution data, the remaining 23 differ from
Steidel's values by a mean of $0.4\sigma$ in wavelength and $0.8\sigma$
in equivalent width.  The automated search procedures have limited
efficiency in complicated troughs; as the intrinsic absorption of BAL
QSOs is not relevant to our program, we have not attempted any analysis
in the region of BAL troughs. It is estimated that there are typical
variations of $10-20$\% in equivalent widths due to continuum
uncertainties.  All wavelengths used were transformed to the vacuum
heliocentric frame.

\subsection{\civ\ Systems}

The method used by SBS and Steidel (1990) was employed to select \civ\
absorbers, so as to use their large sample as a comparison data base:
all line pairs between \Lya\ and \civ\ emission were noted which had
both components significant at the $\geq 3\sigma$ level, with
wavelength and equivalent width ratios consistent with those of the
\civ\ $\lambda\lambda 1550$ doublet.  Complexes with component
separation $\Delta v \leq 150$\kms\ were counted as one system.

{}From among the \civ\ systems identified in the above manner, a subset
was excluded from further analysis.  We first omit ``associated
absorption systems" within 5000 \kms\ of the emission redshift (Foltz
et al. 1986).  Such systems
appear to be associated with the background QSOs themselves.  In order
to use the statistics of SBS and Steidel (1990), we construct our
samples using the same rest equivalent width detection limits as SBS
and Steidel in order to compare our results to their statistics.  The
``weak" \civ\ survey consists of absorbers in regions of spectra where
the signal-to-noise ratio allows the rest equivalent width detection
limit for the \civ\ $\lambda 1550$ line at the 3$\sigma$ level to be
$W_0=0.15$\AA .  In a similar manner, using a rest equivalent width
detection threshold of $W_0=0.30$\AA\ produces a sample of ``strong"
\civ\ systems.

The weak survey contains 22 \civ\ systems along 12 lines of sight with
$\langle z_{abs}\rangle = 2.34$ (Fig. 3).  The strong survey contains
12 \civ\ systems along 18 lines of sight with $\langle z_{abs} \rangle
= 2.18$.  Eight \civ\ systems are common to both samples. In every case
for both surveys where the wavelength coverage permits, there is strong
\Lya\ absorption identified at the absorber redshift.

\subsection{Other Heavy Element Systems}

Doublets of \nv\ $\lambda$ 1240\AA , \siiv\ 1400\AA ,
\aliii\ 1860\AA\ and \mgii\ 2800\AA\ were similarly searched for.  A
complete list of all 4$\sigma$ absorption features found in the
spectra, along with the 3$\sigma$ features used to identify \civ ,
\mgii , \siiv\ and \aliii\ doublets, is in Table 2.  The metal systems
identified by \civ , \mgii , \siiv\ and \aliii\ doublets redward of
\Lya\ emission were cross-correlated with heavy element transition
wavelengths from Morton (1991) in order to find additional metal lines
associated with those absorbers.

Individual metal systems are summarized in Table 3. \civ\ systems in
the strong and weak surveys are listed in Table 4, marked with S, W or
both.

\section{ Results and Discussion}
\label{sec-results}

\subsection{\civ\ Absorber-Absorber Correlations}

It is possible to estimate the expected number of \civ\ systems in our
survey by using Steidel's (1990) sample ES2, which has selection
criteria consistent with our weak survey.  Steidel approximated the
redshift density of ES2 with $d{\cal N}/dz \propto (1+z)^\gamma $,
$\gamma = -1.26\pm0.56$; he calculated the mean number density of
\civ\ systems per unit redshift interval at $\langle z_{abs}\rangle
=2.134$ to be $N_0=2.44\pm 0.29$.  If we take $d{\cal N}/dz$ to be
constant for $z<2.0$, then the expected number of \civ\ systems in our
weak survey is $13.3^{+2.0}_{-1.9}$, using the mean and extreme values
of $\gamma$ and $N_0$.  Thus there is a \civ\ overdensity of
$1.9^{+0.4}_{-0.5}\sigma$ (with the uncertainty being an overestimate) toward
the SGP region.  Using the KS test, the 22 observed \civ\ systems are
consistent with the broken power law redshift distribution found by
Steidel for $\gamma = -1.26 \pm 0.56$.  The probability of the data
arising from the distribution for $-1.82 < \gamma < -0.70$ is $>16$\%
.  The strong sample is too small for the KS test.  Using
a pure power law with $\gamma=-1.84\pm 0.68$, $N_0=1.31\pm 0.16$,
$\langle z_{abs} \rangle = 1.998$ (from Steidel's ES5), we expect
$9.1\pm 1.3$ strong systems and count 12, so we find no significant
overdensity for them.

To test for clustering, we calculate the two point correlation function
in three dimensions.  The estimator is that used by Davis \& Peebles
(1983), in which the observed data are cross-correlated with a randomly
generated dataset to provide the normalization for the distribution of
\civ\ - \civ\ pairs from the observed data.  Synthetic datasets were
generated as follows.  The expected \civ\ absorber redshift density
function $d{\cal N}/dz = {\cal N}_0 (1+z)^\gamma$ for each line of
sight was calculated as in the above paragraph for the strong and weak
samples.  The distributions for each line of sight were then integrated
over redshift, making a cumulative distribution as a function of $z$
for each line of sight which was then
normalized by the sum of the cumulative distributions for all lines of sight
\begin{equation}
g_i(z) = \frac{\int_{z_{0,i}}^z  d{\cal N}/dz}
{\sum_{i=1}^n \int_{z_{0,i}}^z  d{\cal N}/dz}
\end{equation}
Here  $z_{0,i}\leq z \leq z_{1,i}$ for
redshift sensitivity limits $z_{0,i},z_{1,i}$ for line of sight $i$, and $n$
is the total number of lines of sight used in the strong or weak
survey.  The cumulative distributions $g_i(z)$ for
each line of sight
were then summed
\begin{equation}
h_i(z) = \left\{ \begin{array}{ll}
                 g_i(z)  & \mbox{if $i=1$}  \\
                 g_i(z) + \sum_{j=1}^{i-1} g_j(z_{1,j}) & \mbox{if $i\geq 2$}
                  \end{array}
	          \right.
\end{equation}
In this way, the
distributions of the expected \civ\ redshift density along each line of
sight have been concatenated to map into a single 1-dimensional interval
from 0 to 1,
representing the expected number of \civ\ systems as a function of
redshift over the whole set of RA, declination and redshift for which
the data are complete to the samples' equivalent width limits.

Table~5
shows, separately for the strong and weak surveys, the expected number
of systems per line of sight and the range within the cumulative
distribution which is contained in each individual line of sight.
Choosing a random number between 0 and 1 and finding the point
where the cumulative distribution $h_i(z)$ has that value corresponds
to choosing randomly a point at redshift $z$ along line of sight $i$
among the combined lines of sight.  Doing
this 12 (22) times creates a synthetic dataset of the same size as the
observed strong (weak) dataset, but which has a Poisson distribution
about the expected number of systems at each point in redshift space
along a given line of sight.

This synthetic dataset was then
cross-correlated with the observed data by calculating the distance $\ell $
between every absorber in the synthetic dataset and every absorber in
the observed dataset.  The pair separation
$\ell_{12}$ for systems at redshifts $z_1, z_2$ and angular separation
$\alpha$ are calculated by the law of cosines
\begin{equation}
\ell_{12} = \frac{(r_1^2 + r_2^2 - 2 r_1 r_2 \cos \alpha)^{1/2}}{1+z_{12}}
\end{equation}
where
$z_{12} = (z_1+z_2)/2$, and $r$ is the proper distance in a $q_0=0.5$,
$\Lambda = 0$ cosmology
\begin{equation}
r = \frac{c}{H_0 q_0^2 } \{ q_0 z + [q_0 - 1] [(1+2 q_0 z)^{1/2} - 1 ] \}
\end{equation}
as presented by Misner, Thorne, \& Wheeler (1973) and applied by Crotts
(1985).  The distribution of distances between the
absorbers in a particular synthetic dataset and the observed data
$d{\cal N}/d\ell $ was
 recorded, using bins of $5h^{-1}$ proper or $15h^{-1}$ comoving
Mpc. In the nomenclature of Davis \& Peebles, the correlation function is
then calculated as $\xi(\ell) = \frac{DD(\ell)}{DR(\ell)} - 1$, where
$DD(\ell)$ refers to the distribution of number pairs with distance
$d{\cal N}/d\ell $ from the autocorrelation of the observed
\civ\ system pairs, and $DR(\ell)$ refers to that from the
cross-correlation of observed-synthetic \civ\ pairs.

The total number of pairs in the synthetic-observed cross-correlation
is larger than that in the observed-observed autocorrelation ($N_c^2$
{\rm versus} $N_c (N_c-1)/2$, for number of \civ\ absorbers in the
sample $N_c$), so the number of absorber pairs falling in each distance
bin for $d{\cal N}/d\ell $ is normalized by a factor $N_c (N_c
-1)/(2N_c^2)$.  A total of $10^3$ such synthetic datasets were
generated.  In this way the expected mean and first moment about the
mean for each bin in the distance distribution was estimated.

The distribution of the number of absorber pairs with distance, $d{\cal
N}/d\ell $, can be understood by noting that the geometry of the space
which is explored is at first approximation cylindrical (Fig. 3).  The
redshift density is a slowly changing function of $z$ for the redshifts
considered ($1.5<z<2.8$) and each line of sight covers a different
redshift range.  It would be expected that there would be more pairs
where most lines of sight overlap in redshift space, at distances
corresponding to the width of the cylinder, as compared to larger
distances.  The field width is about a degree, which is $13-15h^{-1}$
proper Mpc at $1.5<z<2.8$.  Very small distances can only be probed along the
same line of
sight.  The closest pair of lines of
sight for the weak survey is Q$0042-2656$ ($z_{em}=3.33$) and
Q$0042-2657$ ($z_{em}=2.898$), which both probe $2.4<z<2.8$, separated by
6 arcmin ($1.3-1.4h^{-1}$ proper Mpc);
few \civ-\civ\ pairs are expected at
smaller scales.  This is borne
out in the number of pairs expected in the various bins.  For $N_c=22$
absorbers, 231 pairs are expected, with there being 4.7, 14.6, 17.4,
15.2, 14.0, 13.3, 12.0 pairs expected at separations $0-5$, $5-10$, ...,
$30-35h^{-1}$ proper Mpc.  There is a sharp decrease in the number
of pairs expected for $0-5h^{-1}$ proper Mpc, and the number of
expected pairs peaks at $10-15h^{-1}$ proper Mpc, then decreases
monotonically at larger separations.

The results are shown in Figures 4 and 5, plotted both in proper and
comoving $h^{-1}$ Mpc.  The few isolated peaks in the strong survey
correlation function are $<3\sigma$ and based upon no more than five
pairs in a bin, with the results independent of binning.  For the weak
survey, there are similarly-structured peaks in both plots covering
$15-35h^{-1}$ proper or $60-120h^{-1}$ comoving Mpc, covering
$2.5-3.2\sigma$ or $2.1-2.9\sigma$ per bin respectively.  There is also
an isolated peak in the proper separation plot at $100-105h^{-1}$ Mpc
which does not appear in the comoving separation plot.  It is noted
that the geometry of the region studied is not expected to produce a
smooth distribution in the number of raw pair counts per bin, but this
should be accounted for by the Monte Carlo simulations.

To test the robustness of the results, the two point correlation
function is recalculated for the proper distance case with each of the
twelve lines of sight removed from the sample in turn.  Over
$15-25h^{-1}$ Mpc, there is always a signal for the $15-20$,
$20-25h^{-1}$ Mpc bins of at least $1.1\sigma$,$2.3\sigma$ over the
mean.   The peak at $100-105h^{-1}$ Mpc is removed almost entirely
(only $1.1\sigma$ overdensity) when Q$0041-2607$ is omitted from the
dataset.

A closer examination of the variation of results reveals the causes of
the peaks in the two point correlation function.  The overdensity at
$15-20h^{-1}$ Mpc is reduced the most (to $1.1\sigma$) when Q$0041-2658$
($z_{abs} = 1.8708,2.0217,2.2722,2.3805$) is removed from the sample.
Other QSO removals producing similar effects are Q$0042-2627$ (to
$1.5\sigma$, $z_{abs}=2.4761,2.5070$) and Q$0041-2607$ (to $1.9\sigma$,
$z_{abs}=1.8907,1.9567,2.4012$).  The overdensity at $20-25h^{-1}$ Mpc
is most strongly reduced by taking out Q$0041-2707$ ($2.3\sigma$,
$z_{abs}=2.3429,2.5997$, Q$0042-2656$ ($2.6\sigma$,
$z_{abs}=2.4985,2.6489,2.6853$ and Q$0042-2627$ ($2.7\sigma$).  This
can be interpreted by noting that the absorbers fall into several
groups by redshift:  {\it a}) a pair at $1.8708<z<1.8907$, {\it b}) a
pair at $2.2656<z<2.2722$, {\it c}) three at $2.3122<z<2.3429$, {\it
d}) another pair at $2.3805<z<2.4012$ and {\it e}) five at
$2.4761<z<2.5201$.  The largest extent within these groups is less than
$6h^{-1}$ proper Mpc in redshift space, which dominates pair
separations for distances much larger than the transverse field size
($\sim 15h^{-1}$ proper Mpc).  Interactions or ``beating" between these
groups causes increased signal in the two point correlation function.
For example, the redshift space proper distance range between the two
largest groups {\it c,e} is $19-29h^{-1}$ Mpc, which accounts for 15 of
the 231 absorber-absorber pairs in the entire sample.  Another 30 pairs
are accounted for in beating between groups {\it a,e} ($98-107h^{-1}$
Mpc), {\it b,e} ($29-36h^{-1}$ Mpc and {\it d,e} ($10-19h^{-1}$ Mpc).
Consolidating groups {\it b,c,d} means that 15\% of the
absorber-absorber pairs are accounted for by the beating of two groups
of absorbers, subtending $\sim~ 13 \times 5 \times 21h^{-3}$ and $\sim
7 \times 1 \times 15h^{-3}$ Mpc$^3$ respectively and with a line of
sight proper distance between them of $12-37h^{-1}$ Mpc.  This produces
a noticeable effect in the two point correlation over the range
$15-25h^{-1}$ Mpc, conservatively determined to be significant to at
least the $3.2\sigma$ level.  The feature at $100-105h^{-1}$ Mpc, which
is formally nearly as strong, is interpreted as less significant
because the removal of one line of sight nearly eliminates it, the
total number of pairs upon which the feature is based is much smaller,
and in any case the 1$\sigma$ error bars for the two-point correlation
function tend to be underestimated by Poisson estimates at these large
scales (Hamilton 1993).

SBS and Steidel (1990) found significant correlation at the $200 <
\Delta v < 600$ km s$^{-1}$ scale ($\sim 0.3-1.0h^{-1}$ Mpc at
$z=2.3$), and marginal correlation from $10^3-10^4$ \kms\ ($\sim
2-17h^{-1}$ Mpc at $z=2.3$).  Our survey is less sensitive at small
scales, due to the small number of lines of sight and the minimum
comoving distance between sightlines $\sim 2h^{-1}$ Mpc.  Heisler,
Hogan \& White (1989) found the excess power at large scales in SBS
attributable to the PKS $0237-233$ absorber group.  Conservatively, that group
spans $1.5773 < z < 1.6731$ toward PKS $0237-233$ and Q~$0233-2430$,
subtending $\sim 22 \times 26h^{-2}$ Mpc$^2$, larger than either
association toward the SGP.  It would be extremely useful to combine our data
with that of Jakobsen \& Perryman (1992) and Foltz et al. (1993) to confirm
these results, but their spectra are not of uniformly high enough quality to
make complete spectroscopic searches for \civ\ absorbers as done here.

\subsection{QSO--\civ\ Absorber Correlations}

     It is possible to cross-correlate the strong and weak
\civ\ samples with the QSOs in the field.    We cross-correlate the
weak and strong samples with the 25 QSOs with $z_{em}>1.5$ observed in
the field.   There is no indication for signal on
any scale for either sample (Fig. 6).
We recalculate the absorber-QSO correlation function
neglecting spatial distances (Fig 7).  The weak sample still shows no
significant signal, but the strong sample has a $1.8\sigma$ overdensity
at $0-5h^{-1}$ proper Mpc.

In our strong sample, 14 ($\pm 3.7$) QSO-absorber pairs are found where 8 were
expected.  The total number of adjacent QSOs in the \civ\ fields is only
25.  Therefore,
redshift coincidences or anti-coincidences with \civ\ systems may
be affected by small number statistics.  The fact that including spatial
distance in the
correlation function washes out the signal could mean either that the
structures tend to be sheet-like, or that they are larger than
$5h^{-1}$ proper Mpc and have large peculiar velocities.

To investigate this hypothesis, we count
QSOs within $10h^{-1}$ Mpc projection distance (at
the QSO redshift) of the \civ\ system.  We merged the QSO sample here
with the QSOs from the Hewitt \& Burbidge catalogue (1993) at $z>1.15$
spanning a $3^\circ \times 3^\circ$
field ($00^{\rm h} \, 35^{\rm m} < \alpha < 00^{\rm h} \, 49^{\rm
m}$, $-28^\circ < \delta < -25^\circ$), producing a sample of 63 QSOs
(Table~6).  Making the calculation again with the larger
QSO database makes no significant change in the results.  We then
cross-correlated this list separately with the strong and weak samples
(Fig. 8).  The weak sample has a marginal $2.6\sigma$ signal at +31500
\kms , which arises from beating between the \civ\ group at
$2.47<z<2.52$ and a QSO group of four at $2.15<z<2.18$.  There is a
$2.2\sigma$ overdensity at +4500 \kms\ which can be attributed to
beating between three QSOs at $2.43<z<2.47$ and five \civ\ systems at
$2.47<z<2.52$.  There is no significant signal at 1500 \kms .  This would
be expected if QSO redshifts were systematically blueshifted as a consequence
of the velocity difference between their high and low ionization emission
lines (e.g. Espey et al. 1989).
The strong sample also has  a +28500 \kms\ overdensity
at the $4.1\sigma$ level, arising from 7 QSO-absorber pairs from the
beating of a pair of absorbers at $z_{abs}=2.4761,2.4920$ and QSOs at
$z_{em}=2.15,2.15,2.17,2.18$.  There are $3.2\sigma$ overdensities at
$-6000$ to 0 \kms\ which are attributable
to beating between absorbers at $z_{abs}=2.1285,2.1449$ and QSOs at
$z_{em}=2.15,2.15,2.17$.

M{\o}ller (1995) found a correlation between \civ\ systems and the QSOs
within $10h^{-1}$ Mpc projection distance of \civ\ systems toward QSOs
along widely separated lines of sight.  Specifically, M{\o}ller found a
significant peak in the correlation function 1500 \kms\ blueward of the
QSO emission redshift, but did not find any such effect for weak
\civ\ absorbers.  We have applied M{\o}ller's test to our data and find
a QSO-\civ\ correlation for the strong \civ\ sample, but in the opposite
velocity sense that M{\o}ller found.  Although the number of QSO-strong
\civ\ pairs in our study is larger than in M{\o}ller's survey, we find
a weaker correlation.  The difference is not likely due to the higher
mean redshift of our systems.  Our mean survey redshift for the SGP
strong \civ\ survey is $\langle 2.18 \rangle$; in M{\o}ller's strong
survey it is $\langle 1.85 \rangle$, not a significantly lower
redshift.  Rather, it is likely due to small number statistics, as the
signal at -1500 \kms\ in our sample is dominated by a few QSOs and
strong \civ\ systems, each of which can affect more than one line of sight.

\subsection{\mgii\  Systems}

     In the process of making the \civ\ absorber survey, some data on
\mgii\ systems were also acquired.  As evidence mounts that
\mgii\ systems arise from normal galaxies (Steidel, Dickinson, \& Persson
1994), while
galaxies and QSOs also exhibit correlations (e.g. Komberg \& Lukash
1994), it is useful to examine the \mgii\ absorber distribution.
We
use a detection threshold of 0.6\AA\ for the \mgii\ $\lambda 2796$
line, as in Sargent, Steidel \& Boksenberg (1989).  The 24 lines of
sight and redshift intervals useful
for a $3\sigma$ detection of \mgii~$\lambda 2796$ are listed in
Table~7. Of the 14
\mgii\ systems listed in Table~3, 11 fit the detection criteria; the
$z=1.0532,1.0548$ systems toward Q$0043-2633$ and the $z=0.7805$ system
toward Q$0044-2628$ are excluded.  We employ a \mgii\ system redshift
density of $d{\cal N}/dz \propto (1+z)^\gamma$, $\gamma=1.55\pm 0.52$,
with a mean number of \mgii\ systems per unit redshift of $0.54\pm
0.08$ at a mean redshift of $\langle z_{abs} \rangle = 1.097$ (Sargent, Steidel
\& Boksenberg).  Using
this fit, the number of expected \mgii\ systems is between 4 and 6,
indicating a marginal overdensity of observed absorbers.  The sample size is
smaller than for the strong \civ\ systems while the volume of space
covered is larger.  We find no significant signals in the \mgii-\mgii\ two
point
correlation function, whose results are dominated by small number
statistics.  We note that 6 of the 11 absorbers are members of pairs
with velocity splittings of $\Delta v < 5200$ \kms .
Cross-correlations with QSOs analogous to those for \civ\ systems were
made between the 11 \mgii\ systems in the complete sample and with 116
QSOs from the Hewitt \& Burbidge (1993) catalogue at $00^{\rm h} \,
25^{\rm m} < \alpha < 00^{\rm h} \, 58^{\rm m}$, $-30^\circ < \delta <
-25^\circ$, $0.13<z<1.42$.  No significant indications of structure in
the \mgii-QSO distribution were found, even when expanding the
projection radius search for the M{\o}ller-style calculations to
$20h^{-1}$ and $30h^{-1}$ proper Mpc.

\section{Conclusions}
\label{sec-conclusions}

We have examined $\sim 2$\AA\ resolution
spectra of 25 QSOs with $z_{em}>1.5$ in a $\sim 1^\circ$ region near
the South Galactic Pole, and find the following:

1) In a complete survey to rest equivalent width
$W_0=0.15$\AA\ threshold for the \civ\ $\lambda\lambda1550$ doublet, we
find 22 ``weak" \civ\ absorbers toward 12 QSOs, with velocity separation
at least $\Delta v \geq 5000$ \kms\ from the background QSO.  A similar
``strong"
survey with $W_0=0.30$\AA\ reveals 12 \civ\ absorbers toward 18 QSOs.
We also identify 8 associated absorbers with $\Delta v \leq 5000$ \kms .

2) The weak and strong surveys have been compared to the redshift
distribution of a larger \civ\ survey by Steidel (1990) using widely
scattered lines of sight.  There is a 1.9$\sigma$ overdensity of weak
systems and 0.8$\sigma$ overdensity of strong systems in our sample.
The redshift distribution of both our weak and strong samples is
consistent with that of the Steidel sample.

3) We calculate the two point correlation function in three dimensions
for each \civ\ absorber sample.  There is no significant indication of
clustering in the strong sample.  However, there is an excess of
\civ\-\civ\ pairs in the weak sample at a separation of $15-35h^{-1}$
proper Mpc, at approximately the $3\sigma$ level.  The signal arises
from ``beating" between groups of seven absorbers at $2.26<z<2.40$ and five at
$2.47<z<2.50$.  The scale of the absorber pair excess is up to twice as
large as marginal correlations found by SBS and Steidel (1990).

4) We also cross-correlate the strong and weak \civ\ absorber samples with
63 QSOs in the field, considering at QSOs within $10h^{-1}$ Mpc from
\civ\ systems.  There are several occurrences of beating from
small groups of QSOs and \civ\ absorbers, but these may be attributed to
small number statistics.

5) Using the same QSO spectra, we have made a complete survey for
\mgii\ absorbers with a rest equivalent width detection threshold of
0.6\AA\ for the \mgii\ $\lambda$2796 line, using 24 lines of sight.  We
find 11 \mgii\ systems, a $2.1\sigma$ overdensity over what is expected
using results from a survey by Sargent, Steidel \& Boksenberg using
many widely scattered lines of sight.  There is no evidence for clustering
in the \mgii-\mgii\ two point correlation function, nor with the
\mgii-QSO correlation, though both results may suffer from small statistical
samples.  However, we note that 6 of the \mgii\ systems are members of
pairs with velocity splittings of $\Delta v < 5200$ \kms .

We find evidence at the $3\sigma$ level for groups of \civ\ absorbers
on the $5-10h^{-1}$ Mpc proper distance scale at $2.3\simlt z \simlt 2.5$,
separated by a $15-25h^{-1}$ proper Mpc gap.  We find evidence similar to the
QSO-\civ\ correlations on the $10h^{-1}$ Mpc scale noted by M{\o}ller
(1995), and conjecture that the structures outlined by such
associations are either sheet-like or have large peculiar velocities.
The beamwidth of
$\sim 15h^{-1}$ Mpc for this survey is at the small end of the scale
length where the two point correlation function indicates structure.
We plan to widen the survey to find out whether the apparent groups of
\civ\ absorbers form ``sheets and walls", as is seen in local galaxy
surveys, or more separated structures.

\acknowledgements
{CH acknowledges support from NSF grant AST-86491.  RGM thanks the Royal
Society for support.  CH and RGM acknowledge support from NATO grant
CRG 900300.  We would like to thank R.C.  Smith for help in preparing
the Argus observing program, R.F. Carswell for absorption line
detection software, and N.  Dinshaw, R. Elston, P. M{\o}ller, J.
Peacock, W.L.W.  Sargent, A.  Smette, C.C. Steidel, R. van de Weygaert
and S.D.M. White for useful discussions, and acknowledge use of the
NASA Extragalactic Database and the Penn State Statistical Consulting
Center for Astronomy.  We also thank an anonymous referee for helpful
comments.
}

\clearpage

\noindent{\bf REFERENCES}
\refs

Broadhurst, T. J., Ellis, R. S., Koo, D. C., \& Szalay, A. S.
1990, Nature, 343, 726

Campusano, L. E. 1991, AJ, 102, 502

Coles, P. 1990, Nature, 346, 446

Crotts, A. P. S. 1985, ApJ, 298, 732

Davis, M., \& Peebles, P. J. E. 1983, ApJ, 267, 465

Espey, B. R., Carswell, R. F., Bailey, J. A., Smith, M. G., \&
Ward, M. J. 1989, ApJ, 342, 666

Foltz, C. B., Hewett, P. C., Chaffee, F. H., \& Hogan, C. J. 1993, AJ, 105, 22

Foltz, C. B., Weymann, R. J., Peterson, B. M., Sun, L., Malkan,
M. A., \& Chaffee, F. H. 1986, ApJ, 307, 504

Hamilton, A. J. S. 1993, ApJ, 417, 19

Hazard, C. et al. 1995, in preparation

Heisler, J., Hogan, C. J., \& White, S.D.M. 1989, ApJ, 347, 52

Hewitt, A., \& Burbidge, G. 1993, ApJS, 87, 451

Icke, V. \& van de Weygaert, R. 1987, AA, 184, 16

Jakobsen, P. \& Perryman, M. 1992, ApJ, 392, 432

Kaiser, N. \& Peacock, J. A. 1991, ApJ, 379, 482

Kibblewhite, E. J., Bridgeland, M. T., Bunclark, P. S. \& Irwin, M. J. 1984,
in {\it Astronomical Microdensitometry Conference} NASA-2317, ed.
D. A. Klinglesmith (NASA, Washington D.C.), 277

Komberg, B. V. \& Lukash, V. N. 1994, MNRAS, 269, 277

Lanzetta, K. M., Wolfe, A. M., Turnshek, D. A., Lu, L., McMahon, R. G.,
\& Hazard, C. 1991, ApJS, 77, 1

Misner, C. W., Thorne, K. S. \& Wheeler, J. A. 1973,
{\it Gravitation} (San Francisco: Freeman)

M{\o}ller, P. 1995, A\&A, submitted

Morris, S. L., Weymann, R. J., Anderson, S. F., Hewett, P. C., Foltz, C. B.,
Chaffee, F. H., Francis, P. J., \& MacAlpine, G. M. 1991, AJ, 102, 1627

Morton, D. C. 1991, ApJS, 77, 119

Rauch, M., Carswell, R. F., Chaffee, F. H., Foltz C. B., Webb J. K.,
Weymann R. J., Bechtold J., \& Green R.F. 1992, ApJ, 390, 387

Sargent, W. L. W., Boksenberg, A., \& Steidel, C. C. 1988, ApJS, 68, 539 (SBS)

Sargent, W. L. W., Steidel, C. C., \& Boksenberg, A. 1989, ApJS, 69, 703

Steidel, C. C. 1990, ApJS, 72, 1

Steidel, C. C., Dickinson, M., \& Persson, S.E. 1994, ApJ, 437, L75

SubbaRao, M. U., \& Szalay, A. S. 1992, ApJ, 391, 483

van de Weygaert, R. 1991, MNRAS, 249, 159

Warren, S. J., Hewett, P. C., \& Osmer, P. S. 1991, ApJS 76, 23

Young, P. J., Sargent, W. L. W., Boksenberg, A., Carswell, R. F.,
\& Whelan, J. A. J. 1979, ApJ, 229, 891

\unrefs

\clearpage

{\normalsize
%\begin{planotable}{c cc lr ccc}
\begin{deluxetable}{c cc lr ccc}
\tablecolumns{8}
%\tablewidth{6.5in}
\tablewidth{385pt}
\tablecaption{Observed QSOs\tablenotemark{a}}

\tablehead{\colhead{} &  \colhead{$\alpha_{1950}$} &  \colhead{$\delta_{1950}$} & \colhead{} & \colhead{$z$-ref\tablenotemark{b}} 
& \multicolumn{3}{c}{exposure (s) (\AA )} \cr 
\colhead{Object}	& \colhead{$h$\, $m$\, \, $s$} & \colhead{$\circ$\, \, $'$\, \, $''$} & \multicolumn{1}{c}{$z_{em}$} &   
& \colhead{3835-4954} & \colhead{4594-5713} & \colhead{4810-5926}   }
%\tableline

\startdata
Q$0039-2630$	 & $00\, 39\, 36.32$ & $-26\, 30\, 31.6$ & 1.810 & 1 & 18900	& \nodata  & \nodata  \nl
Q$0039-2636$	 & $00\, 39\, 41.83$ & $-26\, 36\, 04.5$ & 1.56  & 2 & 18900	& \nodata  & \nodata  \nl
Q$0041-2607$	 & $00\, 41\, 31.11$ & $-26\, 07\, 41.7$ & 2.505 & 1 & 18900	& \nodata  & 29400    \nl
Q$0041-2608$	 & $00\, 41\, 17.10$ & $-26\, 08\, 27.5$ & 1.72  & 3 & 18900	& \nodata  & \nodata  \nl
Q$0041-2622$	 & $00\, 41\, 19.58$ & $-26\, 22\, 46.0$ & 2.18  & 4 & 18900 & \nodata  & 29400    \nl
Q$0041-2638$	 & $00\, 41\, 15.19$ & $-26\, 38\, 35.9$ & 3.053 & 1 & 18900	& \nodata  & 32063    \nl
Q$0041-2658$	 & $00\, 41\, 38.38$ & $-26\, 58\, 30.0$ & 2.457 & 1 & 22000	& 20241    & 32063    \nl
Q$0041-2707$	 & $00\, 41\, 24.38$ & $-27\, 07\, 54.3$ & 2.786 & 1 & 22000	& \nodata  & 32063    \nl
Q$0042-2627$	 & $00\, 42\, 06.42$ & $-26\, 27\, 45.3$ & 3.289 & 1 & 18900	& 20241    & 29400    \nl
Q$0042-2632$	 & $00\, 42\, 03.36$ & $-26\, 32\, 23.1$ & 1.79  & 4 & 18900	& 20241    & 44363    \nl
Q$0042-2639$	 & $00\, 42\, 08.20$ & $-26\, 39\, 25.0$ & 2.98  & 5 & \nodata & 20241  & 44363    \nl
Q$0042-2642$	 & $00\, 42\, 43.90$ & $-26\, 42\, 15.0$ & 2.81  & 5 & \nodata & 20241  & \nodata  \nl
Q$0042-2645$	 & $00\, 42\, 52.70$ & $-26\, 45\, 11.2$ & 1.69  & 2 & 22000	& 20241    & 32063    \nl
Q$0042-2651$	 & $00\, 42\, 26.96$ & $-26\, 51\, 44.3$ & 1.71  & 2 & 22000	& 20241    & 32063    \nl
Q$0042-2656$	 & $00\, 42\, 24.89$ & $-26\, 56\, 34.4$ & 3.33  & 5 & 22000	& 20241    & 32063    \nl
Q$0042-2657$	 & $00\, 42\, 52.29$ & $-26\, 57\, 15.3$ & 2.898 & 1 & 22000	& 20241    & 32063    \nl
Q$0042-2712$	 & $00\, 42\, 20.84$ & $-27\, 12\, 07.1$ & 1.79  & 3 & 22000 & \nodata  & 17100    \nl
Q$0042-2714$	 & $00\, 42\, 44.12$ & $-27\, 14\, 56.6$ & 2.36  & 6 & 22000	& \nodata  & \nodata  \nl
Q$0043-2606$	 & $00\, 43\, 48.15$ & $-26\, 06\, 27.1$ & 3.11  & 5 & \nodata & \nodata & 29400   \nl
Q$0043-2633$	 & $00\, 43\, 03.10$ & $-26\, 33\, 33.6$ & 3.44  & 5 & 18900 & 20241    & 61463    \nl
Q$0043-2644$	 & $00\, 43\, 38.43$ & $-26\, 44\, 30.0$ & 1.041 & 7 & \nodata & 20241  & 14963    \nl
Q$0043-2647$	 & $00\, 43\, 11.21$ & $-26\, 47\, 29.6$ & 2.12  & 3\tablenotemark{c} & 22000 & \nodata  & 32063    \nl
Q$0043-2708$	 & $00\, 43\, 28.19$ & $-27\, 08\, 28.6$ & 2.15  & 8 & 22000 & \nodata  & 32063    \nl
Q$0043-2710$	 & $00\, 43\, 47.89$ & $-27\, 10\, 32.5$ & 1.84  & 2 & 22000 & \nodata  & 22000    \nl
Q$0044-2628$	 & $00\, 44\, 14.45$ & $-26\, 28\, 45.3$ & 2.47  & 4 & \nodata & 20241  & 29400    \nl
Q$0044-2701$	 & $00\, 44\, 00.92$ & $-27\, 01\, 07.4$ & 2.16  & 4 & 22000 & 20241    & 32063    \nl
\enddata
\tablenotetext{a}{The object listed as Q$0042-263$ ($z=2.38$) in Hewitt \& Burbidge (1993) 
is identified as an emission line galaxy at $z=0.106$.}
\tablenotetext{b}{Our data are consistent with the following literature
references or are measured using the emission lines indicated.
1)~Morris et al 1991; 
2)~based \civ\ $\lambda \lambda 1548,1550$, \ciii\ $\lambda 1909$;
3)~based on \civ\ $\lambda \lambda 1548,1550$
4)~based on \siiv\ $\lambda \lambda 1394,1403$, \civ\ $\lambda \lambda 1548,1550$
5)~Warren et al. 1991;
6)~based on \Lya\
7)~Campusano 1991;
8)~based on \siiv\ $\lambda \lambda 1394,1403$
}
\tablenotetext{c}{Broad Absorption Line QSO}
%\end{planotable}
\end{deluxetable}
}

\setcounter{table}{2}

{\normalsize
%\begin{planotable}{ccll}
\begin{deluxetable}{ccll}
\tablecolumns{4}
%\tablewidth{6.5in}
\tablewidth{0pc}
\tablecaption{Identified Metal Systems}

\tablehead{  
 \colhead{}          &   \colhead{}         & \multicolumn{1}{c}{metal lines}              &  \colhead{}  \cr
\colhead{QSO}        & \colhead{$z_{abs}$}  & \multicolumn{1}{c}{(?=uncertain, bl=blended)} &\multicolumn{1}{c}{comments}    }
%\tableline

\startdata
Q$0039-2630$	& 1.8157	& CIV 1548, 1550(3.7$\sigma$)	& associated system    \nl
Q$0039-2636$	& \nodata	& \nodata		& \nodata  \nl
Q$0041-2607$	& 1.8907	& CIV 1548,1550	&  \nl
		& 1.9567	& CIV 1548,1550	&  \nl
		& 2.4012	& SiII 1264?, SiIV 1393,1402, CIV 1548,1550	&  \nl
		& 2.5069	& CIV 1548,1550	& associated system \nl
Q$0041-2608$	& 1.6048 	& CIV 1548,1550	&        \nl
		& 1.6947 	& CIV 1548,1550	& associated system       \nl
Q$0041-2622$	& 0.4693	& MgII 2796(bl),2803	&   \nl
		& 0.5118	& MgII 2796,2803(bl)	&   \nl
		& 0.6332	& MgII 2796(bl),2803	&   \nl
		& 0.8070	& MgII 2796,2803	&   \nl
		& 1.6745	& CIV 1548,1550, SI 1807?	&   \nl
		& 2.1973	& SiIV 1393,1402, CIV 1548,1550(bl)	& associated system  \nl
Q$0041-2638$	& 0.8626	& MgII 2796,2803(bl)	&	\nl
		& 2.2656	& CIV 1548,1550		& consistent with Steidel (1990) to 2$\sigma$	\nl
		& 2.3399	& CIV 1548,1550(bl)	& consistent with Steidel (1990) to 2$\sigma$	\nl
		& 2.5177	& CIV 1548,1550		&  Al $\lambda 1670$ at $z=2.2658$ in Steidel	\nl
		& 2.5697	& CIV 1548,1550	& tentative in Steidel 	\nl
		& 2.7409	& SiII 1526(1.6$\sigma$)?, CIV 1548,1550  & 4800 \kms\ from $z=2.7568$ 	\nl
		& 2.7568	& CIV 1548,1550(2.5$\sigma$) & consistent with Steidel (1990) to 2$\sigma$ 	\nl
Q$0041-2658$	& 0.5240	& MgII 2796(bl),2803(bl)	&  \nl
		& 1.8708	& NiII 1467?, CIV 1548,1550	&  \nl
		& 2.0217	& SIII 1425?, CIV 1548,1550, NiII 1709?	&  \nl
		& 2.1548	& CIV 1548(3.2$\sigma$),1550, NiII 1709?	&  \nl
		& 2.2722	& CII 1334, SiIV 1393,1402, 	&  \nl
		&		& CIV 1548,1550, AlII 1670	&  \nl
		& 2.3805	& SII 1253?, CIV 1548(bl),1550	&  \nl
		& 2.4264	& CIV 1548,1550	&  associated complex  \nl
		& 2.4383	& NV 1238?, CIV 1548,1550, CI 1657?	& 3600 \kms\ from $z=2.4264$ \nl
Q$0041-2707$	& 2.3387        & CIV 1548	& 1300 \kms\ from $z=2.3429$ \nl
		& 2.3429        & SiIV 1393,1402, SiII 1526,1533, 	&   \nl
		&               & CIV 1548(bl),1550	&   \nl
		& 2.5997        & CII 1334, SiIV 1393,1402, SiII 1526, 	& CIV lines blended with  \nl
		& 	        &  CIV 1548(bl),1550(bl)	&  telluric OI 5577  \nl
Q$0042-2627$	& 2.0291 	& AlIII 1854,1862 &  \nl
		& 2.4761 	& CIV 1548,1550(bl) &  \nl
		& 2.5070 	& CIV 1548,1550 &  \nl
Q$0042-2632$	& 1.8323	& CIV 1548,1550	&  associated system      \nl
Q$0042-2639$	& 2.1285 	& CIV 1548,1550 &         \nl
		& 2.2271 	& CIV 1548,1550 &         \nl
Q$0042-2642$	& \nodata	& \nodata &  		   \nl
Q$0042-2645$	& \nodata	& \nodata &  		   \nl
Q$0042-2651$	& \nodata	& \nodata &  		   \nl
Q$0042-2656$	& 2.4985	& CIV 1548,1550 &  	 \nl
		& 2.6489	& CIV 1548,1550 &  	 \nl
		& 2.6853	& CIV 1548,1550 &  	 \nl
		& 2.9027	& SiIV 1393,1402 &	 \nl
Q$0042-2657$	& 1.0444	& FeII 2382?,2600? MgII 2796,2803	& 	 \nl
		& 2.3122	& NiII 1467?, CIV 1548,1550	& 	 \nl
		& 2.4920	& CIV 1548,1550	& 	 \nl
		& 2.5201	& SiIV 1393?, SI 1473?, CIV 1548,1550	& 	 \nl
Q$0042-2712$	& \nodata 	& \nodata	&         \nl
Q$0042-2714$	& 0.4693	& MgII 2796,2803 &        \nl
		& 1.8143	& SI 1473?, CIV 1548,1550 &        \nl
Q$0043-2606$	& \nodata	& \nodata &  		   \nl
Q$0043-2633$	& 1.0532	& MgII 2796(3.1$\sigma$),2803 & 480 \kms\ from $z=1.0548$ \nl
		& 1.0548	& MgII 2796,2803 & 	 \nl
		& 3.0450	& SiIV 1393,1402 & 	 \nl
		& 3.4056	& NV 1238,1242 & associated system 	 \tablebreak
Q$0043-2644$	& \nodata	& \nodata & $z_{em}=1.04$  		   \nl
Q$0043-2647$	& 1.0617	& MnII 2594?, MgII2796,2803,	& BAL, no identification      \nl
		& 		& MgI 2852	&  attempts blueward of CIV emission     \nl
Q$0043-2708$	& 0.5782	& FeII 2600?, MgII 2796,2803	& 	       \nl
Q$0043-2710$	& \nodata	& \nodata &  		   \nl
Q$0044-2628$	& 0.6691 	& MgII 2796,2803	&  \nl
		& 0.7805 	& MgII 2796,2803, FeI 3021?, 	&  \nl
		& 	 	& TiII 3242?, NaI 3303?	&  \nl
		& 2.1449 	& CIV 1548,1550	&  \nl
		& 2.2067 	& CIV 1548,1550	&  \nl
		& 2.4683 	& CIV 1548,1550	& associated system \nl
Q$0044-2701$	& \nodata	& \nodata &  		   \nl
\enddata
%\end{tabular}	
%\end{planotable}
\end{deluxetable}
}

%\setcounter{page}{40}
%\setcounter{table}{2}
%\begin{planotable}{llccclccc}
\begin{deluxetable}{llccclccc}
\tablecolumns{9}
%\tablewidth{6.5in}
\tablewidth{0pc}
\tablecaption{Strong and Weak Survey \civ\ Systems}

\tablehead{ \colhead{} &  \colhead{} & \multicolumn{2}{c}{\civ\ $z$ sensitivity} & & \multicolumn{1}{c}{Type} 
& \multicolumn{3}{c}{rest eq width \AA } \cr 
\colhead{Object}		& \colhead{$z_{em}$} & \colhead{strong}   
& \colhead{weak} &
\colhead{$z_{abs}$} & \colhead{} & \colhead{$\lambda$1548} 
& \colhead{$\lambda$1550} & \colhead{Ly$\alpha$} }
%\tableline

\startdata

Q$0039-2630$	& 1.810	 	& $1.478-1.780$	& $1.478-1.780$ &\nodata &	& \nodata &\nodata & \nodata \nl
Q$0041-2607$	& 2.505		& $1.747-2.447$	& $1.747-2.447$ & 1.8907& W	& 0.38	& 0.22	& \nodata \nl
		&  		&		&		& 1.9567& W	& 0.24	& 0.15	& \nodata \nl
		&  		&		&		& 2.4012& W	& 0.29	& 0.18	&  0.66  \nl
Q$0041-2608$	& 1.72	 	& $1.548-1.673$	& \nodata	& 1.6048& S	& 0.88	& 0.44	& \nodata \nl
Q$0041-2622$	& 2.18 		& $1.488-2.127$	& $1.835-2.127$ & 1.6745& S	& 1.48	& 1.35	& \nodata \nl
Q$0041-2638$	& 3.053	 	& $2.181-2.830$	& $2.181-2.830$ & 2.2656& W	& 0.26	& 0.18	&  0.71  \nl
 		&  	        & 		&		& 2.3399& W	& 0.22	& 0.34	&  0.75  \nl
		&  		&		&		& 2.7409& W	& 0.38	& 0.23	&  1.51  \nl
Q$0041-2658$	& 2.457		& $1.713-2.398$	& $1.713-2.398$ & 1.8708& SW	& 1.04	& 0.80	& \nodata \nl
		&  		&		&		& 2.0217& W	& 0.46	& 0.28	& \nodata \nl
		&  		&		&		& 2.2722& SW	& 1.83	& 0.56	&  3.85  \nl
		&  		&		&		& 2.3805& W	& 0.57	& 0.15	&  2.08  \nl
Q$0041-2707$	& 2.786         & $1.969-2.723$	& $1.969-2.723$ & 2.3429& SW	& 1.79	& 1.24	&  2.65  \nl
		&  		&		&		& 2.5997& SW	& 0.95  & 0.68 	&  2.58  \nl
Q$0042-2627$	& 3.289	 	& $2.373-2.826$ & $2.373-2.826$ & 2.4761& SW	& 0.30  & 0.34	&  1.56  \nl
		&  		&		&		& 2.5070& W	& 0.16  & 0.19	&  1.70  \nl
Q$0042-2632$	& 1.79		& $1.540-1.742$	& \nodata	&\nodata & 	&\nodata &\nodata & \nodata \nl
Q$0042-2639$	& 2.98	 	& $2.123-2.202$ & \nodata	& 2.1285& S	& 0.57	& 0.33	& \nodata \nl
		& 	 	& $2.564-2.614$ & \nodata	&\nodata & 	&\nodata &\nodata &\nodata \nl
Q$0042-2656$	& 3.33		& $2.398-2.798$ & $2.398-2.798$ & 2.4985& W	& 0.15	& 0.15	&  1.12  \nl
		&  		&		&		& 2.6489& W	& 0.42	& 0.28	&  1.39  \nl
	 	& 		&		&		& 2.6853& SW	& 0.52	& 0.37	&  1.70  \nl
Q$0042-2657$	& 2.898		& $2.061-2.830$	& $2.061-2.809$ & 2.3122& W	& 0.56  & 0.28	&  1.17  \nl
		& 		&		&		& 2.4920& SW	& 0.49  & 0.42	&  1.59  \nl
		&  		&		&		& 2.5201& W	& 0.65  & 0.24	&  1.43  \nl
Q$0042-2712$	& 1.79	 	& $1.505-1.742$	& $1.616-1.742$ &\nodata &	&\nodata &\nodata &\nodata \nl
Q$0042-2714$	& 2.36		& $1.637-2.135$ & \nodata	& 1.8143& S	& 0.54	& 0.61	& \nodata \nl
Q$0043-2606$	& 3.11		& $2.225-2.255$	& \nodata	&\nodata &	&\nodata &\nodata &\nodata \nl
Q$0043-2633$	& 3.44		& $2.485-2.830$	& $2.485-2.805$ &\nodata &	&\nodata &\nodata &\nodata \nl
Q$0044-2628$	& 2.47	 	& $1.968-2.410$	& $2.022-2.410$ & 2.1449& SW	& 0.69	& 0.52 	&\nodata \nl
Q$0044-2701$	& 2.16		& $1.826-1.934$	& \nodata	&\nodata &	&\nodata &\nodata &\nodata \nl
		&  		&		&		& 2.2067& W	& 0.25	& 0.18 	&\nodata \nl
%\end{tabular}	
\enddata
%\end{planotable}
\end{deluxetable}

{\small 
%\begin{planotable}{llccclccc}
\begin{deluxetable}{llccclccc}
\tablecolumns{9}
%\tablewidth{6.5in}
\tablewidth{0pc}
\tablecaption{Expected \civ\ Systems for Synthetic Dataset Construction}

\tablehead{\colhead{} & \colhead{}  &  \multicolumn{3}{c}{STRONG}                      & \multicolumn{3}{c}{WEAK}                     \cr 
\colhead{}            &      \colhead{}                         
& \colhead{\civ\ $z$}   & \colhead{expected}  & \colhead{cumulative}   
& \colhead{\civ\ $z$}   & \colhead{expected}  & \colhead{cumulative} \cr 
 \colhead{QSO}        & \multicolumn{1}{c}{$z_{em}$}  
& \colhead{sensitivity} & \colhead{systems}  & \colhead{distribution}
 & \colhead{sensitivity} & \colhead{systems}  & \colhead{distribution} }
%\tableline

\startdata

Q$0039-2630$	& 1.810	 	& $1.478-1.780$	& 0.506 & $0.000-0.056$ & $1.478-1.780$ & 0.779 & $0.000-0.059$ \nl
Q$0041-2607$	& 2.505		& $1.747-2.447$	& 0.872 & $0.056-0.152$ & $1.747-2.447$ & 1.705 & $0.059-0.187$ \nl
Q$0041-2608$	& 1.72	 	& $1.548-1.673$	& 0.212 & $0.152-0.175$ & \nodata	& 0.000 & \nodata       \nl
Q$0041-2622$	& 2.18 		& $1.488-2.127$	& 0.955 & $0.175-0.280$ & $1.835-2.127$ & 0.744 & $0.187-0.243$ \nl
Q$0041-2638$	& 3.053	 	& $2.181-2.830$	& 0.643 & $0.280-0.350$ & $2.181-2.830$ & 1.382 & $0.243-0.347$	\nl
Q$0041-2658$	& 2.457		& $1.713-2.398$	& 0.876 & $0.350-0.447$ & $1.713-2.398$ & 1.688 & $0.347-0.474$ \nl
Q$0041-2707$	& 2.786         & $1.969-2.723$	& 0.816 & $0.447-0.536$ & $1.969-2.723$ & 1.704 & $0.474-0.603$ \nl
Q$0042-2627$	& 3.289	 	& $2.373-2.826$ & 0.426 & $0.536-0.583$ & $2.373-2.826$ & 0.930 & $0.603-0.673$ \nl
Q$0042-2632$	& 1.79		& $1.540-1.742$	& 0.336 & $0.583-0.620$ & \nodata	& 0.000 & \nodata       \nl
Q$0042-2639$	& 2.98	 	& $2.123-2.202$ & 0.094 & $0.620-0.630$ & \nodata	& 0.000 & \nodata       \nl
		& 	 	& $2.564-2.614$ & 0.047 & $0.630-0.636$ & \nodata	& 0.000 & \nodata       \nl
Q$0042-2656$	& 3.33		& $2.398-2.798$ & 0.402 & $0.636-0.680$ & $2.398-2.798$ & 0.821 & $0.673-0.734$ \nl
Q$0042-2657$	& 2.898		& $2.061-2.830$	& 0.789 & $0.680-0.767$ & $2.061-2.809$ & 1.637 & $0.734-0.858$ \nl
Q$0042-2712$	& 1.79	 	& $1.505-1.742$	& 0.399 & $0.767-0.811$ & $1.616-1.742$ & 0.325 & $0.858-0.882$ \nl
Q$0042-2714$	& 2.36		& $1.637-2.135$ & 0.705 & $0.811-0.888$ & \nodata	& 0.000 & \nodata       \nl
Q$0043-2606$	& 3.11		& $2.225-2.255$	& 0.034 & $0.888-0.892$ & \nodata       & 0.000 & \nodata       \nl
Q$0043-2633$	& 3.44		& $2.485-2.830$	& 0.314 & $0.892-0.926$ & $2.485-2.805$ & 0.647 & $0.882-0.931$ \nl
Q$0044-2628$	& 2.47	 	& $1.968-2.410$	& 0.519 & $0.926-0.983$ & $2.022-2.410$ & 0.918 & $0.931-1.000$ \nl
Q$0044-2701$	& 2.16		& $1.826-1.934$	& 0.152 & $0.983-1.000$ & \nodata	& 0.000 & \nodata       \nl
\enddata
\end{deluxetable}
}

{\small
%\begin{planotable}{llccclccc}
\begin{deluxetable}{lllllllllll}
\tablecolumns{11}
%\tablewidth{6.5in}
\tablewidth{0pc}
\tablecaption{Merged QSO sample used for QSO -- \civ\ cross-correlations}

\tablehead{\colhead{QSO} & \colhead{$z$} &\colhead{\phantom{000}}& \colhead{QSO} & \colhead{$z$} 
&\colhead{\phantom{000}}& \colhead{QSO} & \colhead{$z$} &\colhead{\phantom{000}}& \colhead{QSO} & \colhead{$z$} }

\startdata
Q$0035-2515$ & 1.196 && Q$0042-2639$ & 2.98  && Q$0043-2728$ & 1.89  && Q$0047-2621$ & 2.29  \nl
Q$0039-2630$ & 1.810 && Q$0042-2642$ & 2.81  && Q$0043-2734$ & 3.46  && Q$0047-2649$ & 3.16  \nl
Q$0039-2636$ & 1.56  && Q$0042-2645$ & 1.69  && Q$0044-2628$ & 2.47  && Q$0047-2759$ & 2.143 \nl
Q$0039-2727$ & 1.407 && Q$0042-2750$ & 1.49  && Q$0044-2701$ & 2.16  && Q$0048-2526$ & 2.11  \nl
Q$0040-2606$ & 2.47  && Q$0042-2651$ & 1.71  && Q$0044-2721$ & 3.16  && Q$0048-2545$ & 2.082 \nl
Q$0040-2737$ & 2.43  && Q$0042-2656$ & 3.33  && Q$0044-2725$ & 2.18  && Q$0048-2608$ & 2.249 \nl
Q$0040-2758$ & 3.23  && Q$0042-2657$ & 2.90  && Q$0044-2754$ & 1.88  && Q$0048-2624$ & 2.82  \nl
Q$0041-2607$ & 2.505 && Q$0042-2712$ & 1.79  && Q$0045-2551$ & 2.521 && Q$0048-2626$ & 1.86  \nl
Q$0041-2608$ & 1.72  && Q$0042-2714$ & 2.36  && Q$0045-2604$ & 1.64  && Q$0048-2645$ & 3.17  \nl
Q$0041-2612$ & 1.72  && Q$0042-2739$ & 2.43  && Q$0045-2606$ & 1.242 && Q$0048-2656$ & 2.36  \nl
Q$0041-2622$ & 2.18  && Q$0043-2555$ & 3.31  && Q$0045-2614$ & 2.35  && Q$0048-2659$ & 3.26  \nl
Q$0041-2638$ & 3.053 && Q$0043-2606$ & 3.11  && Q$0046-2616$ & 1.41  && Q$0048-2709$ & 1.79  \nl
Q$0041-2658$ & 2.457 && Q$0043-2633$ & 3.44  && Q$0046-2643$ & 3.52  && Q$0048-2728$ & 2.43  \nl
Q$0041-2707$ & 2.786 && Q$0043-2647$ & 2.12  && Q$0046-2645$ & 2.54  && Q$0048.2-2734$ & 1.39  \nl
Q$0042-2627$ & 3.298 && Q$0043-2708$ & 2.15  && Q$0047-2522$ & 1.184 && Q$0048.4-2734$ & 1.87  \nl
Q$0042-2632$ & 1.79  && Q$0043-2710$ & 1.84  && Q$0047-2538$ & 1.969 &&              &       \nl
\enddata
\end{deluxetable}
}

\centering

%\begin{planotable}{llccclccc}
\begin{deluxetable}{cclcccc}

\tablecolumns{7}
%\tablewidth{6.5in}
\tablewidth{6.5in}
\tablecaption{Expected \mgii\ Systems for Synthetic Dataset Construction}

\tablehead{ \colhead{\phantom{0000000}} & \colhead{} & \colhead{} & 
\multicolumn{3}{c}{$W_{0, \lambda 2796}\geq 0.6$\AA } & \colhead{\phantom{0000000}} \cr
 \colhead{} & \colhead{} & \colhead{} & 
\colhead{\mgii\ $z$}   & \colhead{expected}  & \colhead{cumulative}  & 
\colhead{} \cr
\colhead{} & \colhead{QSO} & \multicolumn{1}{c}{$z_{em}$}   & 
\colhead{sensitivity} & \colhead{systems}   & \colhead{} 
}
\startdata
 & Q$0039-2630$	& 1.810	 	& $0.373-0.772$	& 0.153 & $0.000-0.033$ & \nl
 & Q$0039-2636$	& 1.55	 	& $0.699-0.761$	& 0.028 & $0.033-0.039$ & \nl
 & Q$0041-2607$	& 2.505		& $0.523-1.120$	& 0.289 & $0.041-0.102$ & \nl
 & Q$0041-2608$	& 1.72	 	& $0.421-0.772$	& 0.138 & $0.102-0.132$ & \nl
 & Q$0041-2622$	& 2.17 		& $0.382-1.075$	& 0.310 & $0.132-0.199$ & \nl
 & Q$0041-2638$	& 3.04	 	& $0.763-1.122$	& 0.192 & $0.199-0.240$ & \nl
 & Q$0041-2658$	& 2.457		& $0.504-1.122$	& 0.297 & $0.240-0.305$ & \nl
 & Q$0041-2707$	& 2.786         & $0.645-1.122$	& 0.243 & $0.305-0.357$ & \nl
 & Q$0042-2627$	& 3.289	 	& $0.869-1.120$ & 0.140 & $0.357-0.388$ & \nl
 & Q$0042-2632$	& 1.79		& $0.447-1.097$	& 0.302 & $0.388-0.453$ & \nl
 & Q$0042-2639$	& 2.98	 	& $0.731-0.783$ & 0.024 & $0.453-0.458$ & \nl
 &		& 	 	& $0.975-0.999$ & 0.013 & $0.458-0.461$ & \nl
 & Q$0042-2645$	& 1.69		& $0.722-1.076$	& 0.182 & $0.482-0.500$ & \nl
 & Q$0042-2651$	& 1.71		& $0.646-1.100$ & 0.229 & $0.500-0.550$ & \nl
 & Q$0042-2656$	& 3.33		& $0.883-1.122$ & 0.134 & $0.550-0.579$ & \nl
 & Q$0042-2657$	& 2.898		& $0.696-1.122$	& 0.221 & $0.579-0.627$ & \nl
 & Q$0042-2712$	& 1.79	 	& $0.388-1.122$	& 0.336 & $0.627-0.700$ & \nl
 & Q$0042-2714$	& 2.36		& $0.461-0.738$ & 0.109 & $0.700-0.723$ & \nl
 & Q$0043-2606$	& 3.11		& $0.780-0.805$	& 0.012 & $0.723-0.726$ & \nl
 & Q$0043-2633$	& 3.44		& $0.931-1.122$	& 0.109 & $0.726-0.749$ & \nl
 & Q$0043-2644$	& 1.041		& $0.645-1.008$	& 0.176 & $0.749-0.787$ & \nl
 & Q$0043-2647$	& 2.12		& $0.736-1.121$\tablenotemark{a}	& 0.203 & $0.787-0.831$ & \nl
 & Q$0043-2708$	& 2.15		& $0.425-1.122$	& 0.324 & $0.831-0.902$ & \nl
 & Q$0044-2628$	& 2.47	 	& $0.645-1.098$	& 0.228 & $0.902-0.951$ & \nl
 & Q$0044-2701$	& 2.15		& $0.709-1.043$	& 0.227 & $0.951-1.000$ & \nl
\enddata
%\end{tabular}	
\tablenotetext{a}{In the case of the BAL Q$0043-2647$, we search only
redward of \civ\ emission in order to avoid the vast majority of
absorption lines associated with the BAL region.}
\end{deluxetable}

\clearpage

\flushleft

\centerline{\bf FIGURES}

 Fig. 1:  The SGP QSO field.  Symbols indicate QSOs used in the \civ\
surveys: stars for the weak survey (see \S~3.2), squares for the strong
survey and triangles for other QSOs observed.  Emission redshifts are
listed adjacent to the symbols.  The field is centered at $\alpha =
00^{\rm h}\, 42^{\rm m} \, 00^{\rm s}$, $\delta=-26^\circ \, 40' \,
0''$ (1950).

 Fig. 2:  Spectra for the observed QSOs in relative flux units
{\it vs.} \AA .  A flux unit is nominally $10^{-16}$ erg cm$^{-2}$
s$^{-1}$ \AA$^{-1}$, but these should be regarded as lower limits
only.  The $1\sigma$ error array based on photon counting statistics is
also shown.  Ticks indicate absorption lines in Table~2.  Dashed ticks
indicate doublet components which are present but at $<4\sigma$
significance.  Wavelengths of QSO frame emission lines are labelled.

 Fig. 3:  The SGP field in RA-$z$ space, centered at $\alpha = 00^{\rm
h}\, 42^{\rm m} \, 00^{\rm s}$ (1950).  Symbols represent observed QSOs
as in Fig. 1:  stars for the weak survey, squares for the strong survey
and triangles for other QSOs observed.  Open circles show strong and
filled circles weak \civ\ absorption systems.  Solid and dashed lines
indicate sensitivity to the weak and strong surveys respectively.

 Fig. 4:  The two point correlation function for the strong \civ\
survey, shown in proper and comoving $h^{-1}$ Mpc.  The dotted lines
shows the $1\sigma$ uncertainty limits.  At $z=2.5$, $1^\circ$ on the
sky is $14h^{-1}$ proper Mpc, and $\Delta z = 0.1$ is $13h^{-1}$ proper
Mpc.

 Fig. 5:  The two point correlation function for the weak \civ\ survey,
shown in proper and comoving $h^{-1}$ Mpc.  The dotted lines shows the
$1\sigma$ uncertainty limits.  Note the $3\sigma$ overdensity at
$15-25h^{-1}$ proper $(60-90h^{-1}$ comoving) Mpc.

 Fig. 6:  The QSO-\civ\ absorber two point correlation function for the
strong (above) and weak (below) \civ\ surveys.  The dotted lines shows
the $1\sigma$ uncertainty limits.

 Fig. 7:  The QSO-\civ\ absorber two point correlation function for the
strong (above) and weak (below) \civ\ surveys, using only line of sight
separations and neglecting spatial separations.  The dotted lines shows
the $1\sigma$ uncertainty limits.

 Fig. 8:  The QSO-\civ\ absorber two point correlation function in velocity
space for the
strong (above) and weak (below) \civ\ surveys, using only line of sight
separations and neglecting spatial separations.  QSOs from this sample
and the Hewitt \& Burbidge (1993) QSO catalogue with projected distances
of $<10h^{-1}$ proper Mpc from the \civ\ systems were used.
The dotted lines shows
the $1\sigma$ uncertainty limits.  There is a $4\sigma$ overdensity
of strong \civ\ absorbers at $+1500$ \kms\ in relation to QSOs; no such
overdensity exists for weak systems.

\vfill\eject

\begin{figure}[]
\epsscale{1.0}
\plotfiddle{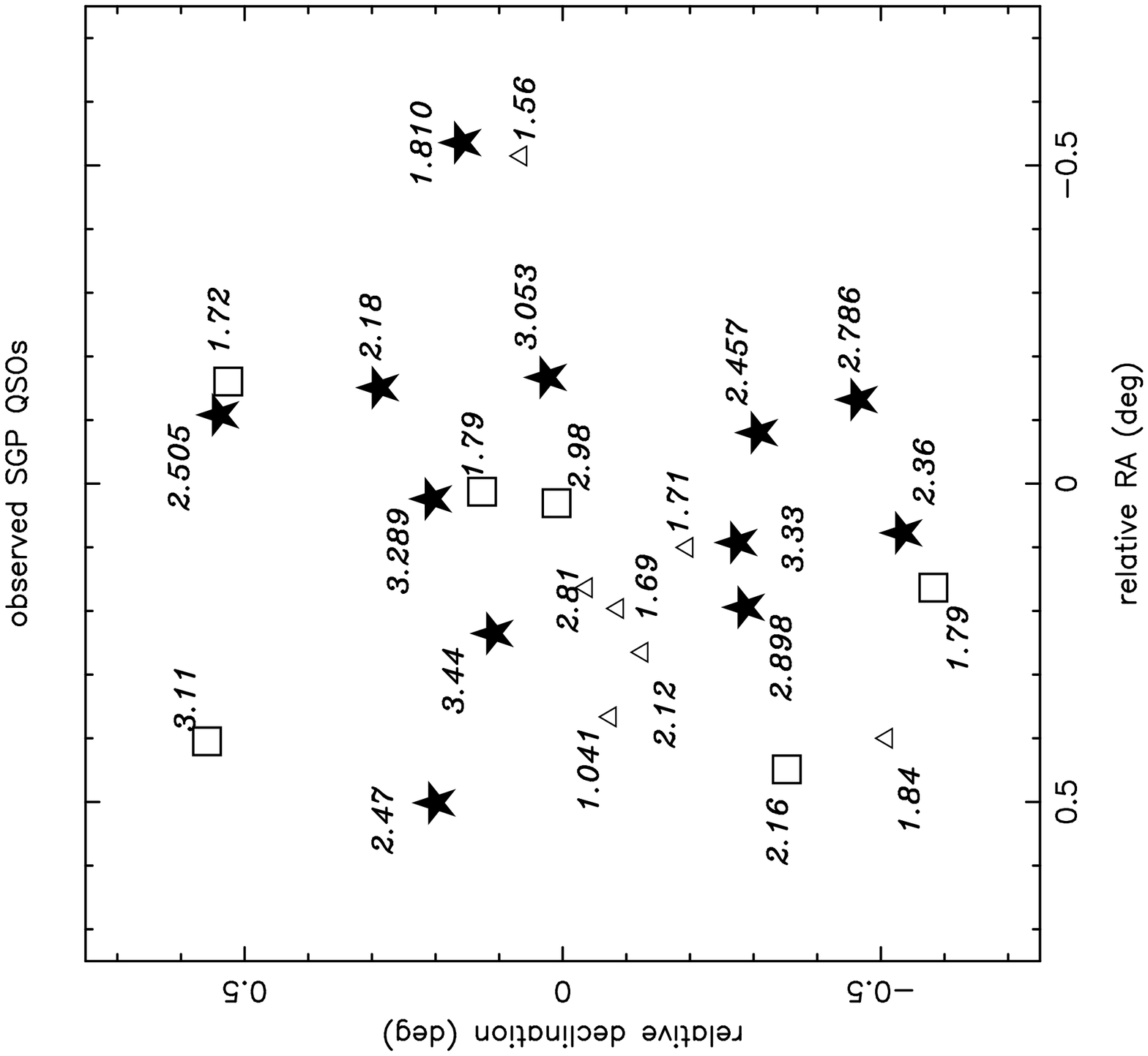 }{7.0in}{-90}{100.}{100.}{-370}{560}
\caption{}
\end{figure}

\begin{figure}[]
\epsscale{1.0}
\plotfiddle{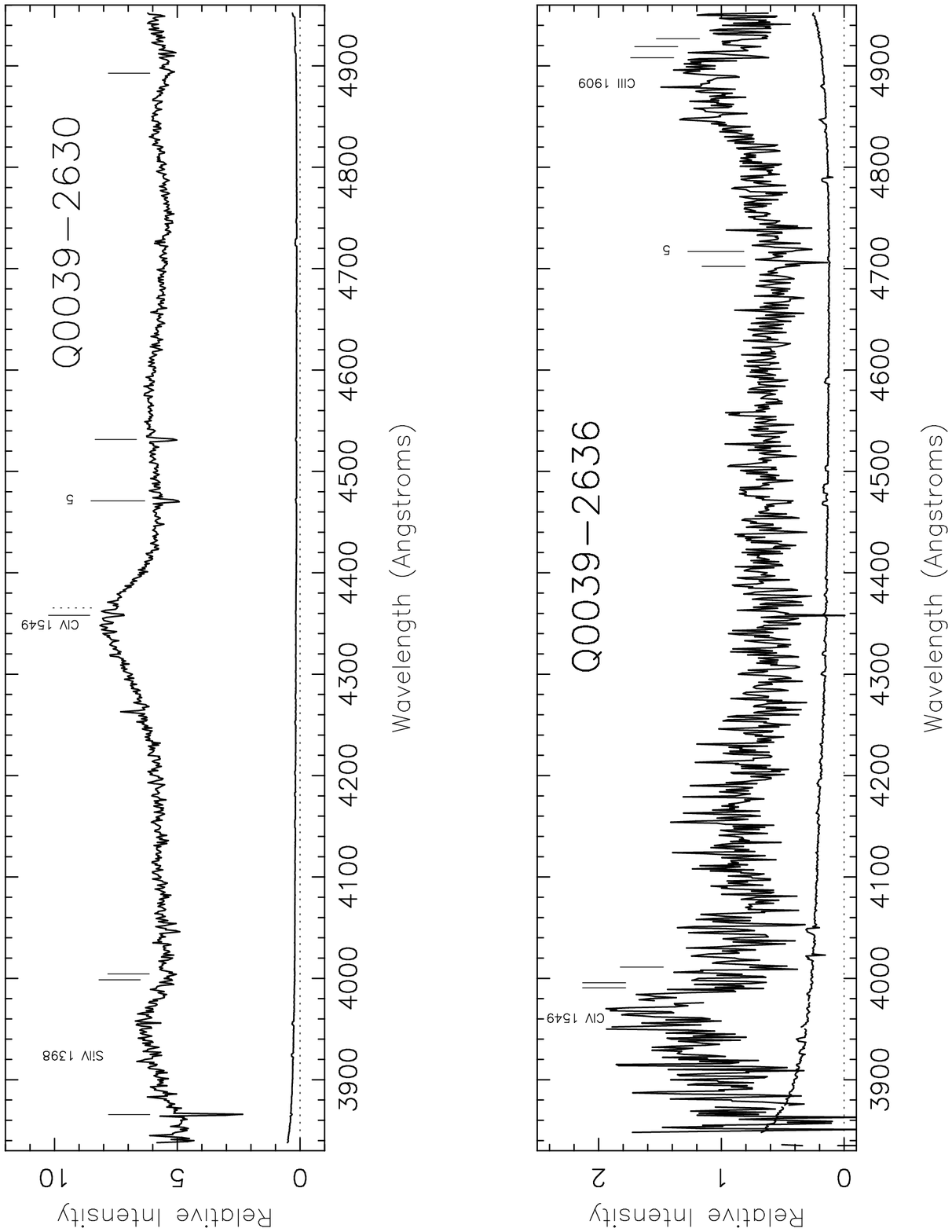 }{6.0in}{0.}{90.}{90.}{-270}{-160}
\vspace{4cm}
\caption{1 of 25}
\end{figure}

\setcounter{figure}{1}
\begin{figure}[]
\epsscale{1.0}
\plotfiddle{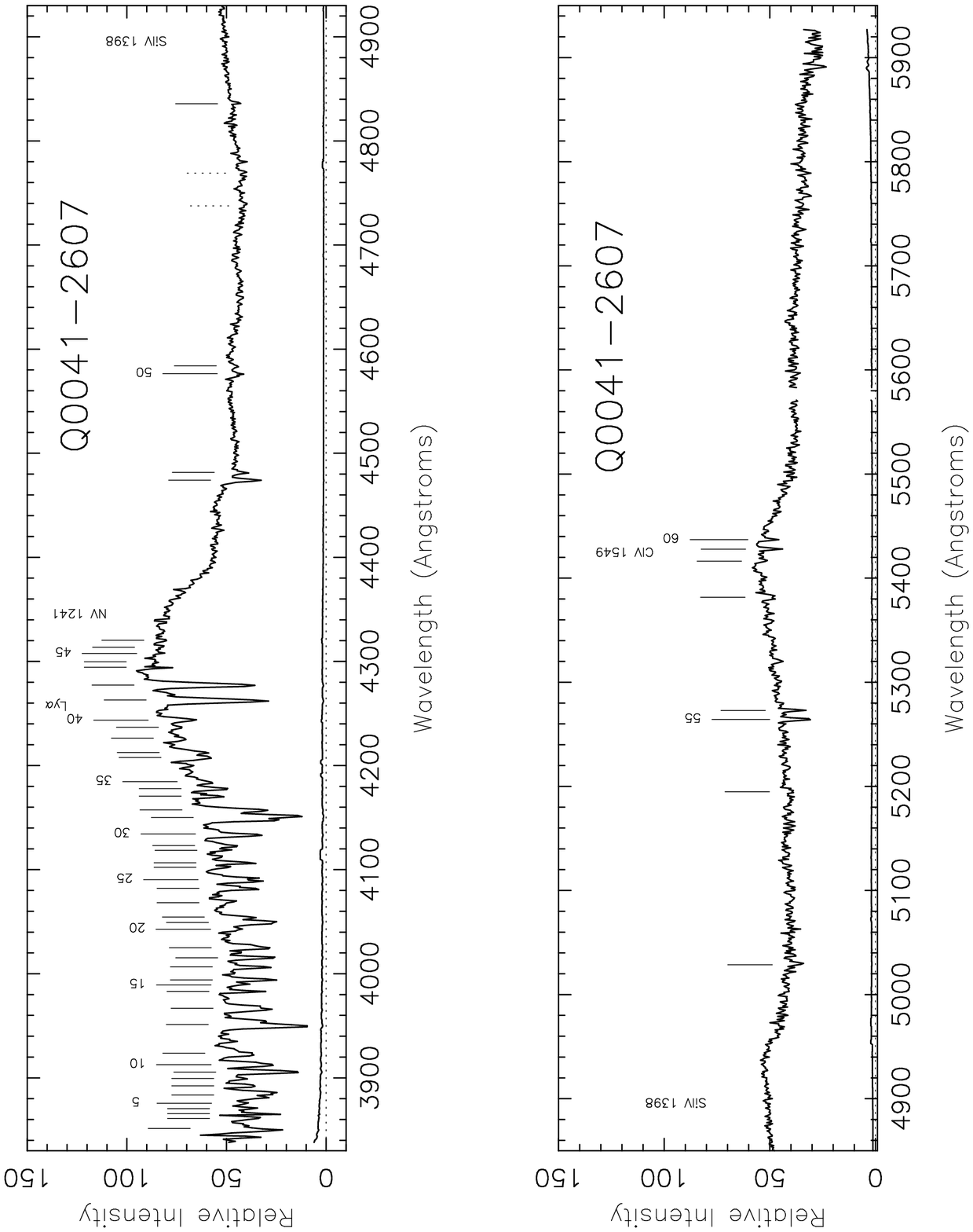 }{6.0in}{0.}{90.}{90.}{-270}{-160}
\vspace{4cm}
\caption{2 of 25}
\end{figure}

\setcounter{figure}{1}
\begin{figure}[]
\epsscale{1.0}
\plotfiddle{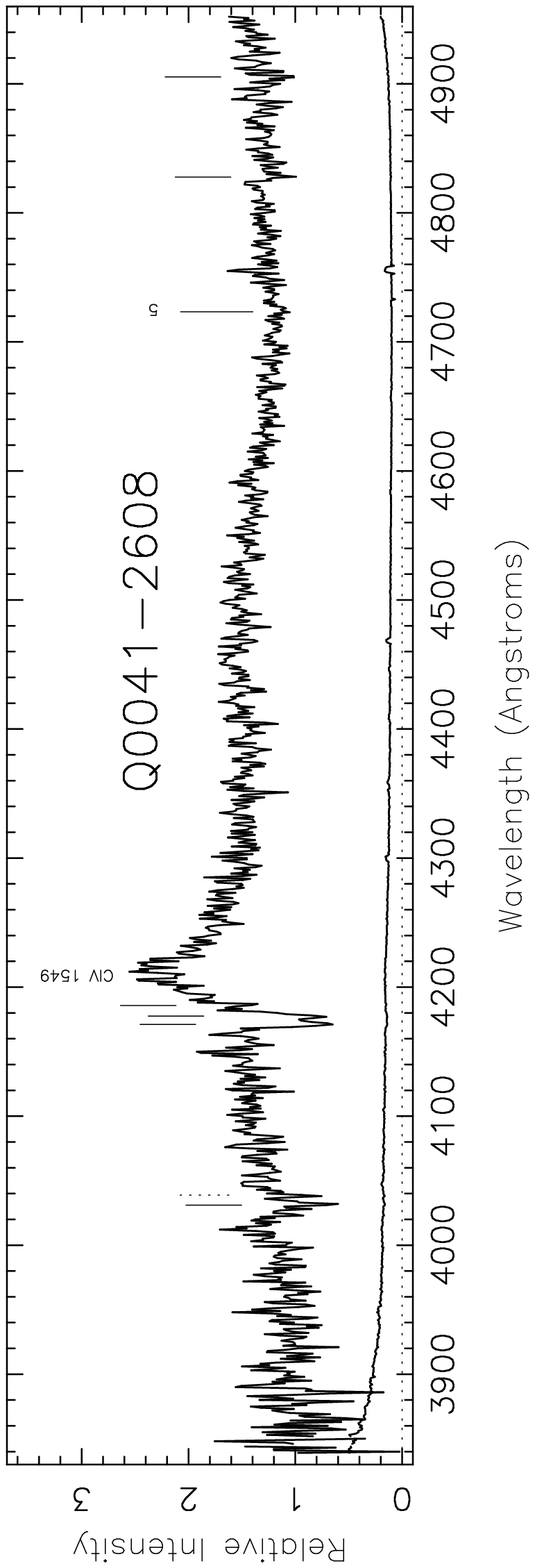 }{6.0in}{0.}{90.}{90.}{-270}{-160}
\vspace{4cm}
\caption{3 of 25}
\end{figure}

\setcounter{figure}{1}
\begin{figure}[]
\epsscale{1.0}
\plotfiddle{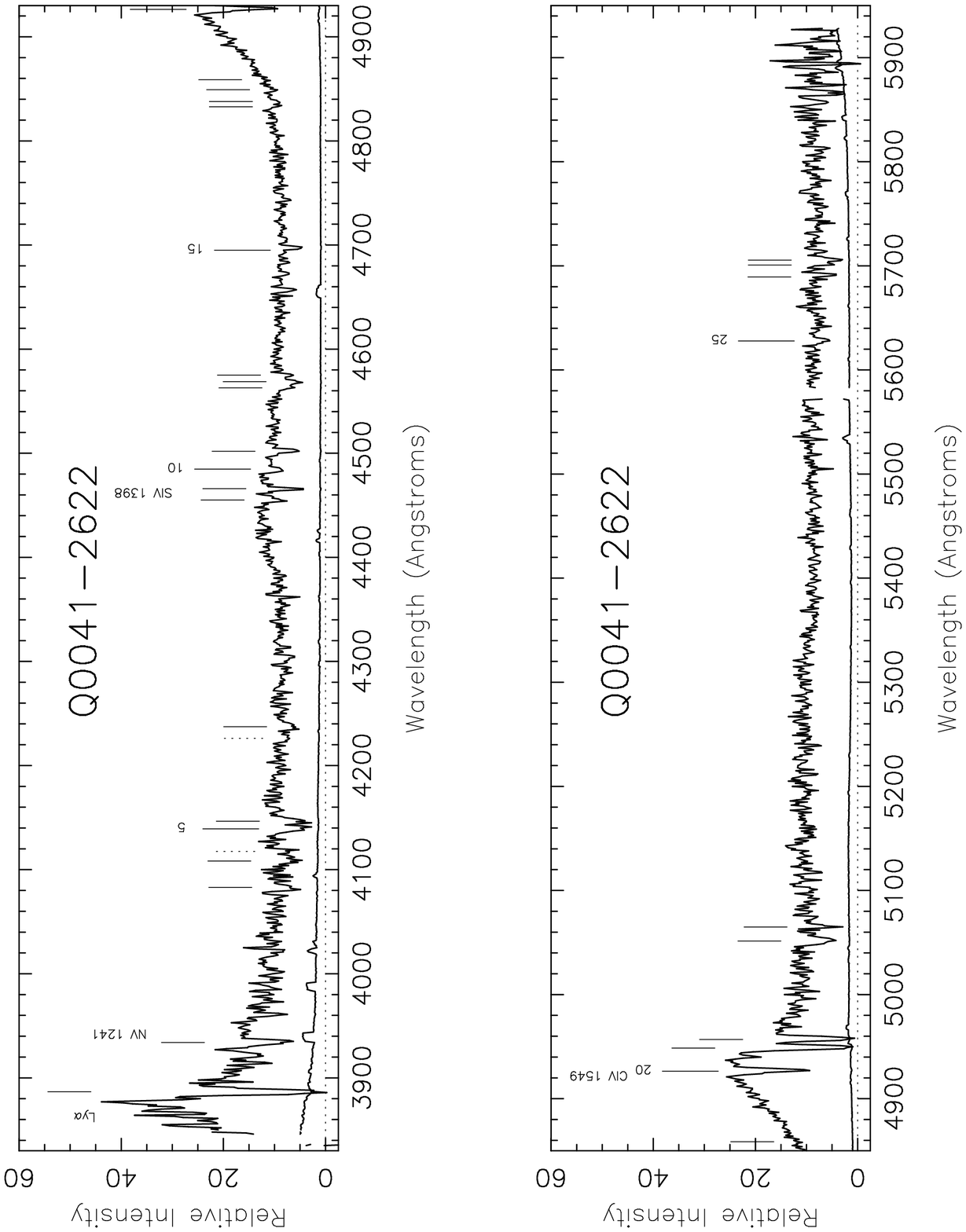 }{6.0in}{0.}{90.}{90.}{-270}{-160}
\vspace{4cm}
\caption{4 of 25}
\end{figure}

\setcounter{figure}{1}
\begin{figure}[]
\epsscale{1.0}
\plotfiddle{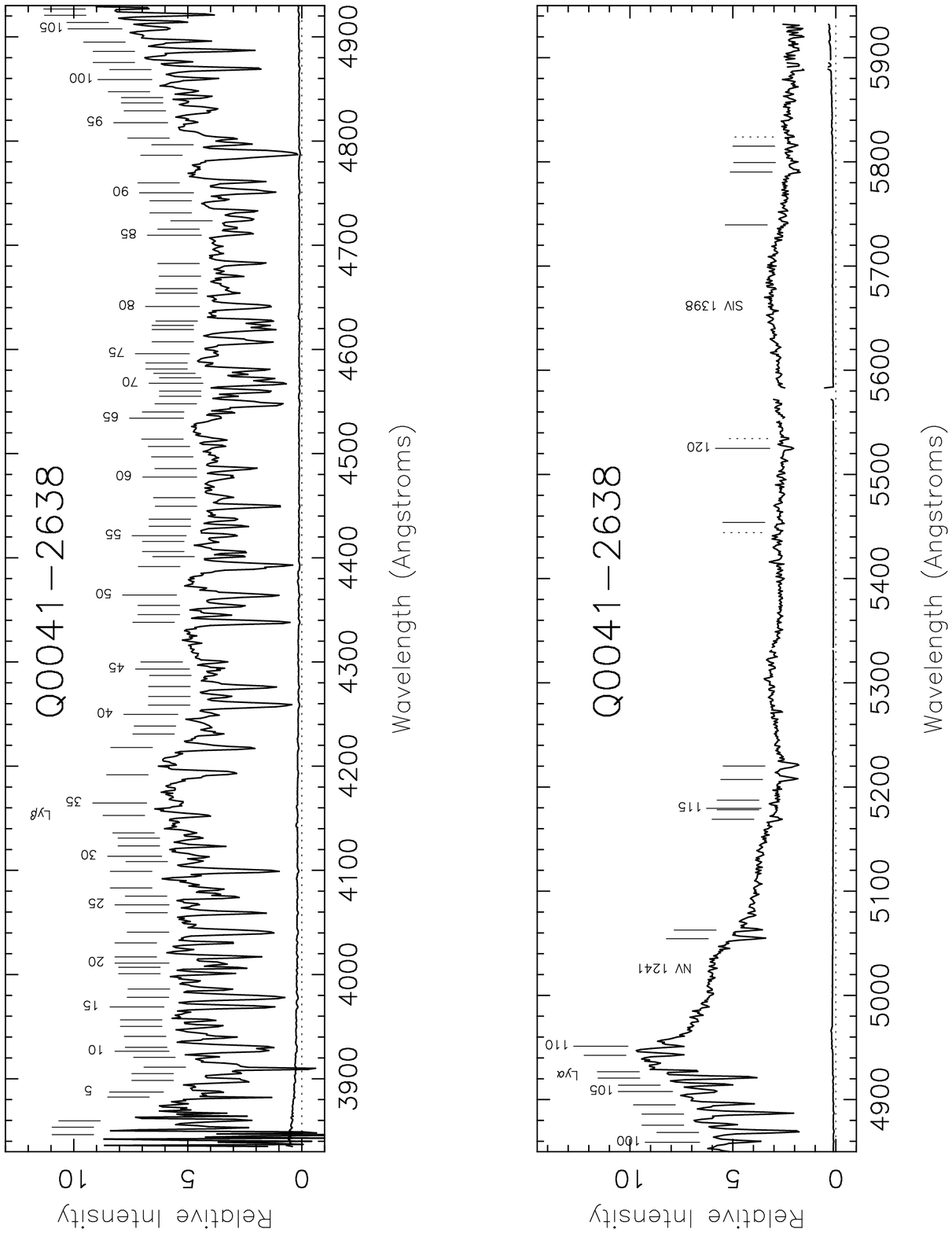 }{6.0in}{0.}{90.}{90.}{-270}{-160}
\vspace{4cm}
\caption{5 of 25}
\end{figure}

\setcounter{figure}{1}
\begin{figure}[]
\epsscale{1.0}
\plotfiddle{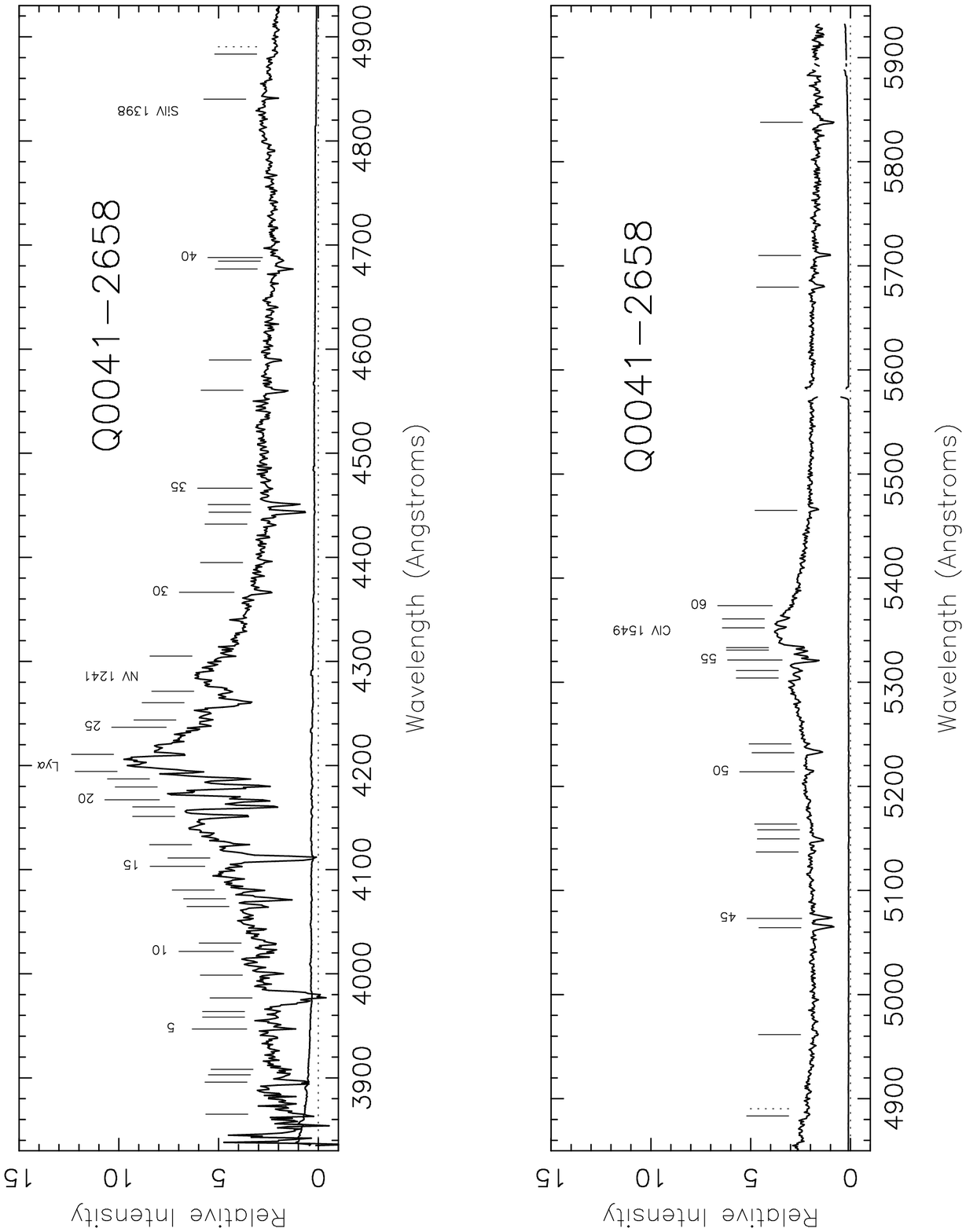 }{6.0in}{0.}{90.}{90.}{-270}{-160}
\vspace{4cm}
\caption{6 of 25}
\end{figure}

\setcounter{figure}{1}
\begin{figure}[]
\epsscale{1.0}
\plotfiddle{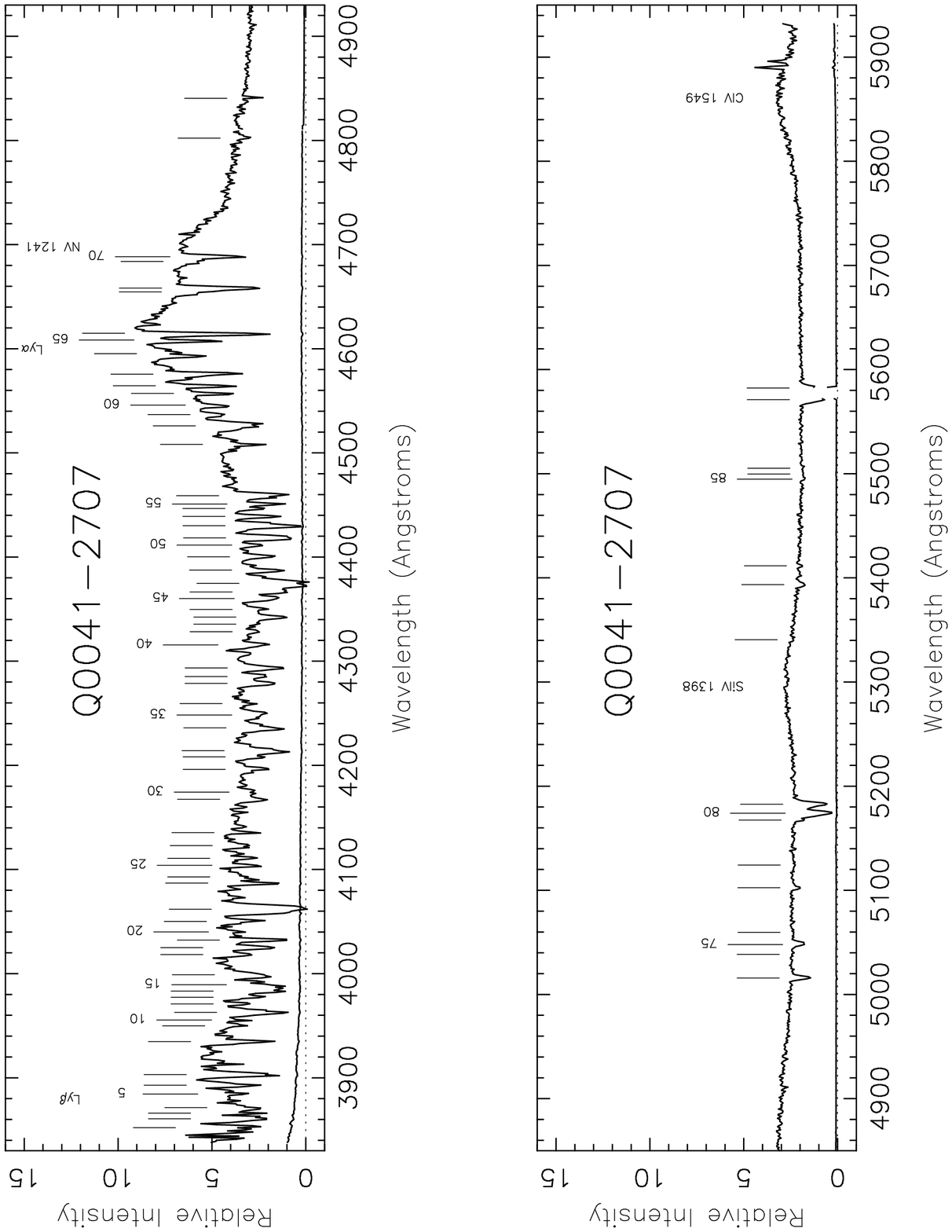 }{6.0in}{0.}{90.}{90.}{-270}{-160}
\vspace{4cm}
\caption{7 of 25}
\end{figure}

\setcounter{figure}{1}
\begin{figure}[]
\epsscale{1.0}
\plotfiddle{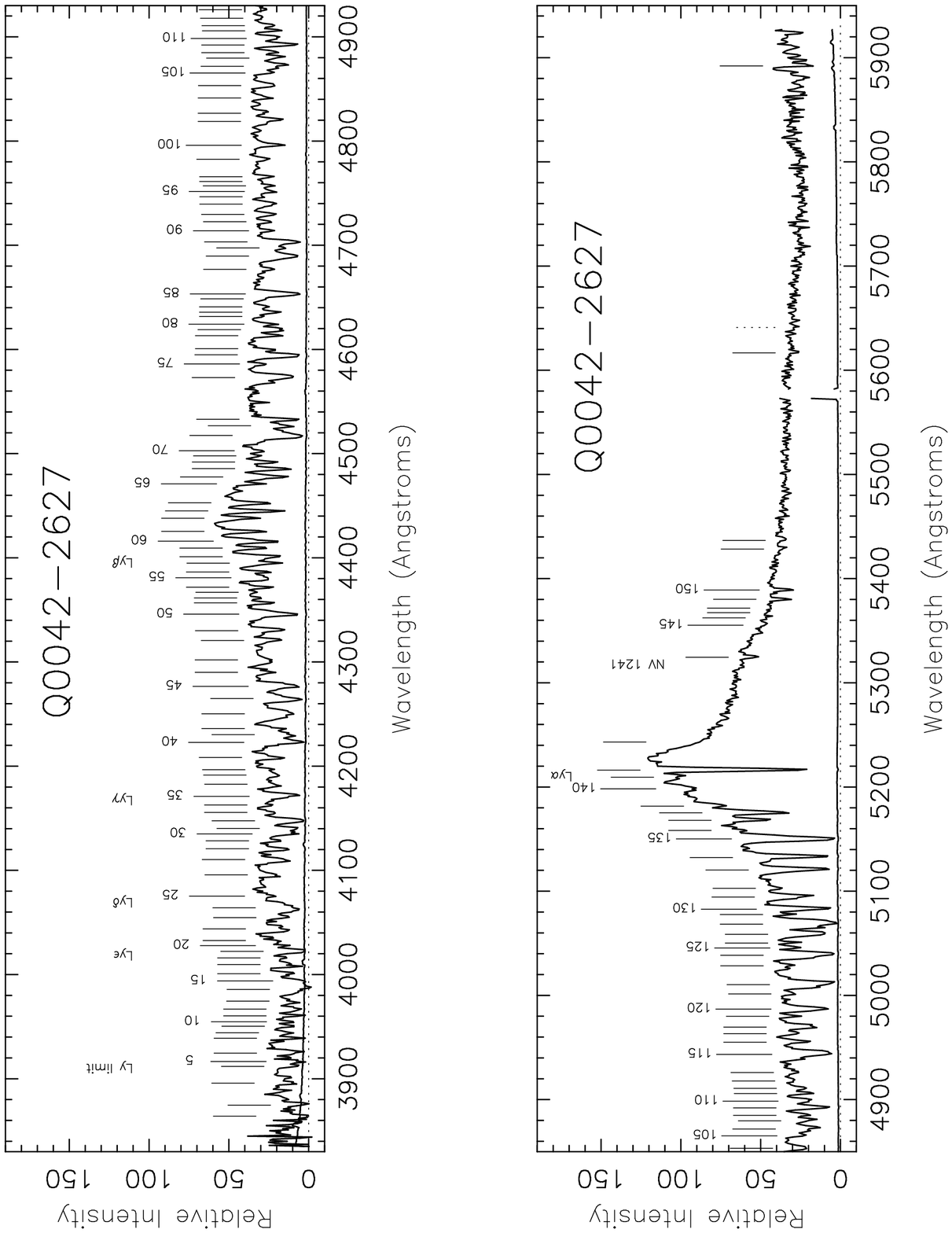 }{6.0in}{0.}{90.}{90.}{-270}{-160}
\vspace{4cm}
\caption{8 of 25}
\end{figure}

\setcounter{figure}{1}
\begin{figure}[]
\epsscale{1.0}
\plotfiddle{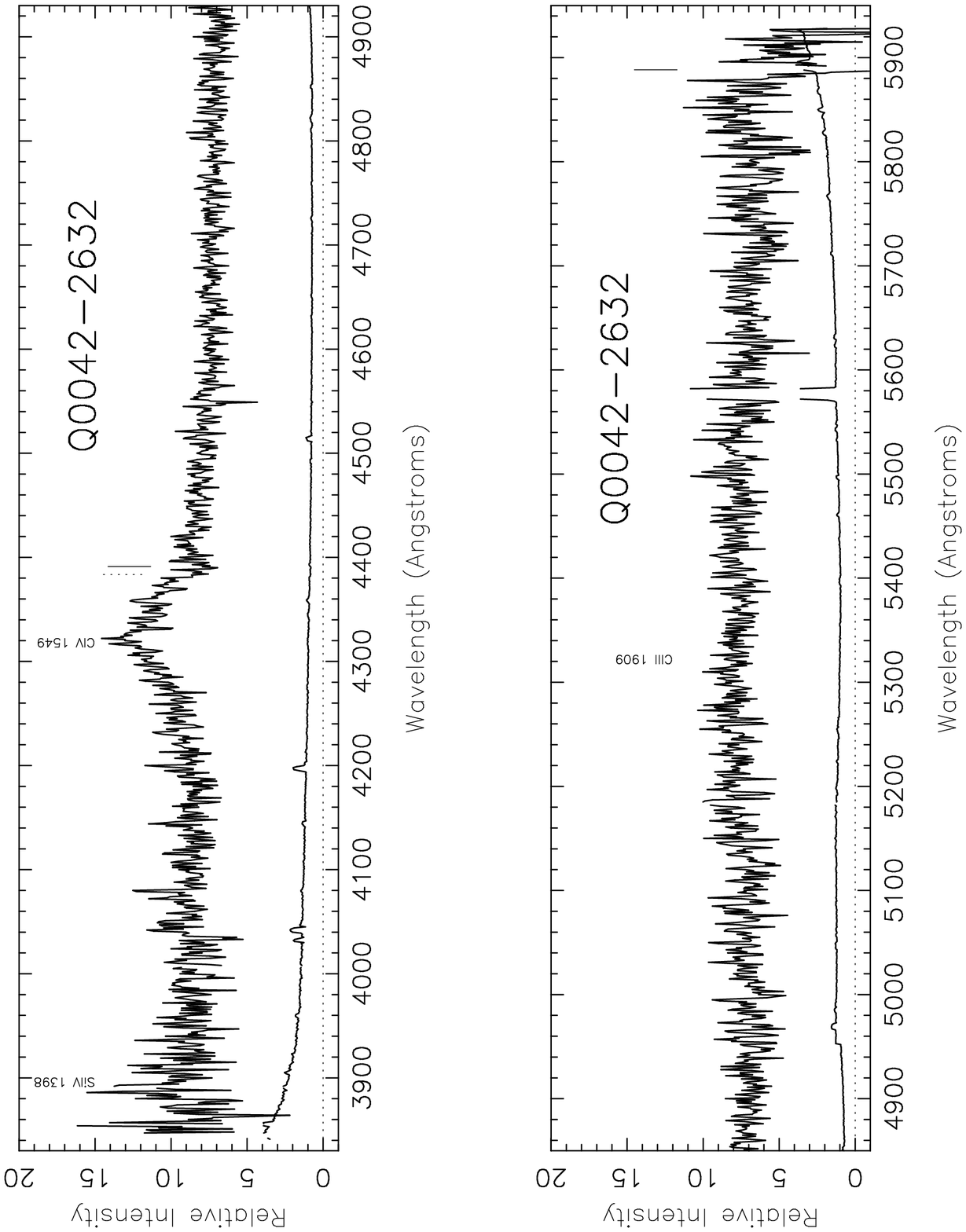 }{6.0in}{0.}{90.}{90.}{-270}{-160}
\vspace{4cm}
\caption{9 of 25}
\end{figure}

%\include{arguspapertab}

%\end{document}

\setcounter{figure}{1}
\begin{figure}[]
\epsscale{1.0}
\plotfiddle{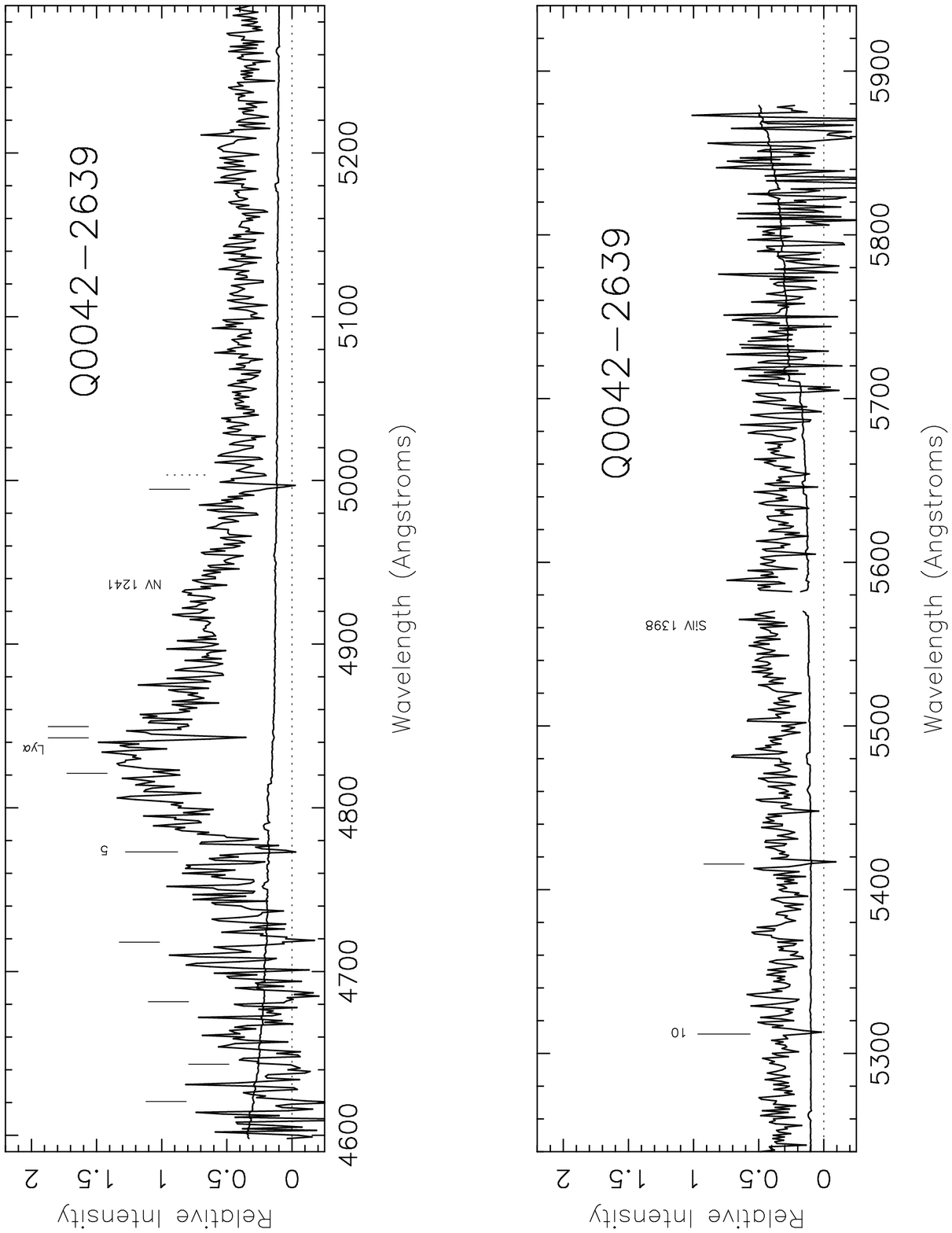 }{6.0in}{0.}{90.}{90.}{-270}{-160}
\vspace{4cm}
\caption{10 of 25}
\end{figure}

\setcounter{figure}{1}
\begin{figure}[]
\epsscale{1.0}
\plotfiddle{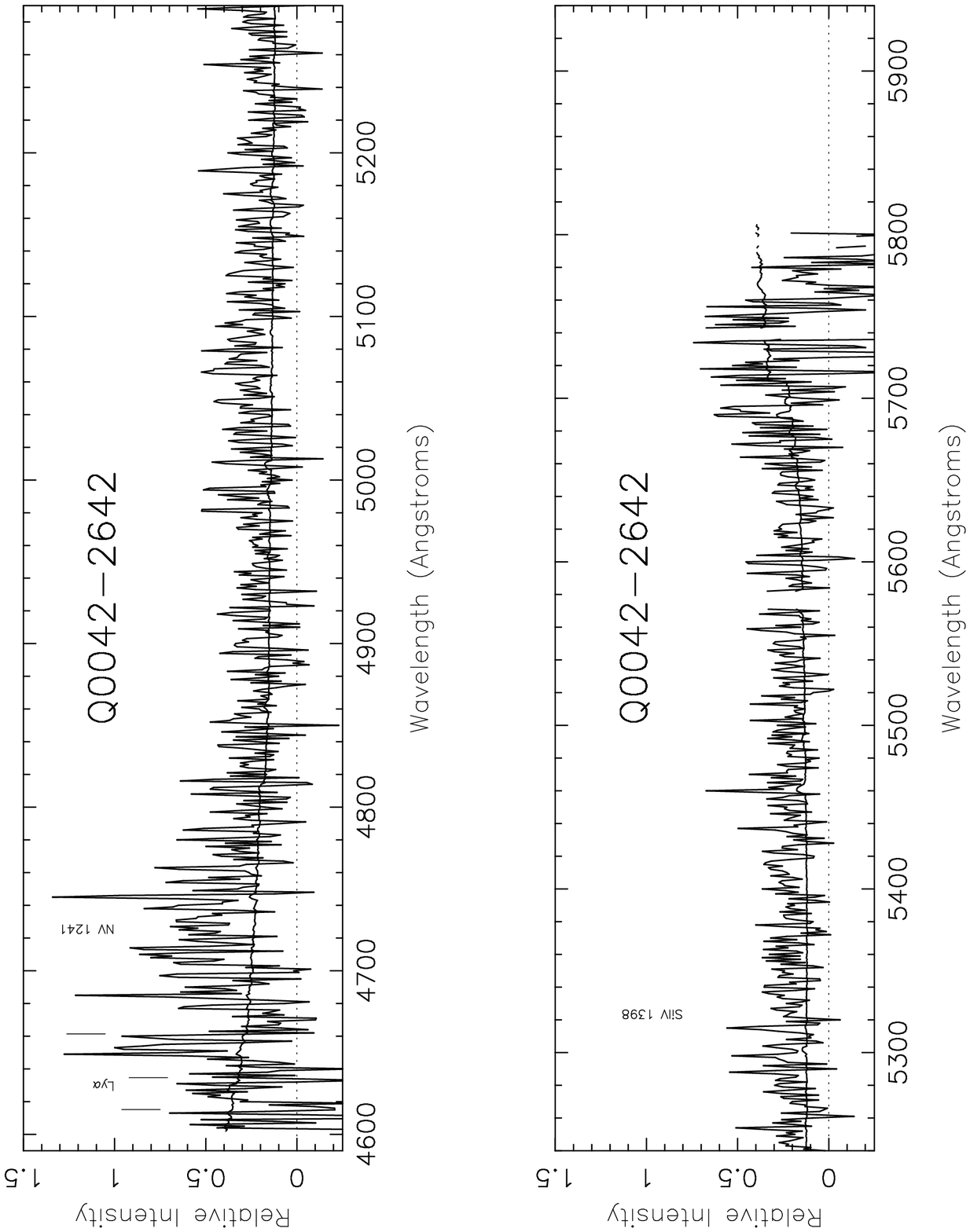 }{6.0in}{0.}{90.}{90.}{-270}{-160}
\vspace{4cm}
\caption{11 of 25}
\end{figure}

\setcounter{figure}{1}
\begin{figure}[]
\epsscale{1.0}
\plotfiddle{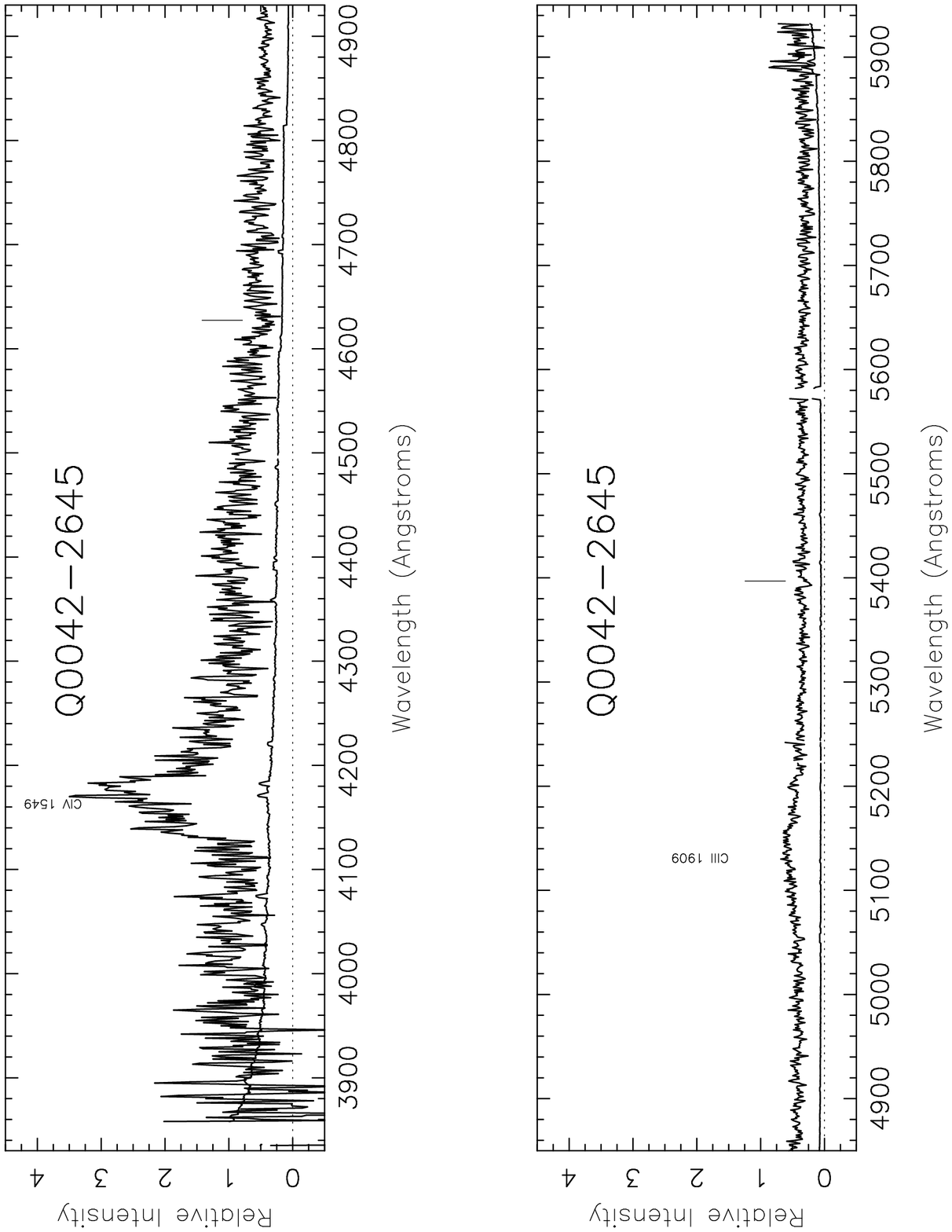 }{6.0in}{0.}{90.}{90.}{-270}{-160}
\vspace{4cm}
\caption{12 of 25}
\end{figure}

\setcounter{figure}{1}
\begin{figure}[]
\epsscale{1.0}
\plotfiddle{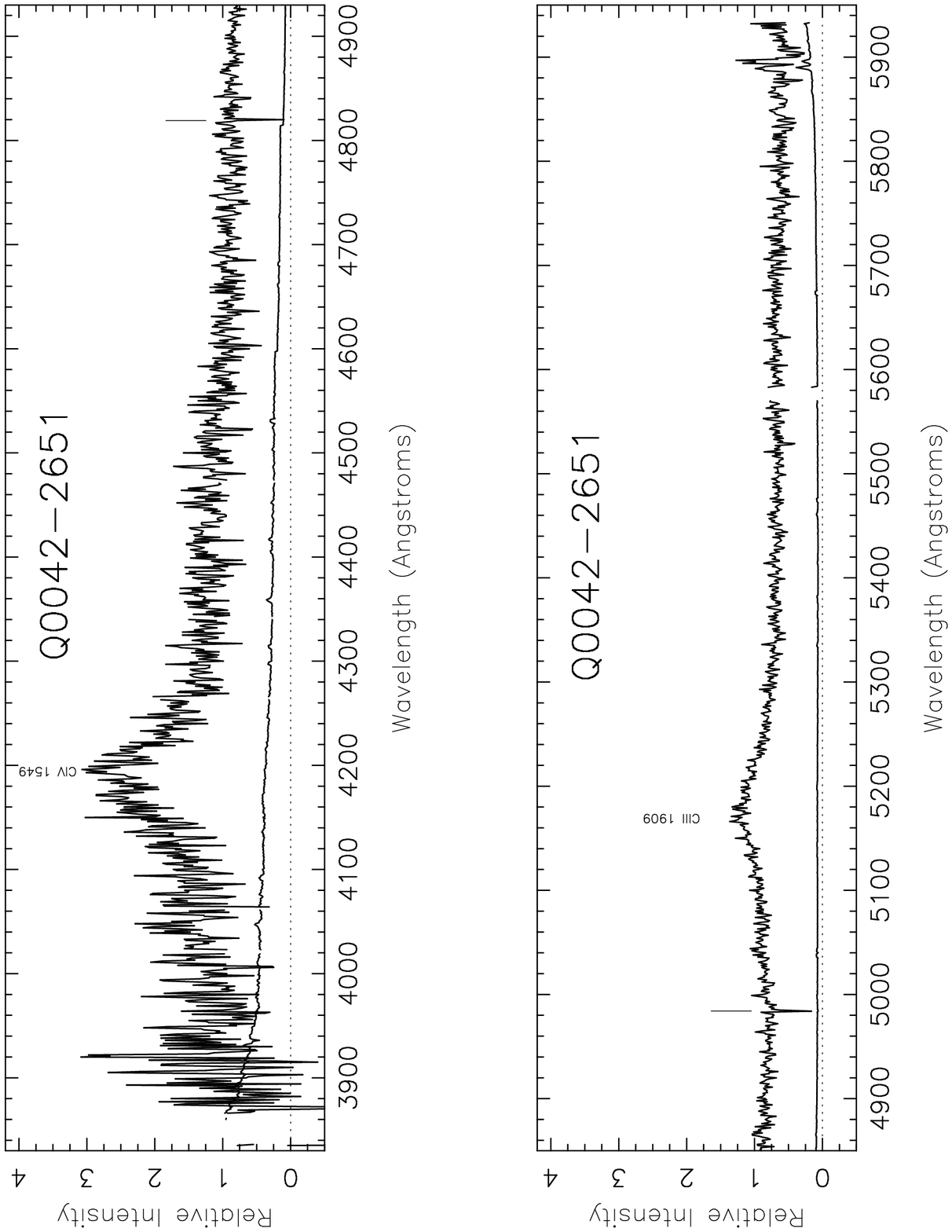 }{6.0in}{0.}{90.}{90.}{-270}{-160}
\vspace{4cm}
\caption{13 of 25}
\end{figure}

\setcounter{figure}{1}
\begin{figure}[]
\epsscale{1.0}
\plotfiddle{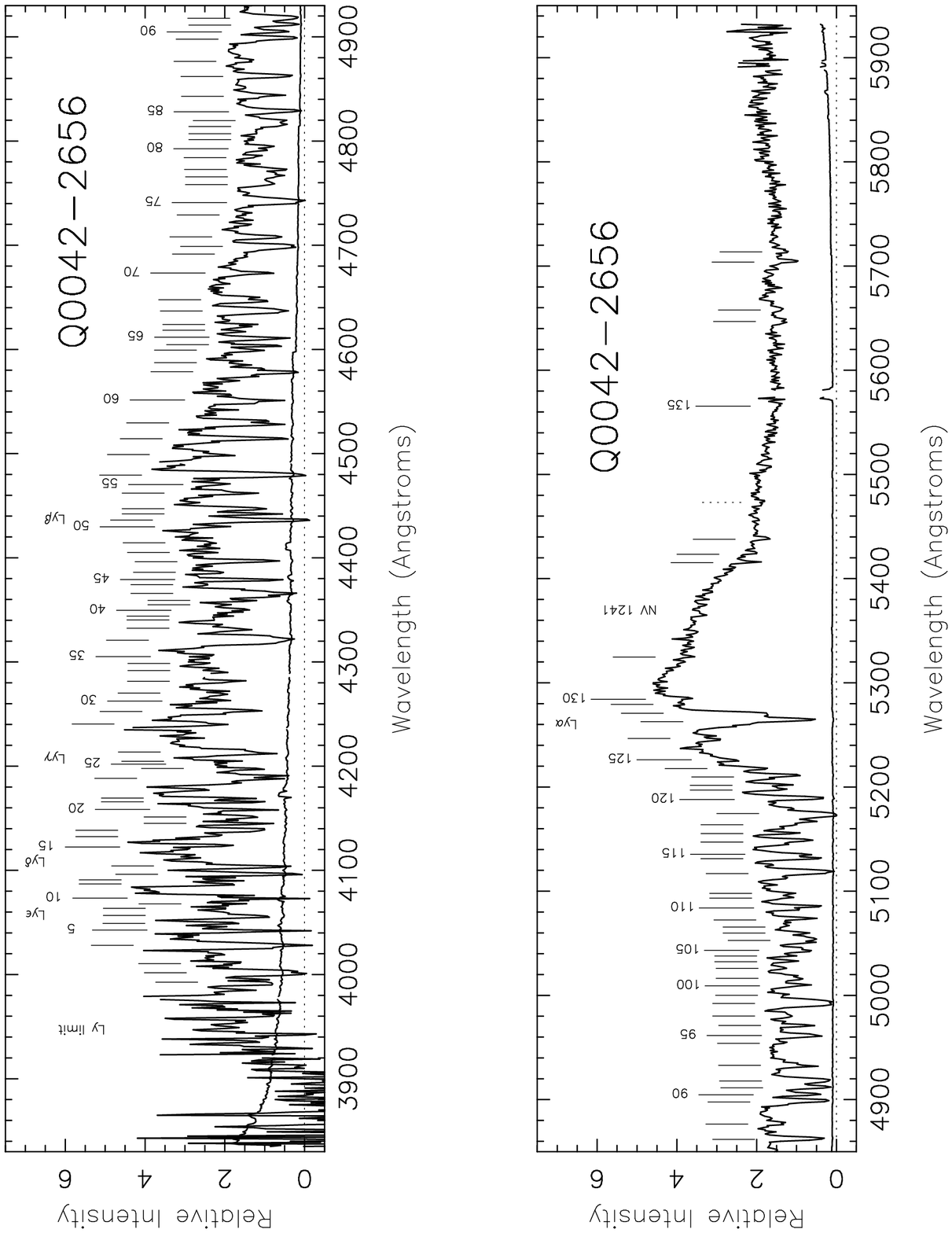 }{6.0in}{0.}{90.}{90.}{-270}{-160}
\vspace{4cm}
\caption{14 of 25}
\end{figure}

\setcounter{figure}{1}
\begin{figure}[]
\epsscale{1.0}
\plotfiddle{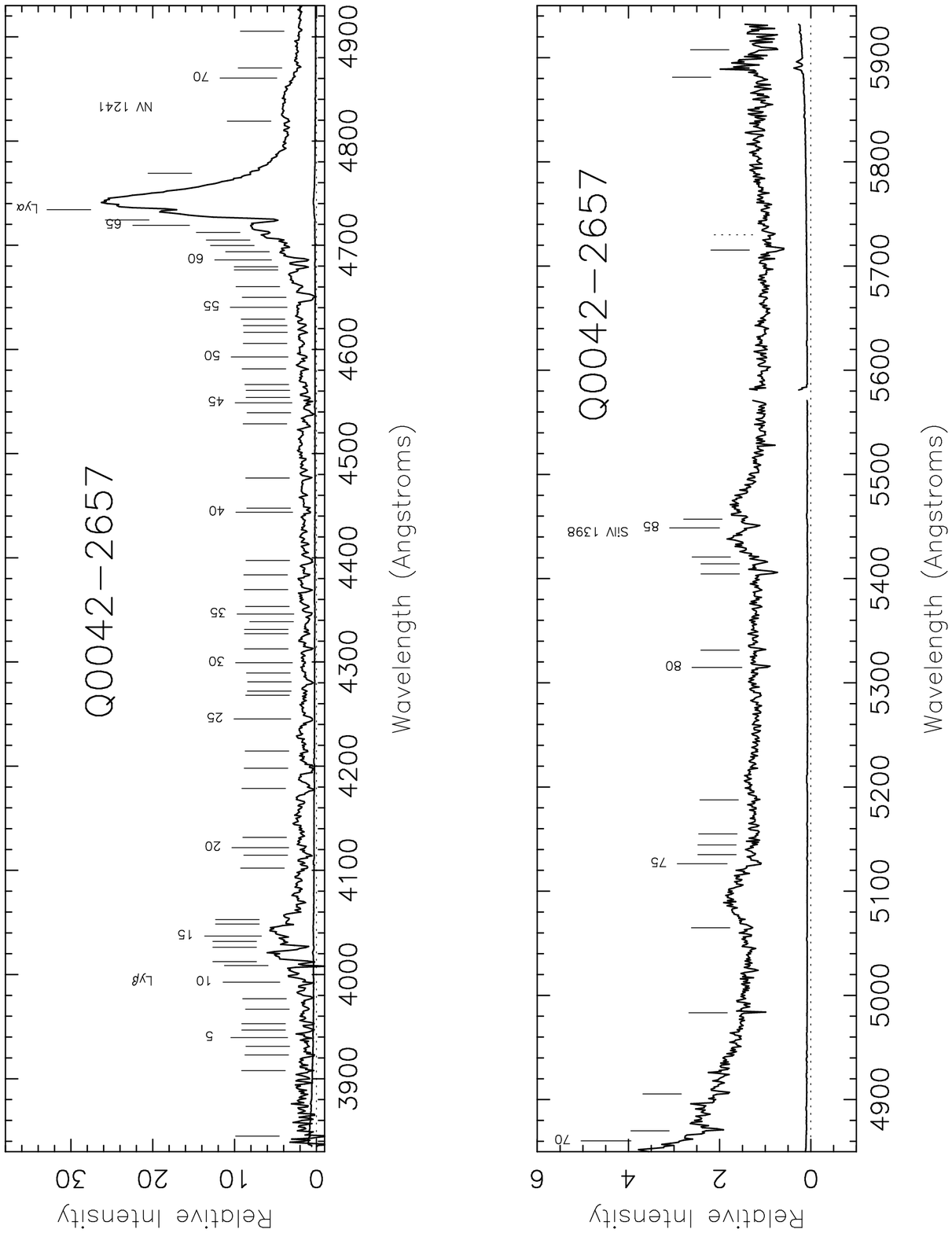 }{6.0in}{0.}{90.}{90.}{-270}{-160}
\vspace{4cm}
\caption{15 of 25}
\end{figure}

\setcounter{figure}{1}
\begin{figure}[]
\epsscale{1.0}
\plotfiddle{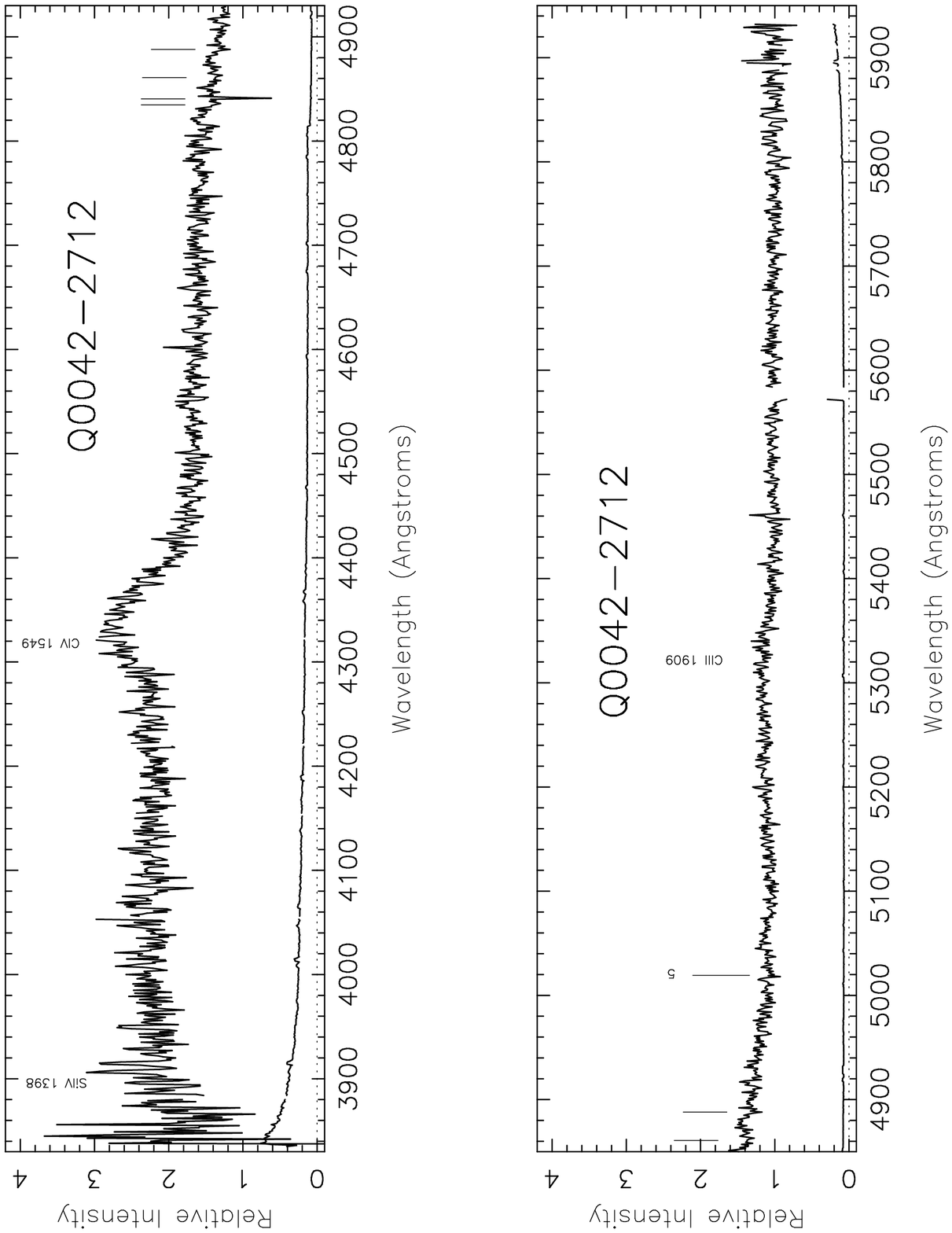 }{6.0in}{0.}{90.}{90.}{-270}{-160}
\vspace{4cm}
\caption{16 of 25}
\end{figure}

\setcounter{figure}{1}
\begin{figure}[]
\epsscale{1.0}
\plotfiddle{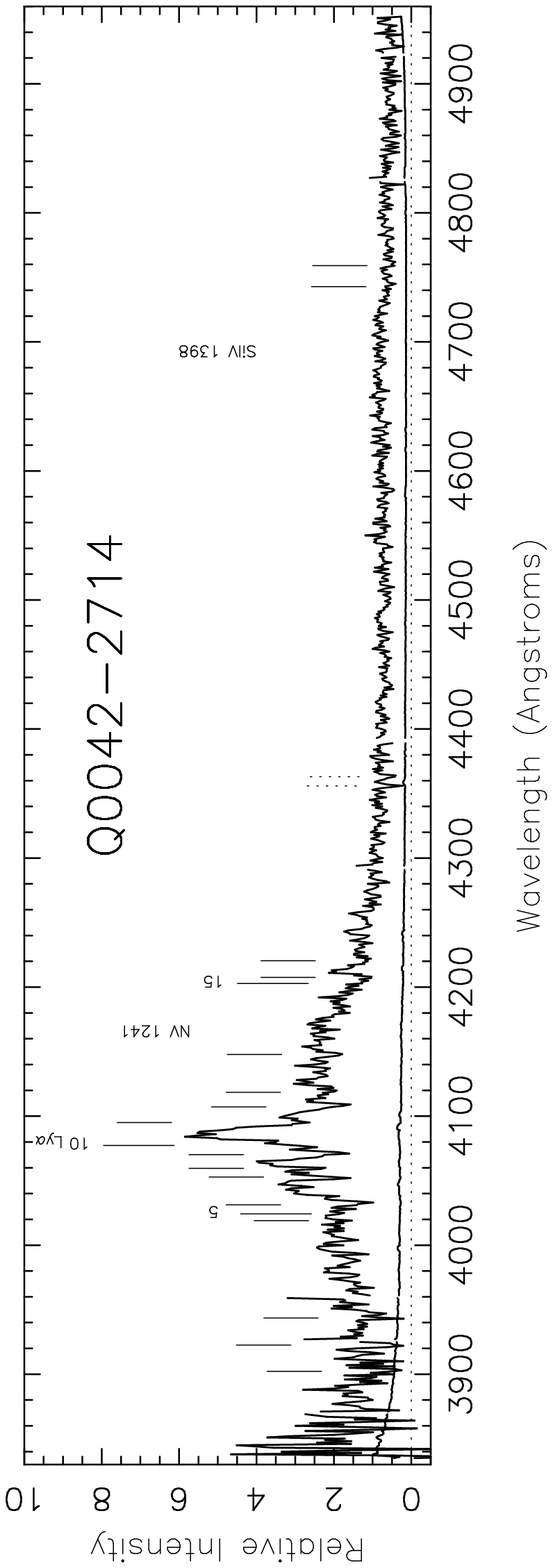 }{6.0in}{0.}{90.}{90.}{-270}{-160}
\vspace{4cm}
\caption{17 of 25}
\end{figure}

\setcounter{figure}{1}
\begin{figure}[]
\epsscale{1.0}
\plotfiddle{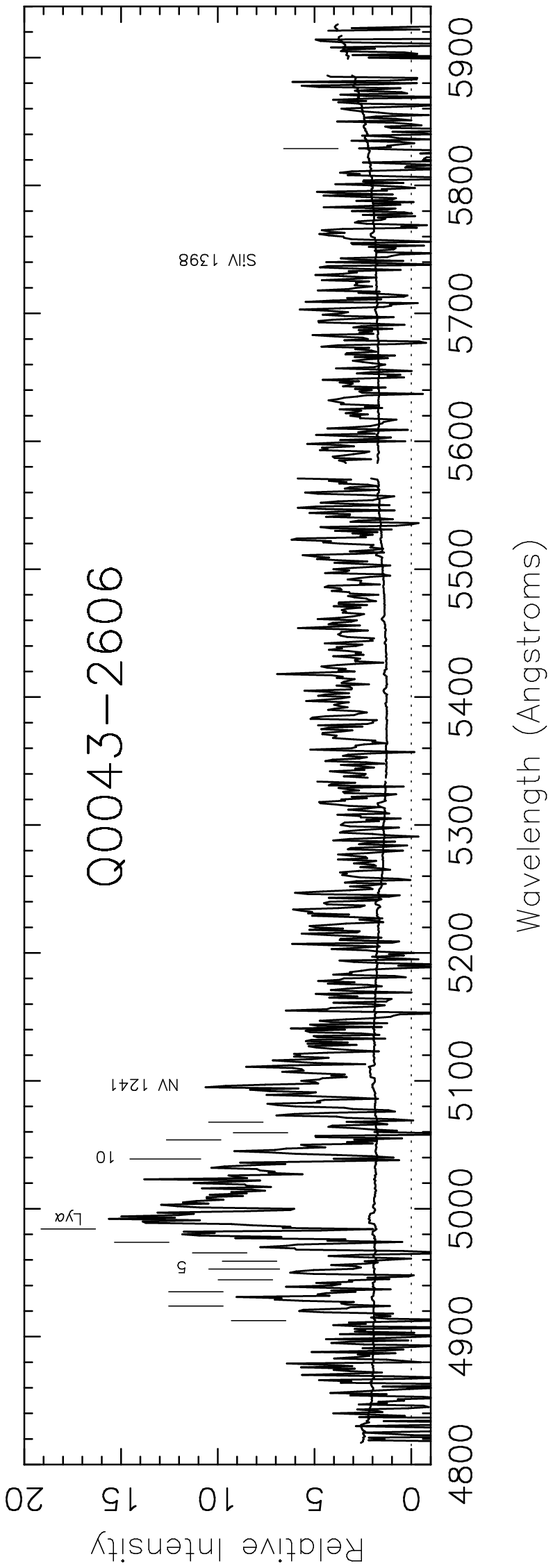 }{6.0in}{0.}{90.}{90.}{-270}{-160}
\vspace{4cm}
\caption{18 of 25}
\end{figure}

\setcounter{figure}{1}
\begin{figure}[]
\epsscale{1.0}
\plotfiddle{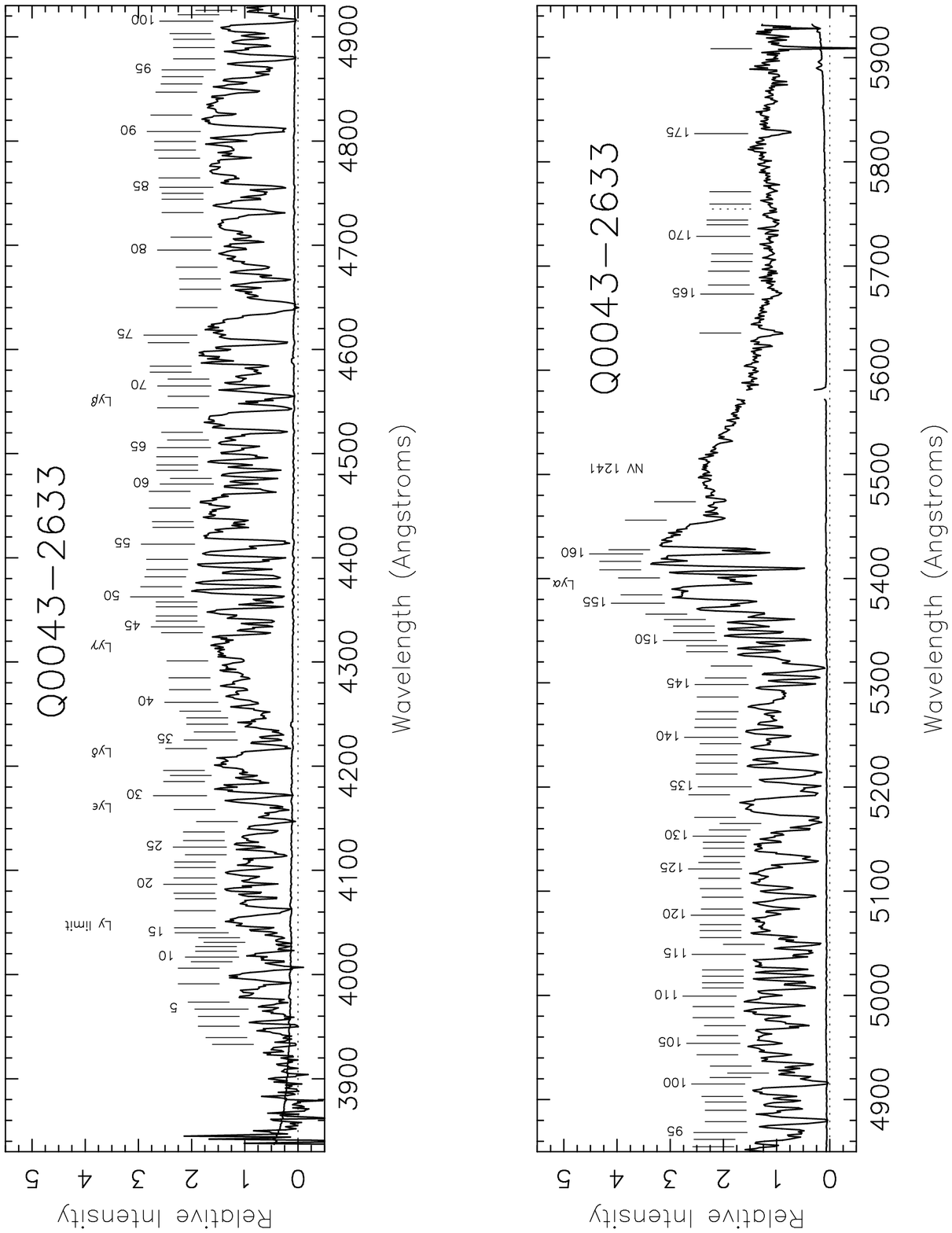 }{6.0in}{0.}{90.}{90.}{-270}{-160}
\vspace{4cm}
\caption{19 of 25}
\end{figure}

\setcounter{figure}{1}
\begin{figure}[]
\epsscale{1.0}
\plotfiddle{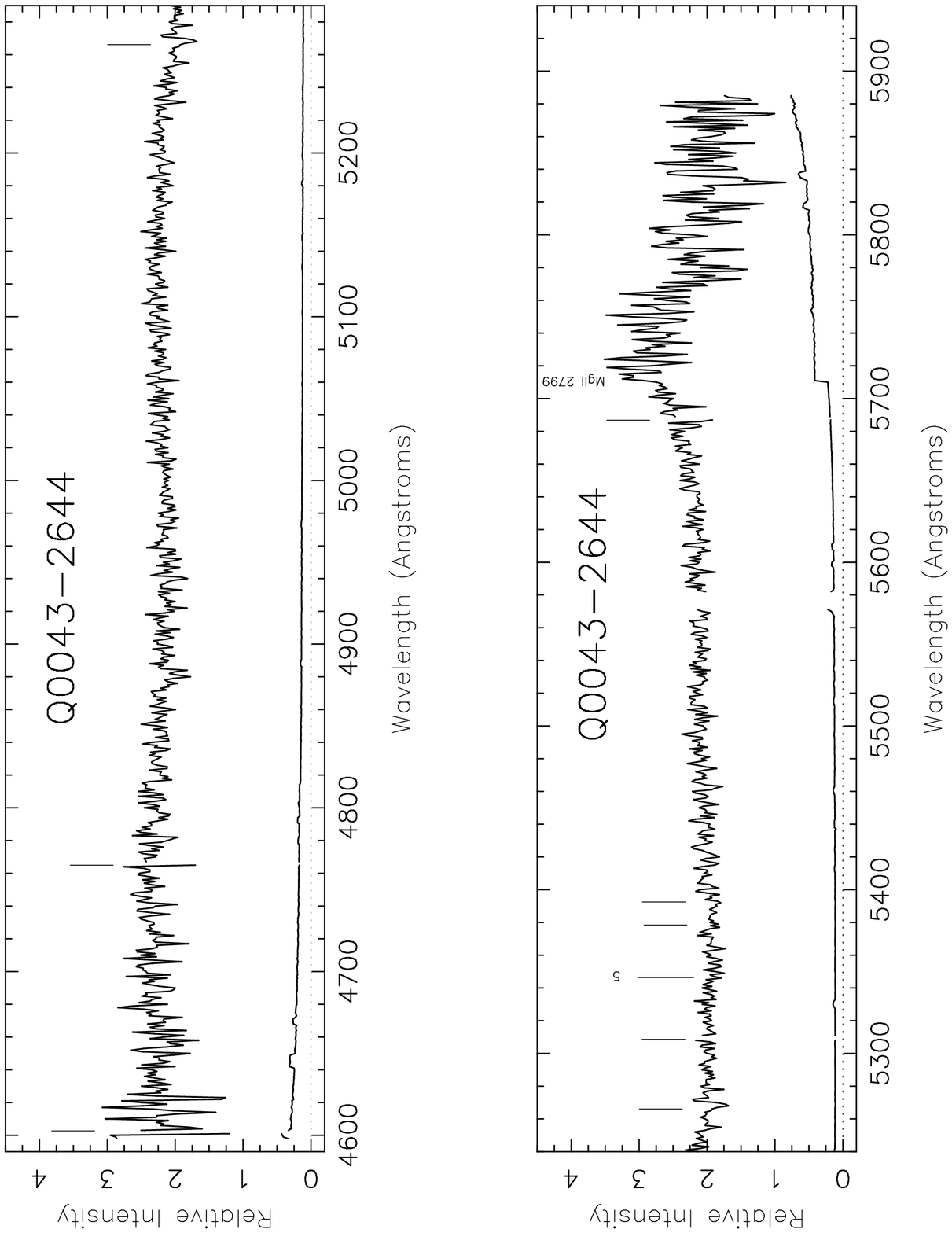 }{6.0in}{0.}{90.}{90.}{-270}{-160}
\vspace{4cm}
\caption{20 of 25}
\end{figure}

\setcounter{figure}{1}
\begin{figure}[]
\epsscale{1.0}
\plotfiddle{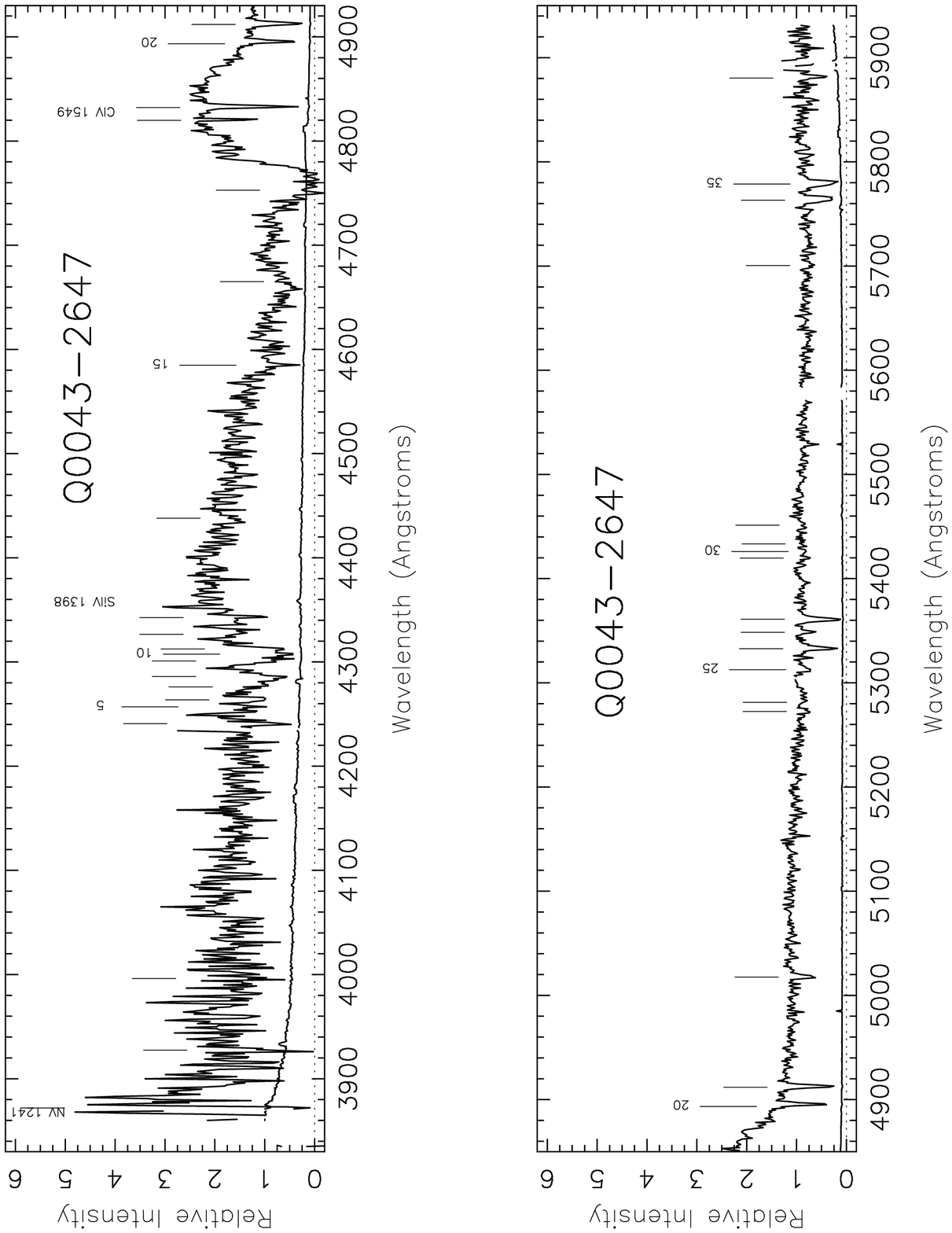 }{6.0in}{0.}{90.}{90.}{-270}{-160}
\vspace{4cm}
\caption{21 of 25}
\end{figure}

\setcounter{figure}{1}
\begin{figure}[]
\epsscale{1.0}
\plotfiddle{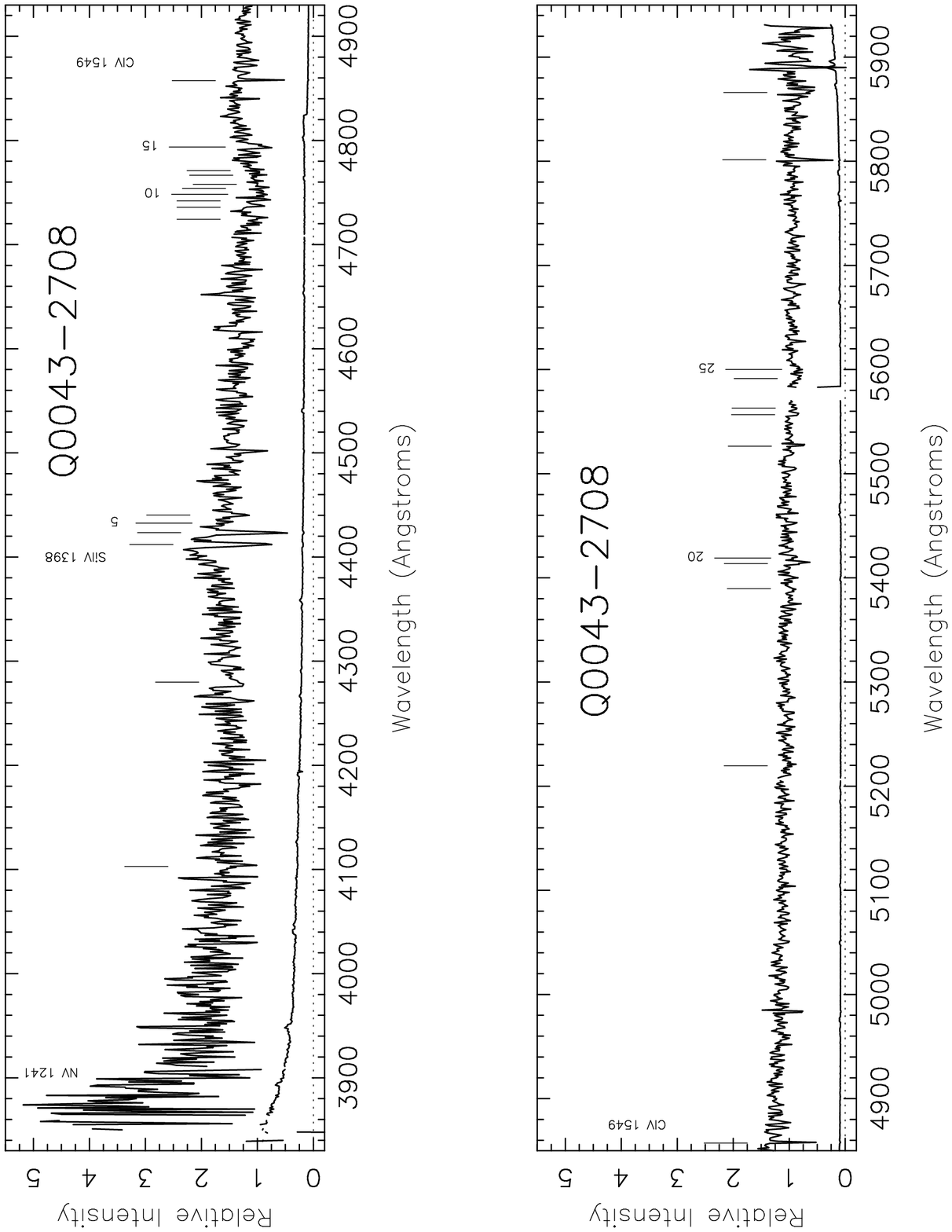 }{6.0in}{0.}{90.}{90.}{-270}{-160}
\vspace{4cm}
\caption{22 of 25}
\end{figure}

\setcounter{figure}{1}
\begin{figure}[]
\epsscale{1.0}
\plotfiddle{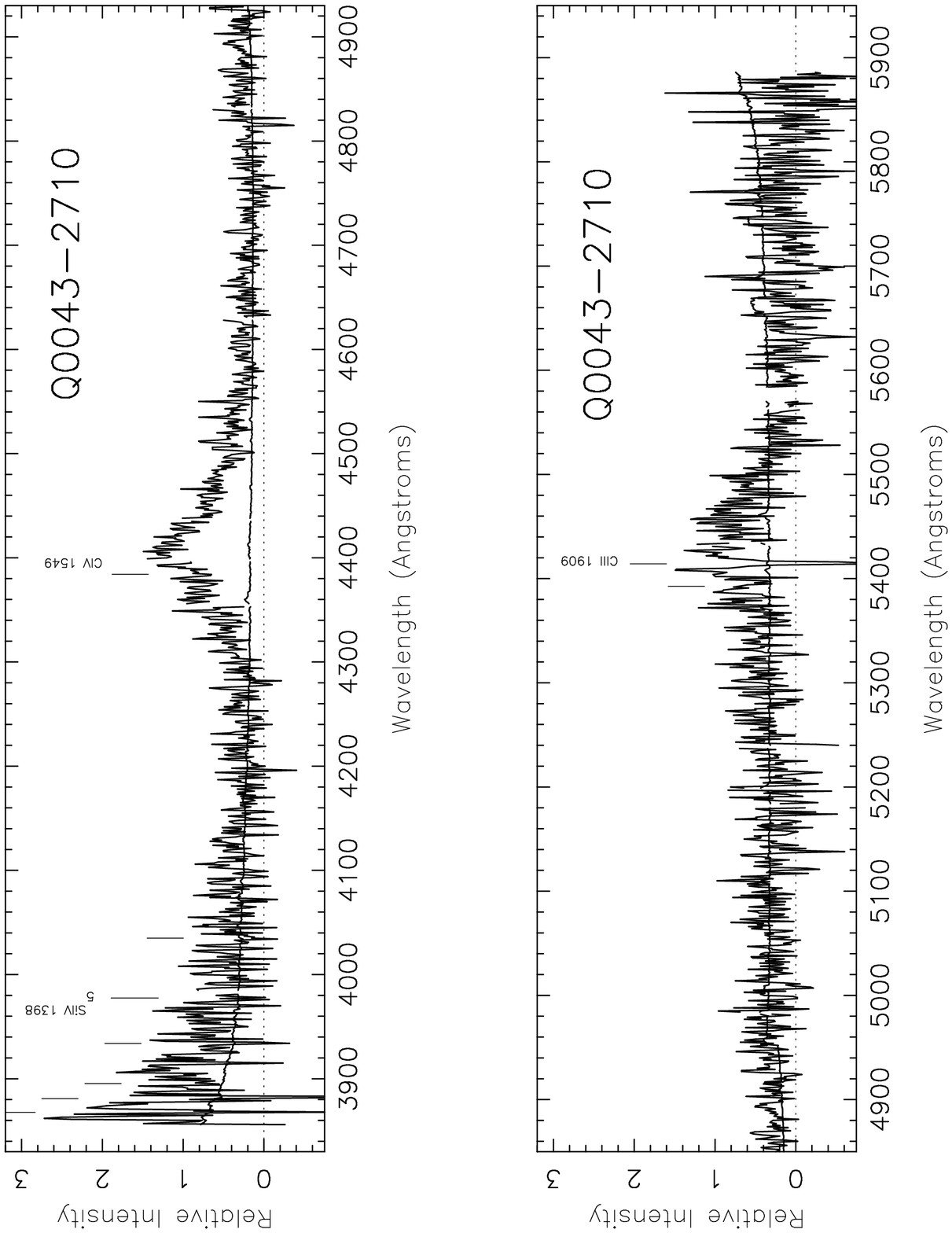 }{6.0in}{0.}{90.}{90.}{-270}{-160}
\vspace{4cm}
\caption{23 of 25}
\end{figure}

\setcounter{figure}{1}
\begin{figure}[]
\epsscale{1.0}
\plotfiddle{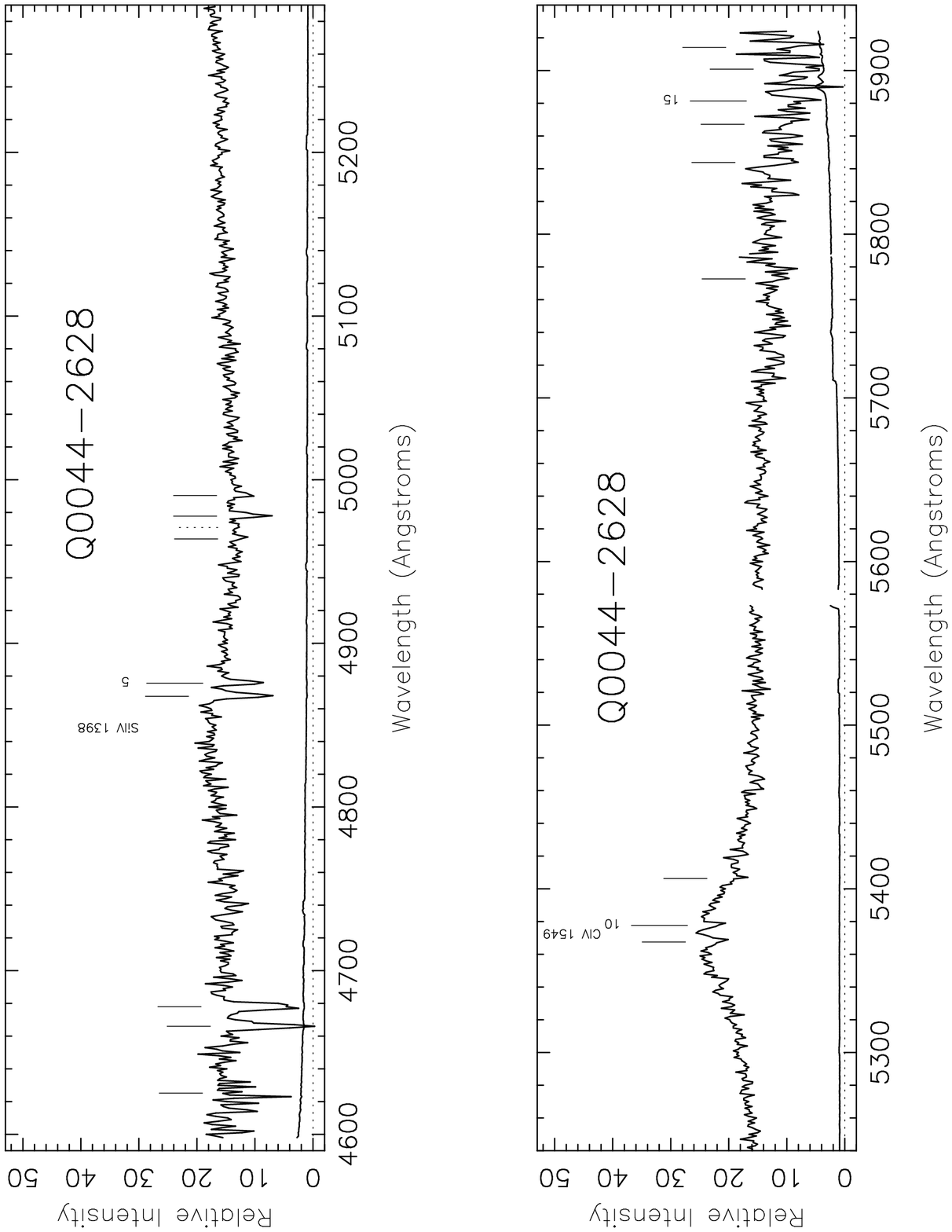 }{6.0in}{0.}{90.}{90.}{-270}{-160}
\vspace{4cm}
\caption{24 of 25}
\end{figure}

\setcounter{figure}{1}
\begin{figure}[]
\epsscale{1.0}
\plotfiddle{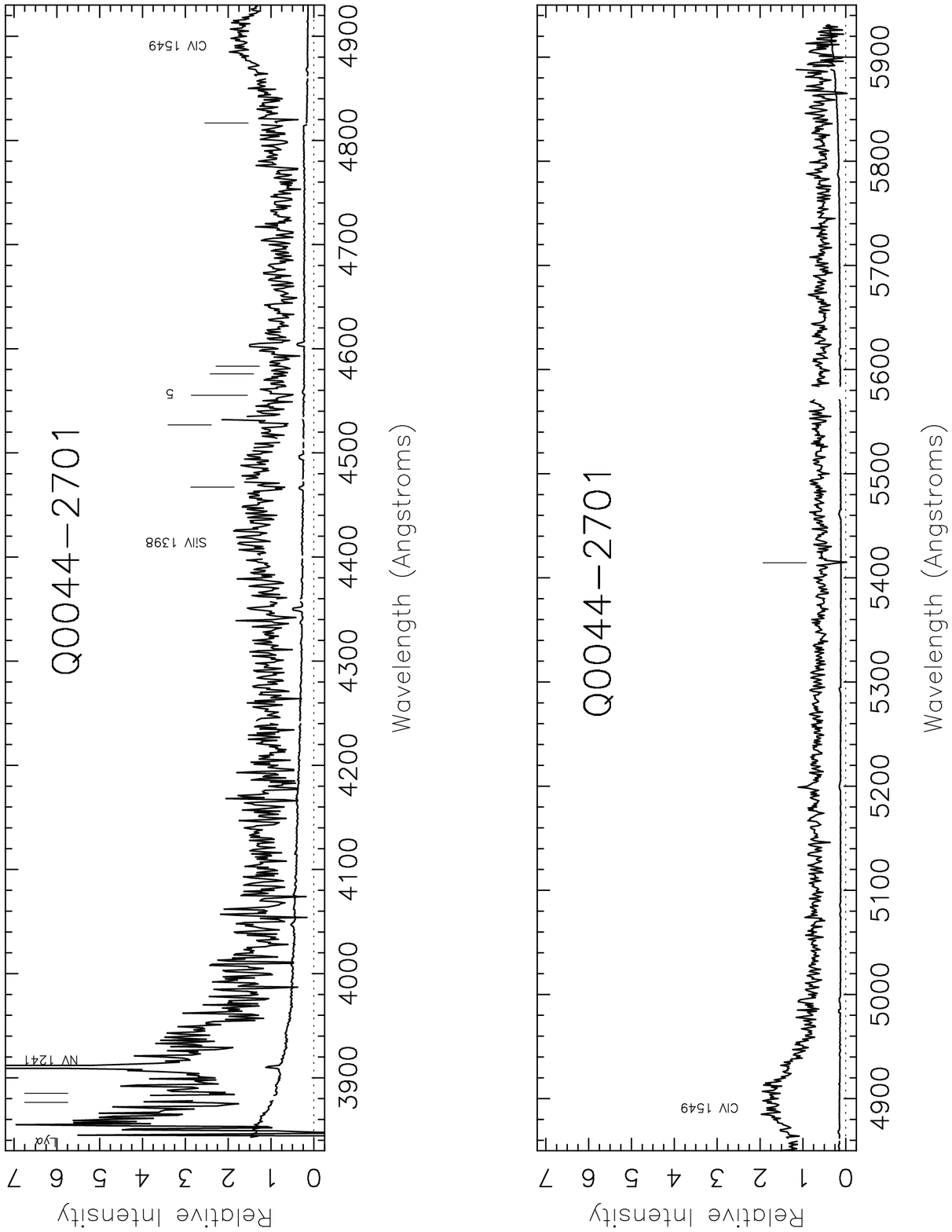 }{6.0in}{0.}{90.}{90.}{-270}{-160}
\vspace{4cm}
\caption{25 of 25}
\end{figure}

%\include{arguspapertab}

%\end{document}

\begin{figure}[]
\epsscale{1.0}
\plotfiddle{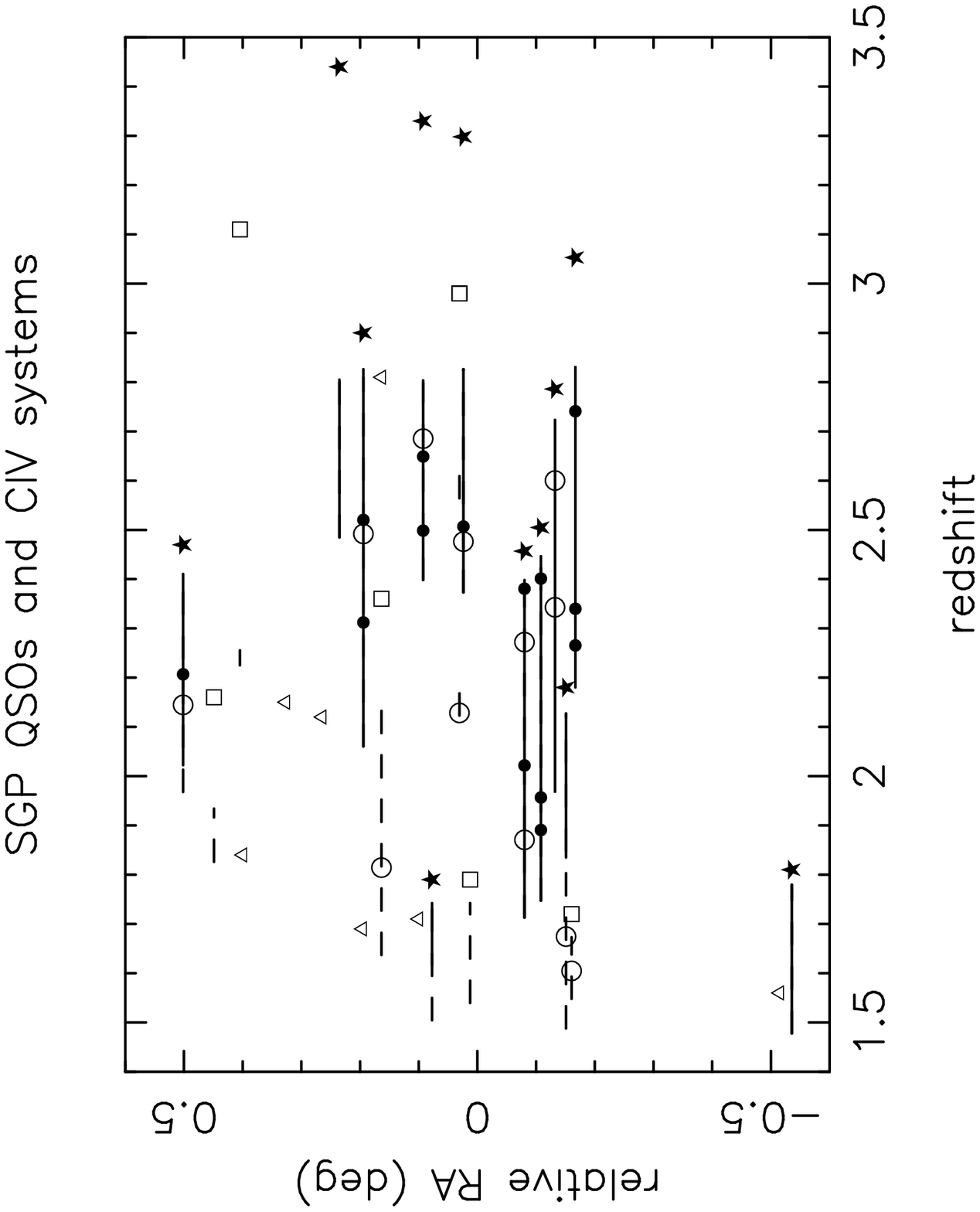}{7.0in}{0.}{110.}{110.}{-380}{-190}
\vspace{2cm}
\caption{}
\end{figure}

\begin{figure}[phtb]
\epsscale{1.0}
\plotfiddle{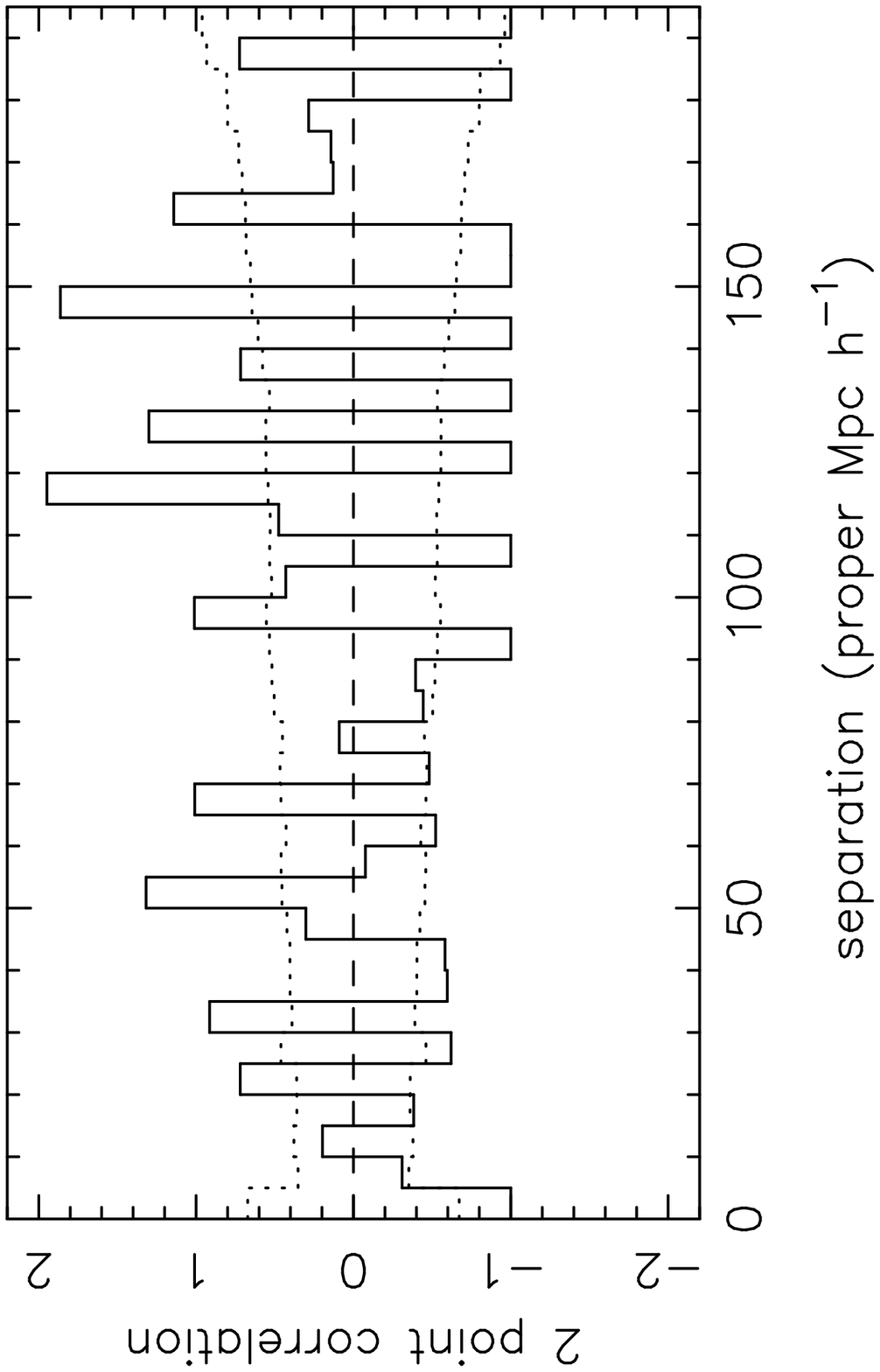}{3.0in}{-90.}{80.}{80.}{-230}{375}
\vskip 0.50in
\plotfiddle{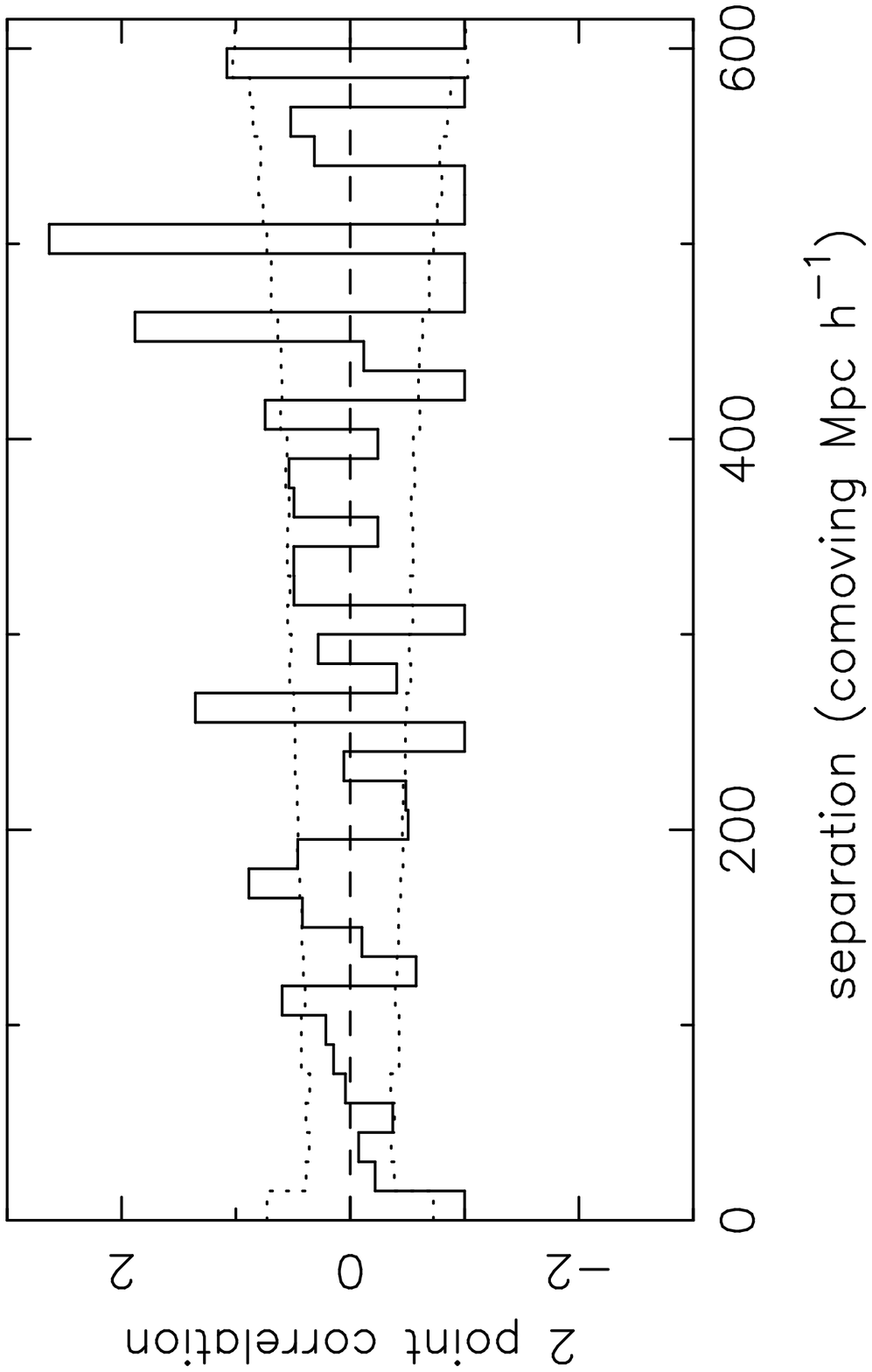}{3.0in}{-90.}{80}{80}{-230}{300}
\vspace{4cm}
\caption{}
\end{figure}

\begin{figure}[phtb]
\epsscale{1.0}
\plotfiddle{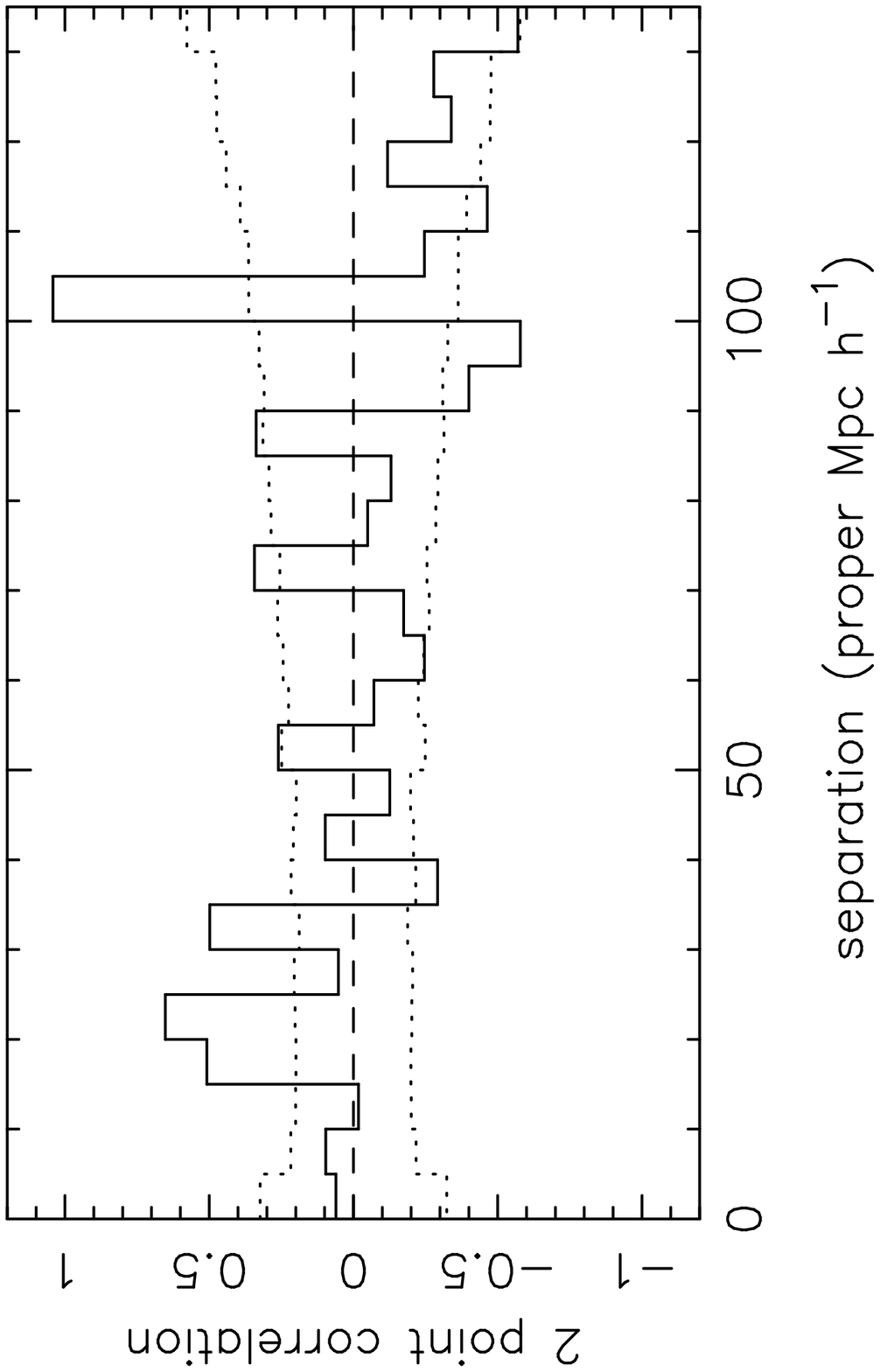}{3.0in}{-90.}{80.}{80.}{-230}{375}
\vskip 0.50in
\plotfiddle{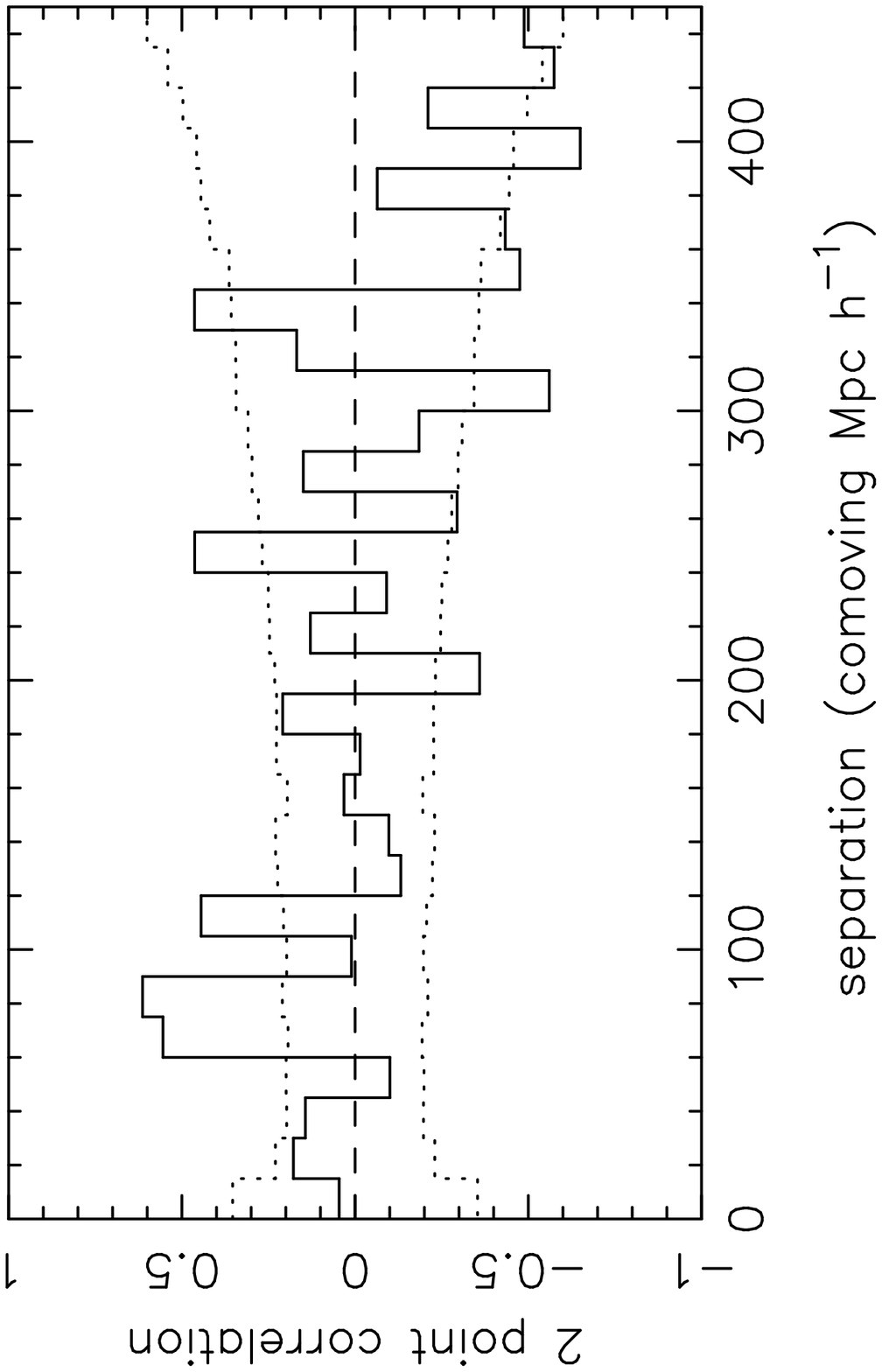}{3.0in}{-90.}{80}{80}{-230}{300}
\vspace{4cm}
\caption{}
\end{figure}

\begin{figure}[phtb]
\epsscale{1.0}
\plotfiddle{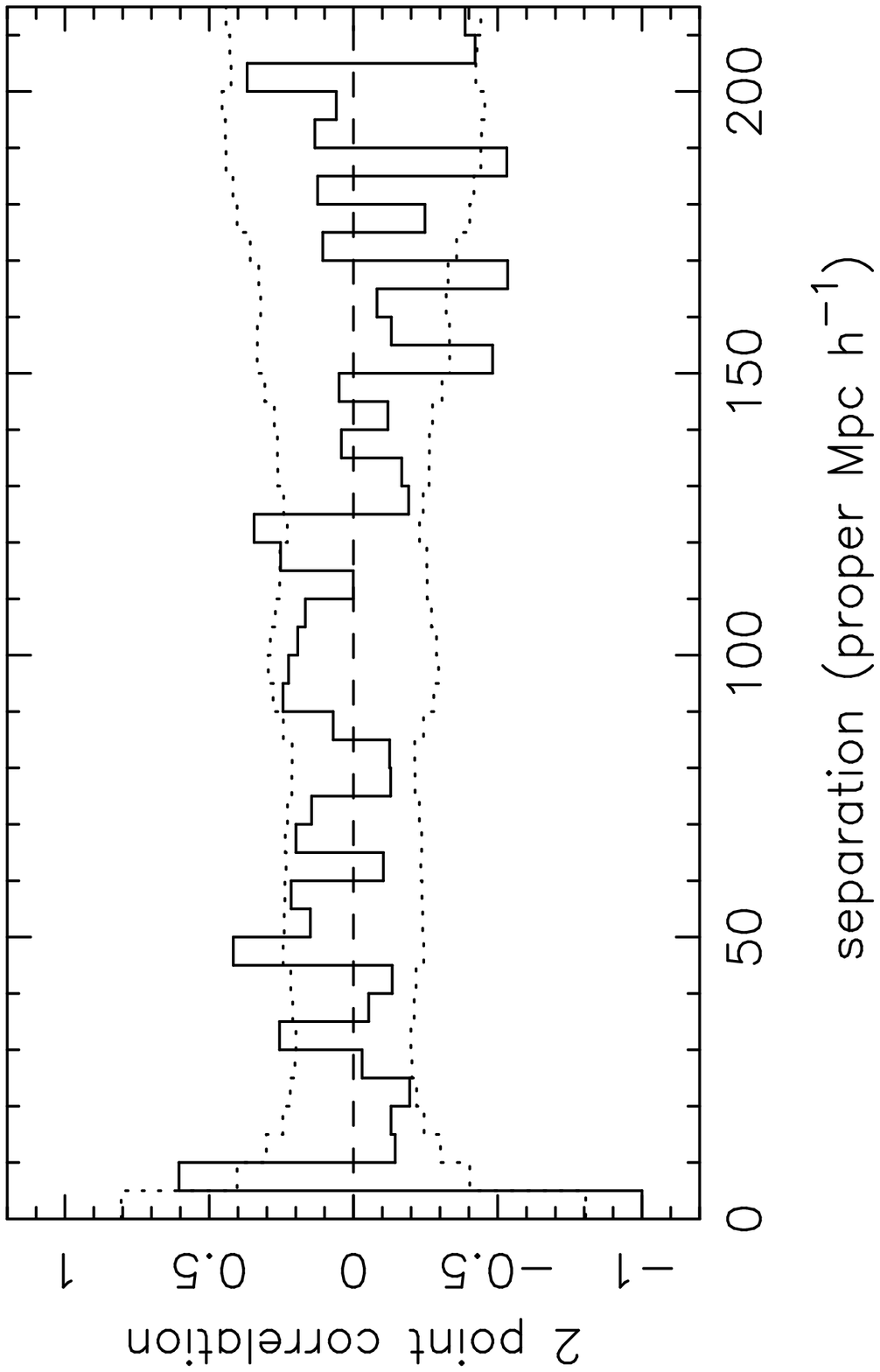}{3.0in}{-90.}{80.}{80.}{-230}{375}
\vskip 0.50in
\plotfiddle{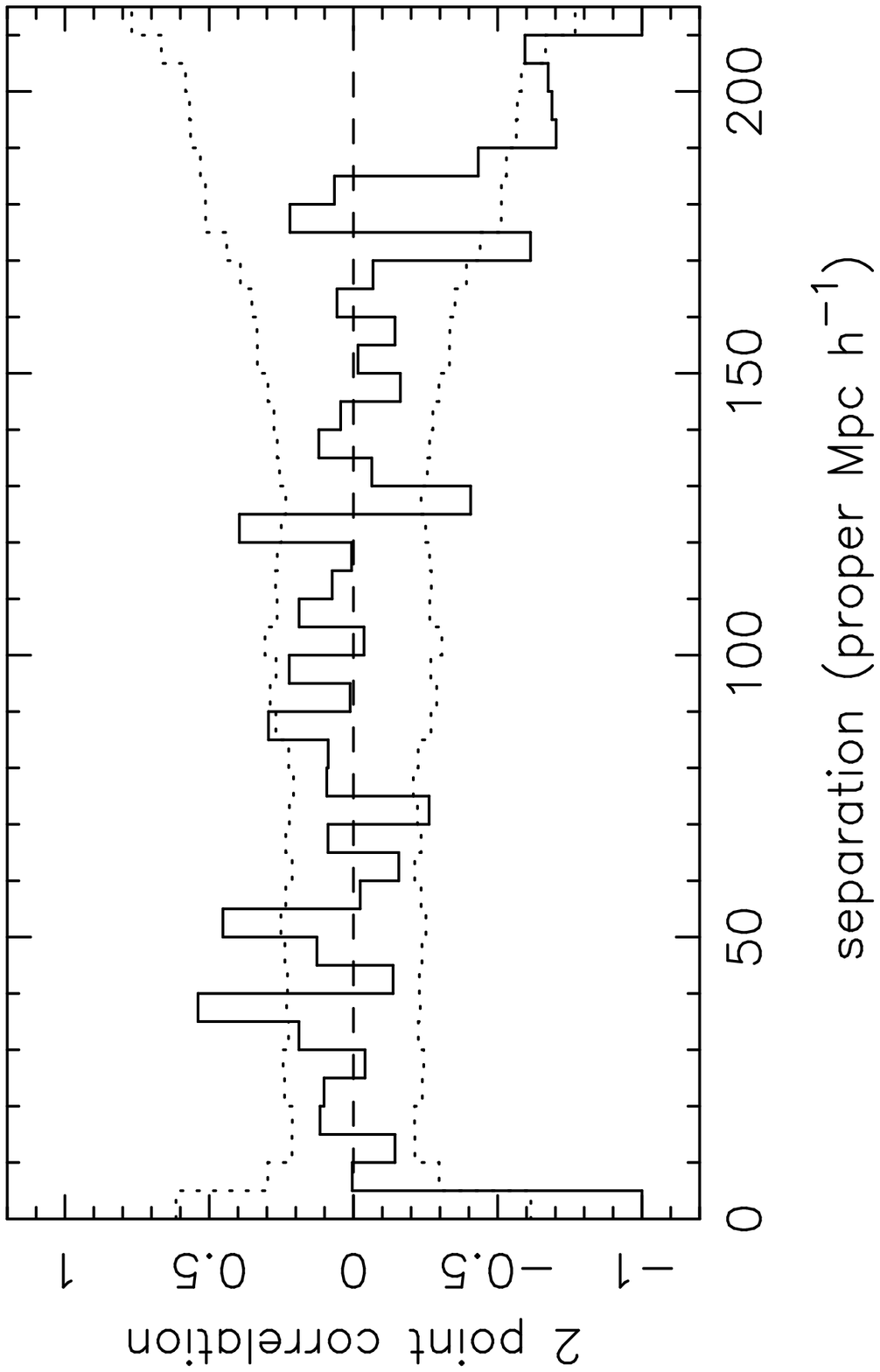}{3.0in}{-90.}{80}{80}{-230}{300}
\vspace{4cm}
\caption{}
\end{figure}

\begin{figure}[phtb]
\epsscale{1.0}
\plotfiddle{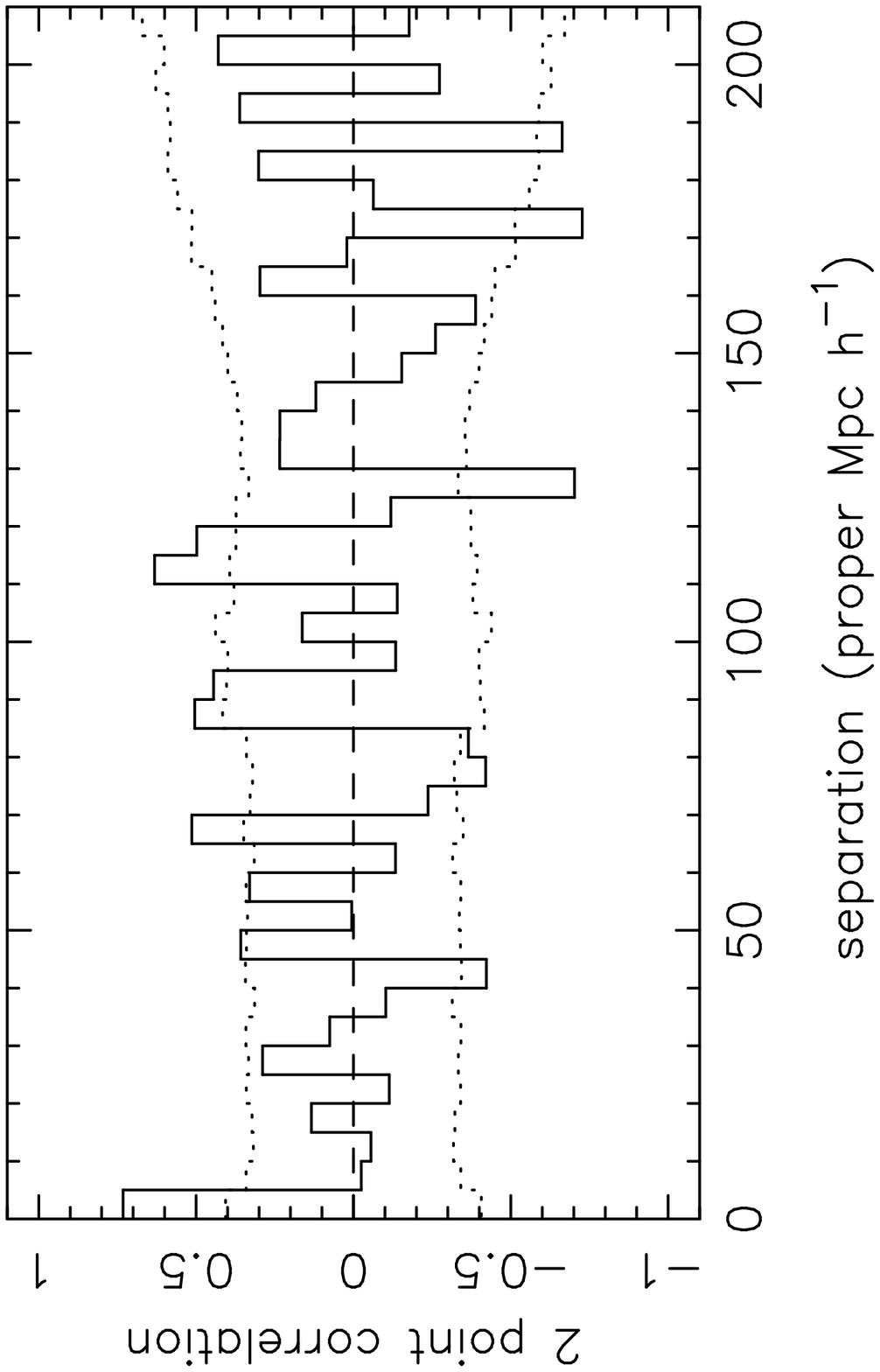}{3.0in}{-90.}{80.}{80.}{-230}{375}
\vskip 0.50in
\plotfiddle{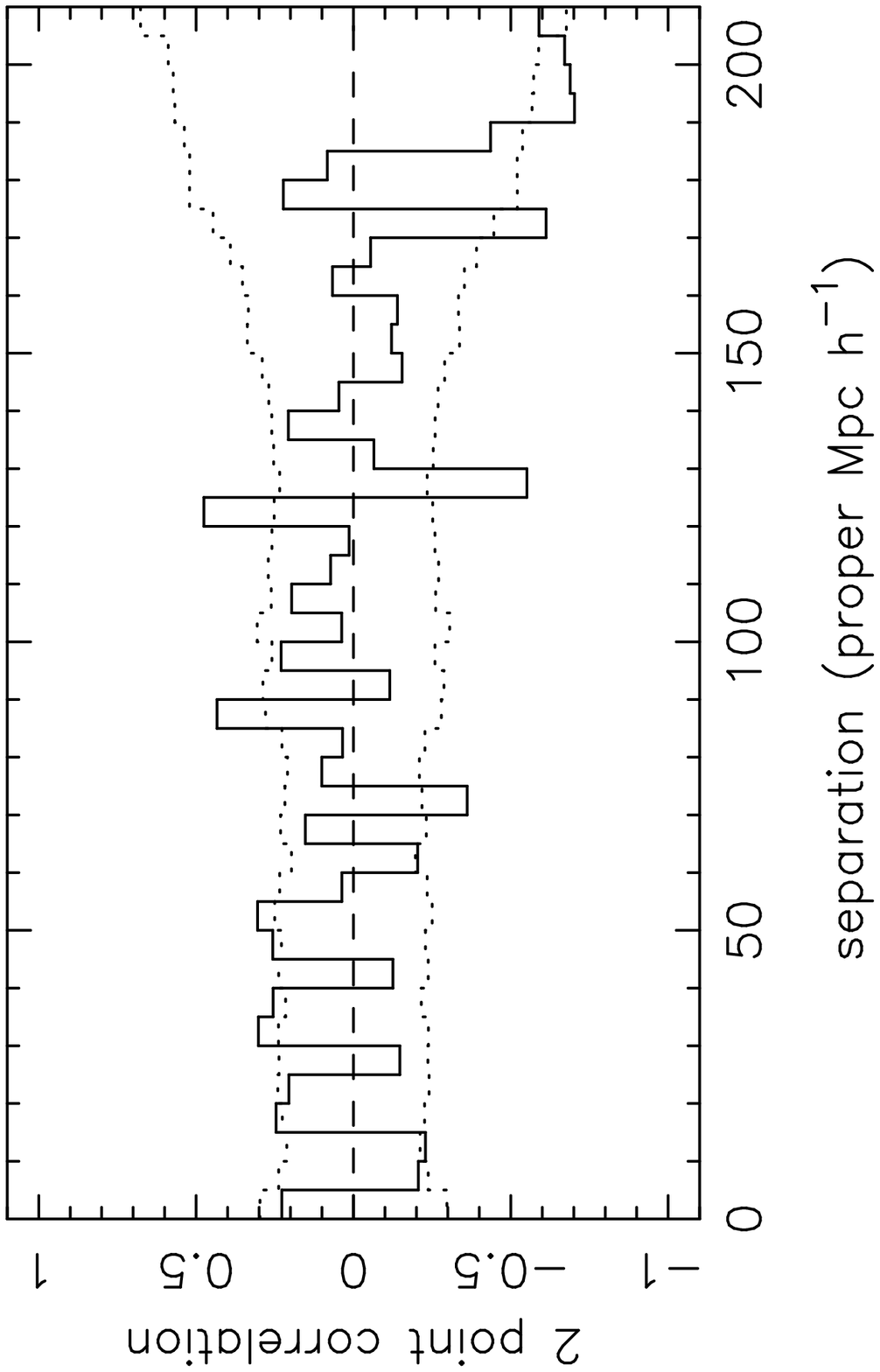}{3.0in}{-90.}{80}{80}{-230}{300}
\vspace{4cm}
\caption{}
\end{figure}

\begin{figure}[phtb]
\epsscale{1.0}
\plotfiddle{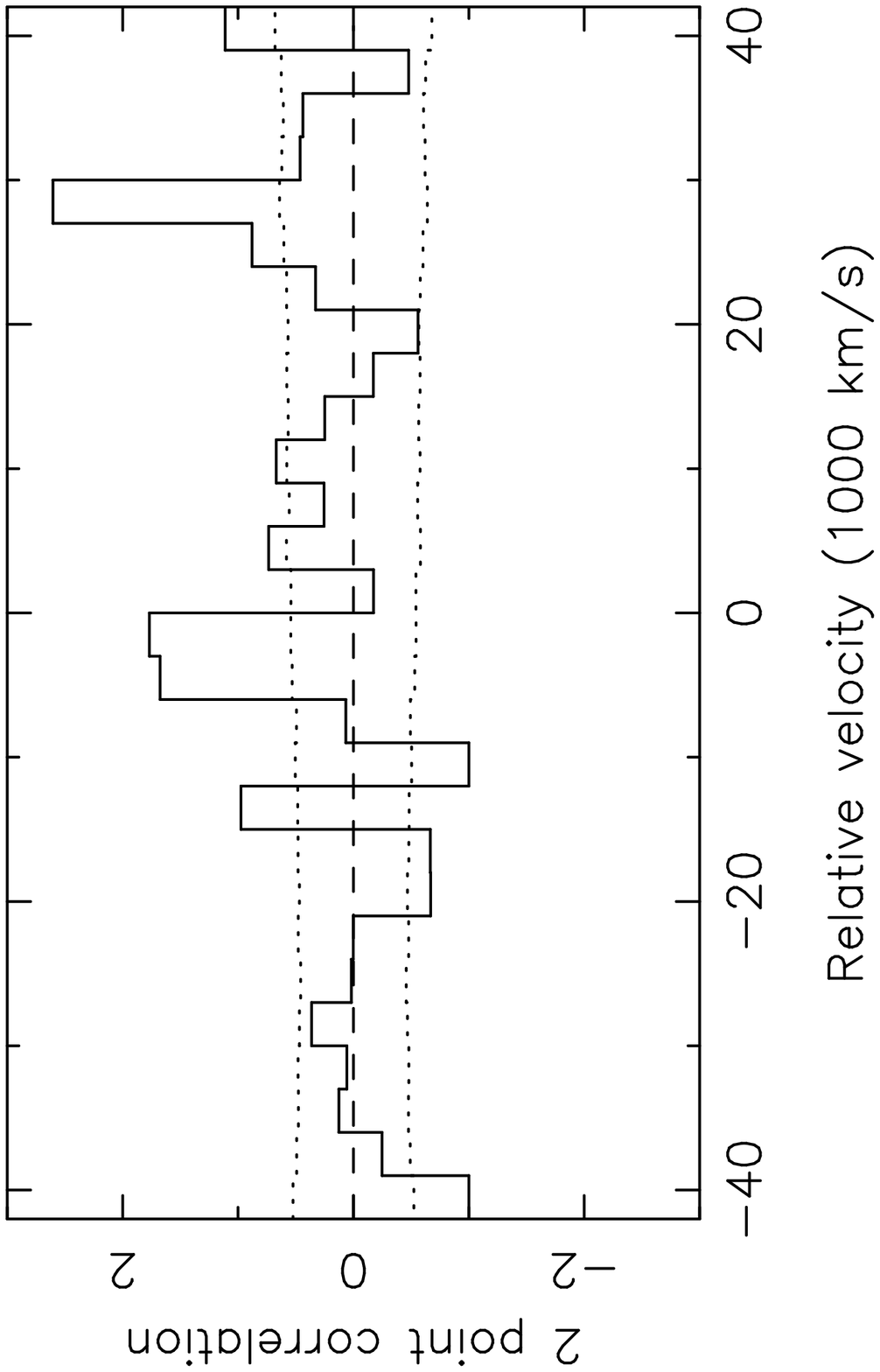}{3.0in}{-90.}{80.}{80.}{-230}{375}
\vskip 0.50in
\plotfiddle{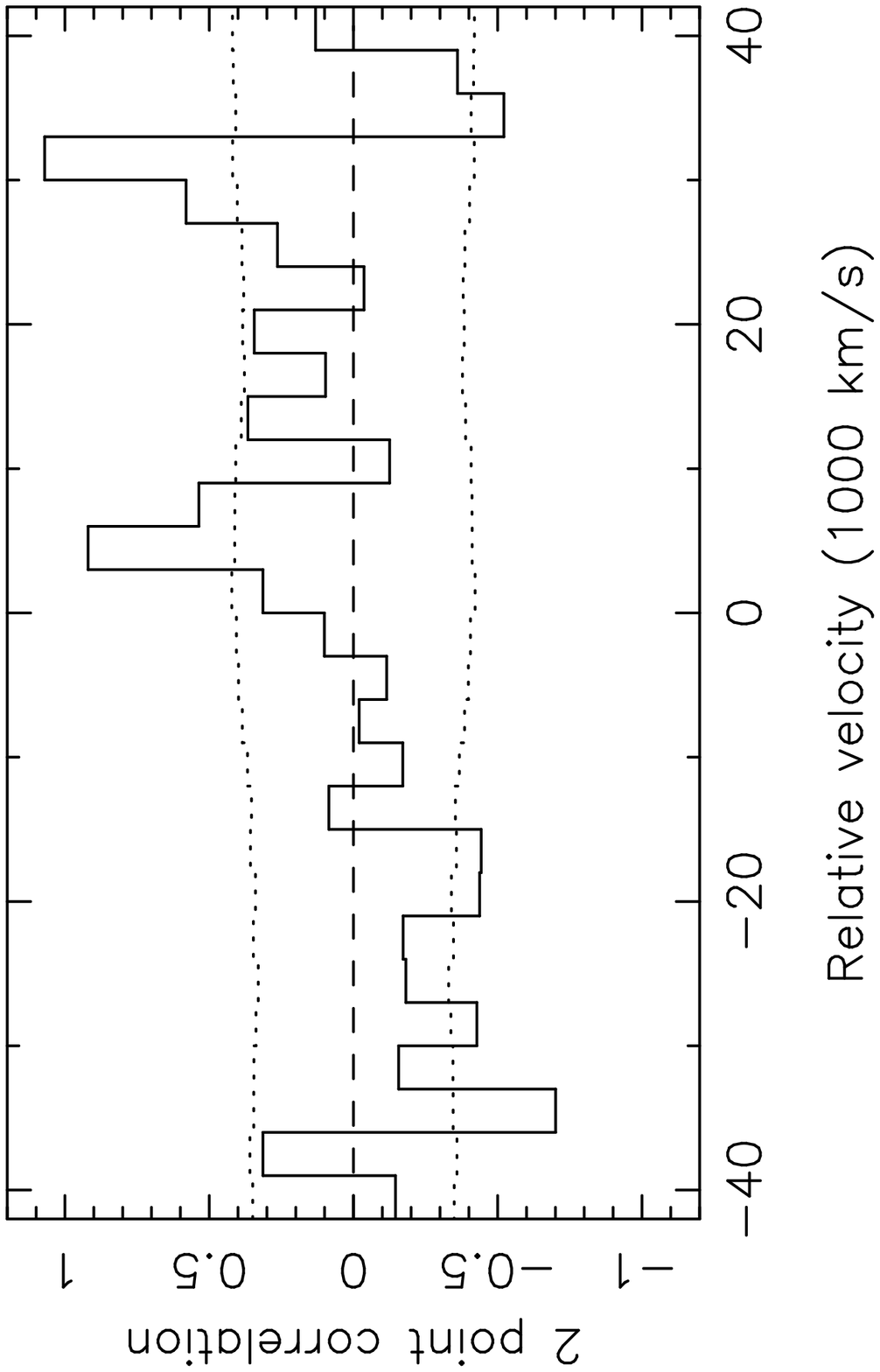}{3.0in}{-90.}{80}{80}{-230}{300}
\vspace{4cm}
\caption{}
\end{figure}

\end{document}